\documentclass[twocolumn,
showpacs,
showkeys,
preprintnumbers,
nofootinbib,
superscriptaddress,
amsmath,
amssymb,
floatfix,
secnumarabic,
aps,
pra,
a4paper,
notitlepage,
final,
]{revtex4}%

\usepackage[colorlinks=true,urlcolor=blue]{hyperref}
\usepackage{graphicx}
\usepackage{epsfig}

\newcommand{\beq}{\begin{equation}}
\newcommand{\eeq}{\end{equation}}
\newcommand{\beqar}{\begin{eqnarray}}
\newcommand{\eeqar}{\end{eqnarray}}
\newcommand{\bal}{\begin{aligned}}
\newcommand{\eal}{\end{aligned}}

\def\tr{\hbox{ tr }}
\def\dalam{\hbox
{\vrule\vbox{\hrule\hbox to 1ex{ \hfill}\kern 1 ex\hrule}\vrule}}

\def\1/2{\hbox{$ {1 \over 2}$ }}
\def\tr{\hbox{Tr}}

\def\cM{\mathcal{M}}
\def\cW{\mathcal{W}}
\def\cK{\mathcal{K}}
\def\cI{\mathcal{I}}

\def\h{\hbar}
\def\i/h{{i \over \h}}

\def\inf{\infty}
\def\pd{\partial} 
\def\v{\vec}

\def\a{\alpha} 
\def\b{\beta} \def\wtB{\widetilde {B}}
 \def\wtC{\widetilde {C}}

\def\g{\gamma} \def\G{\Gamma} 
\def\d{\delta} \def\D{\Delta}
\def\l{\lambda} 
\def\e{\epsilon} \def\E{\hbox{$\cal E $}}

\def\ve{\varepsilon}
\def\s{\sigma}
\def\r{\rho} 
\def\x{\xi}
 
\def\vf{\varphi}
 \def\F{\Phi}
 
\def\p{\psi}

\def\n{\nu}

  \def\vk{\varkappa}

\def\z{\zeta}

\def\tt{\theta}

\def\<{\langle}
\def\>{\rangle}

\def\({\left(}
\def\[{\left[}
\def\){\right)}
\def\]{\right]}

\usepackage{subfigure}

\usepackage[notcite,notref,color]{showkeys}  %Для отображения содержимого label{} к формулам, рисункам, таблицам, литературе  %(чтобы не переименовывать формулы, перед отправкой в журнал пакет отключить)

\usepackage{tikz}
\usetikzlibrary{decorations.pathreplacing}
\usetikzlibrary{decorations.pathmorphing}    %Для рисования контуров интегрирования 2
\usepackage{multirow}
\usepackage{dcolumn}		%выравнивание по точке в таблицах, использовать d вместо c в спецификации колонки
\newcolumntype{.}{D{.}{.}{-1}}
\newcolumntype{i}[1]{D{.}{.}{#1}}

\newcommand{\myfrac}[2]{{\ifmmode{}^{#1}\!/_{\!#2}\else${}^{#1}\!/_{\!#2}$\fi}}

\begin{document}
\sloppy

\title{Essentially non-perturbative and peculiar polarization effects in planar QED with strong coupling}

\author{Yu.~Voronina}
\email{voroninayu@physics.msu.ru} \affiliation{Department of Physics and
Institute of Theoretical Problems of MicroWorld, Moscow State
University, 119991, Leninsky Gory, Moscow, Russia}

%%remember to change the email and affiliation!!!!!
\author{K.~Sveshnikov}
\email{costa@bog.msu.ru} \affiliation{Department of Physics and
Institute of Theoretical Problems of MicroWorld, Moscow State
University, 119991, Leninsky Gory, Moscow, Russia}

%%remember to change the email and affiliation!!!!!
\author{P.~Grashin}
\email{grashin.petr@physics.msu.ru} \affiliation{Department of Physics and
Institute of Theoretical Problems of MicroWorld, Moscow State
University, 119991, Leninsky Gory, Moscow, Russia}

%%remember to change the email and affiliation!!!!!
\author{A.~Davydov}
\email{davydov.andrey@physics.msu.ru} \affiliation{Department of Physics and
Institute of Theoretical Problems of MicroWorld, Moscow State
University, 119991, Leninsky Gory, Moscow, Russia}

%Remember to change the date
\date{\today}

%%%%%%%%%%%%%%%%%%%

\begin{abstract}

The essentially non-perturbative  polarization effects   are considered for  a planar supercritical Dirac-Coulomb system with strong coupling (similar to graphene and graphene-based heterostructures) in terms of  induced charge density $\r_{VP}(\v r)$. The main attention is paid to the  renormalization, convergence of the partial expansion  and the behavior of $\r_{VP}(\v r)$ and the integral induced charge  $Q_{VP}$ in the overcritical region.  The dependence of the induced density  on the screening of the Coulomb asymptotics of the external source  is also explored in detail. Some peculiar effects in the discrete spectrum with the lowest rotational numbers $m_j=\pm 1/2\, , \pm3/2$ in the screened case are detected and their possible role in the transition through corresponding $Z_{cr}$ is also discussed.

\end{abstract}

%%pacs  numbers should be changed - see http://publish.aps.org/PACS
%\pacs{12.20.Ds, 31.15.A-, 31.30.J-, 34.10.+x}

\maketitle
%\tableofcontents
%%%%%%%%%%%%%%%%%%%%

%%%%%%%%%%%%%%%%%%%%%%%
\section{Introduction}\label{sec:intro}

There is now a lot of interest to the study of various 2+1 QED-effects in graphene-based planar heterostructures. It is known that the charge carriers in graphene are described as massless (or massive on a substrate) relativistic fermions, what leads to an intriguing analogy between the physics of graphene and that of QED. Moreover, the effective fine-structure constant $ \a_g \sim 1$ in graphene turns out to be much larger than in the ``normal'' 3+1 QED \cite{CastroNeto2009},\cite{Reed2010}. Due to such a large value of $ \a_g$ it is much easier to observe many non-trivial QED-effects experimentally. In particular,  the critical charges of atomic collapse in graphene are subject of condition $Z \a_g > 1/2$ \cite{Shytov2009},\cite{Shytov2007}, the observation of the Klein paradox requires electric fields $\sim 10^5\,\, \text{V/cm}$ (eleven orders of magnitude less than the fields necessary for the observation of
the Klein paradox for elementary particles) \cite{Katsnelson2006a}, the quantum Hall effect  can be observed for much higher temperatures and lower magnetic fields than in the conventional semiconductors \cite{Gusynin2005, Giesbers2008, Cobaleda2011}. Some effects turn out to be strong enough to affect the transport properties of graphene. For instance, highly charged impurities in graphene exhibit resonances which should manifest themselves via various transport properties, such as the transport scattering cross-section \cite{Shytov2009}. The Klein paradox plays an important role in transport properties of different graphene systems \cite{Katsnelson2006a}: graphene p-n-p junctions \cite{Shytov2009},\cite{Shytov2008}, twisted graphene bilayer \cite{He2013}. The Dirac-like dynamics of graphene  results also in an unconventional form of the Hall quantization \cite{Gusynin2005}.  The main feature inherent in all these effects is that they are essentially non-perturbative due to the large value of $ \a_g$ and therefore cannot be described within the perturbation theory (PT).

In this work we explore another essentially non-perturbative effect in the two-dimensional strongly coupled  QED with application  to graphene-like planar systems, namely, the vacuum polarization, caused by diving of discrete levels into the lower continuum in the  supercritical static or adiabatically slowly varying Coulomb fields, which are created by localized extended sources with $Z> Z_{cr}$.  Such effects have  attracted a considerable amount of theoretical and experimental activity in  3+1 D heavy ions collisions, where for $Z > Z_{cr,1} \simeq 170$ a non-perturbative reconstruction of the vacuum state is predicted, which should be accompanied by a number of nontrivial effects including the vacuum positron emission (\cite{Greiner1985a, Plunien1986, Ruffini2010, Greiner2012, Rafelski2016} and refs. therein).

 Similar phenomena could occur in graphene with the charge impurities acting as atomic nuclei, while the graphene itself -- as the QED vacuum and its  electrons and holes  --- as the relativistic virtual particles which populate the vacuum. A remarkable circumstance here is that due to the large value of the effective fine-structure constant these effects should take place for relatively small impurity charges $Z \simeq 1-10$. Since for these effects the charge carriers in graphene play the role of the virtual QED-particles, the induced charge density can be measured directly. In Ref.~\cite{Wang2013}, the five-dimer cluster consisting of Ca-atoms was used as a charge impurity and the induced density was measured via STM. Polarization effects in graphene, caused by charged impurities, have also been considered by many authors  (\cite{Katsnelson2006b, Biswas2007, Pereira2007, Kotov2008, Terekhov2008, Nishida2014, Khalilov2017} and refs. therein). Here it should be noted that in most cases the impurity is  modeled  as a point-like charge, what causes some problems in the supercritical case. Our work is aimed mainly at the study of vacuum polarization effects, caused by extended supercritical Coulomb sources with non-zero size $R_0$, which provide a physically clear and unambiguous problem statement like in Refs.~\cite{Shytov2007, Fogler2007, Milstein2010}, where the charge is assumed to be displaced away or smeared over a finite region of the graphene plane.

The external Coulomb field $A^{ext}_{0}(\v r)$ is chosen in the form of a projection onto a plane of the potential of the uniformly charged sphere with the radius  $R_0$ and a cutoff of the Coulomb asymptotics at some $R_1>R_0$
\beq \label{1.0}
A^{ext}_{0}(\v r)=Z |e| \[\frac{1}{R_0}\tt\(R_0-r\)+\frac{1}{r}\tt \(R_0 \leq r \leq R_1\) \] \ ,
\eeq
what leads to the potential energy
\beq
\label{1.1}
V(r)= - Z \a \[\frac{1}{R_0}\tt\(R_0-r\)+\frac{1}{r}\tt \(R_0 \leq r \leq R_1\) \] \ .
\eeq
The radius of the source is taken as $R_0=a$, where $a \simeq 1.42\, A$ is the approximate C-C distance in the  graphene lattice. Such cutoff of the Coulomb potential at small distances  has been used  in \cite{Pereira2008}. The cutoffs $R_0=a/2$ and $R_0=2a$ are also considered. The screening of the Coulomb asymptotics  is taken  in the form  the simplest shielding via vertical wall for $r \geqslant R_1> R_0$ , which allows to perform the most part of calculations in the analytical form. However, even such type of  screening reveals some peculiar features, which are quite different from the unscreened one and are absent in the similar one- or three-dimensional DC systems. The external cutoff $R_1$ will be taken as $R_1=2 R_0\, , 5R_0\, , 10 R_0\, , \inf$ for the study of screening effects and  to establish a smooth transition into the unscreened case, which  will  be considered at first.

The effective fine-structure constant is defined as
\beq
\label{1.2}
\a=e^2/(\hbar v_F \ve_{eff}) \ , \quad \ve_{eff}=(\ve +1)/2 \ ,
\eeq
with $\ve$ being  the substrate dielectric constant and $v_F = 3ta/2 \hbar$ -- the Fermi velocity in graphene. In its turn, $t$ is the hopping amplitude, while  $\l_c = {\hbar / m v_F}$  is the effective Compton length \cite{Goerbig2011}. Here  $m$ denotes the effective fermion mass, which is related to the local energy mismatch in the tight-binding formulation through the relation $\D=2mv_F^2$. These definitions lead to the relation $\l_c/a \simeq 3t/\D$. In this work we consider $\a=0.4$ (which corresponds to  graphene on the SiC substrate \cite{Pereira2008}) and $\a=0.8$ (graphene on the h-BN substrate \cite{Goerbig2011, Sadeghi2015}).

Henceforth the system of units in which $\hbar=v_F=m=1$ is used, and so the distances are measured in units of $\l_c$, while the energy --- in units of $mv_F^2$.   For $\a=0.4$ the local energy mismatch is $\D=0.26$ eV and therefore for $R_0=a/2,a,2a$ one obtains  $R_0=1/60,1/30,1/15$ in the  units chosen, while for  $\a=0.8$ one has $\D=0.056$ eV and so $R_0=1/350,1/175,2/175$.

\section{The perturbative approach to 2+1 QED for an extended Coulomb source with unscreened asymptotics }\label{sec:PT}

In 2+1 QED the induced charge density to the lowest order of PT is
 determined from the vacuum polarization (Uehling) potential
\beq
\label{2.2}
\r^{(1)}_{VP}(\vec{r})=-\frac{1}{4 \pi} \D_{2}\, A^{(1)}_{VP,0}(\vec{r}) \ ,
\eeq
where $\D_2$ is the two-dimensional Laplace operator.
In its turn, the Uehling potential $A^{(1)}_{VP,0}$ is expressed in terms of the renormalized polarization function $\Pi_{R}(-\vec{q}\,^2)$ and the Coulomb potential of external source in the momentum space  $\widetilde{A}_{0}(\vec{q})$ ~\cite{Greiner2012}
\beq
\label{2.3}
\bal
A^{(1)}_{VP,0}(\vec{r})&=\frac{1}{(2 \pi)^2} \int d^2q\,\mathrm{e}^{i \vec{q} \vec{r}} \Pi_{R}(-q^2)\widetilde{A}_{0}(\vec{q}), \\
\widetilde{A}_{0}(\vec{q})&=\int d^2 r'\,\mathrm{e}^{-i \vec{q} \vec{r\,}' }A^{ext}_{0}(\vec{r}\,' ), \quad q=|\vec{q}| \ ,
\eal
\eeq
where
\beq
\label{2.4}
\Pi_R(-q^2)=\frac{\a}{2q} \[\frac{2 }{q} + \(1-\frac{4}{q^2}\) \arctan \( \frac{q}{2}\)\] \ ,
\eeq
 and the two-dimensional representation of the  Dirac matrices has been used (for the choice of the Dirac matrices see below in Section 3).

From (\ref{2.3}), (\ref{2.4}) for the external source (\ref{1.0}) one obtains the following expression for the Uehling potential
\begin{multline}\label{2.5}
A_{VP,0}^{(1)}(r)=\frac{Z \a |e|}{4}\int\limits_{0}^{\infty}dq\,\frac{J_0(qr)}{q} \[\frac{2 }{q} + \(1-\frac{4}{q^2}\) \times \right. \\
 \left. \times \arctan \( \frac{q}{2}\)\]
\times \(2\[1+J_1(q R_0)-q R_0 J_0(q R_0)\]+ \right. \\
 \left. + \pi q R_0\[ J_0(q R_0) \mathbf{H}_1(q R_0)- J_1(q R_0) \mathbf{H}_0(q R_0)\]\) \
\end{multline}
with  $J_{\n}(z)$ and  $\mathbf{H}_{\n}(z)$ being the Bessel and Struve functions, correspondingly.
From (\ref{2.2}) and (\ref{2.5}) the first-order perturbative  induced  density $\r^{(1)}_{VP}$ can be calculated. In order to figure out whether it is possible to insert the Laplace operator under the integral over $dq$ in (\ref{2.5}) let us consider the asymptotical behavior of the integrand for large $q$. The leading term of the integrand asymptotics in (\ref{2.5}) equals to
\beq \label{2.5a}
\frac{ \sin (q (r+R_0))+\cos (q (r-R_0))}{\sqrt{r}R_0^{3/2}\,q^{3}} \ , \quad q \to \inf \ .
\eeq
Applying the Laplace operator to (\ref{2.5a}), to the leading order one obtains
\beq\label{2.5b}
 - \frac{\sin (q (r+R_0))+\cos (q (r-R_0))}{\sqrt{r}R_0^{3/2}\,q} + O(1/q^2) \ .
\eeq
Therefore, at  $r=R_0$ the possibility of  inserting the Laplace operator under the sign of the integral in (\ref{2.5}) is absent, since in this case the integral over $d q$ in (\ref{2.5}) diverges logarithmically.

So the vacuum density, obtained from (\ref{2.2})
\begin{multline}
\label{2.5c}
\r_{VP}^{(1)}(r)=\frac{Z\a |e|}{16\pi}\int\limits_{0}^{\infty}dq\,q J_0(q r)\[\frac{2 }{q} + \(1-\frac{4}{q^2}\) \times \right. \\
 \left. \times \arctan \( \frac{q}{2}\)\]
\times \(2\[1+J_1(q R_0)-q R_0 J_0(q R_0)\] + \right. \\ \left. +\pi q R_0\[ J_0(q R_0) \mathbf{H}_1(q R_0)- J_1(q R_0) \mathbf{H}_0(q R_0)\]\) \ ,
\end{multline}
is finite for all  $r\neq R_0$ with the logarithmic singularity at $r \to  R_0$.

By means of the QED-renormalization condition ${\tilde  \Pi_R} (q^2) \sim  q^2$ for $q \to 0$ it is easy to verify that within PT to the leading order the total induced charge  vanishes exactly
\beq
\label{2.7}
\int \! d^2r \ \r^{(1)}_{VP}(r)= 0 \
\eeq
(for more details see~\cite{Davydov2018a}, App.B). The relation (\ref{2.7}) confirms the assumption that for the  external background like (\ref{1.0})  in the subcritical region with $Z<Z_{cr,1}$ the correctly renormalized total induced charge should vanish, while the  polarization effects could only distort its spatial density ~\cite{Greiner2012, Mohr1998}.  However, it is not a theorem, but just a plausible statement, which in any concrete case should be verified via direct calculations. In the case under consideration the direct check confirms (see~\cite{Davydov2018a}, App.B) that upon renormalization the induced  charge turns out to be  non-vanishing only for $Z>Z_{cr,1}$ due to non-perturbative effects, caused by diving of discrete levels into the lower continuum in accordance with Refs.~\cite{Greiner1985a, Plunien1986, Greiner2012, Rafelski2016}. This circumstance significantly affects the behavior of the Casimir energy in the overcritical region, which has been recently shown for a toy 2+1 D model in \cite{Davydov2018b}, and will be considered for DC system with current parameters in a separate work.

\section{The Wichmann-Kroll method for the induced density in the unscreened Coulomb background}\label{sec:WK}

The most effective non-perturbative approach to calculation of the induced  density $\r_{VP}(\vec{r})$ is based on the Wichmann-Kroll (WK) method~\cite{Wichmann1956}.
The starting point of the WK method is the following expression for the  induced density
\beq \label{3.1}
\r_{VP}(\vec{r})=-\frac{|e|}{2}\(\sum\limits_{\e_{n}<\e_{F}} \p_{n}(\vec{r})^{\dagger}\p_{n}(\vec{r})-\sum\limits_{\e_{n}\geqslant \e_{F}} \p_{n}(\vec{r})^{\dagger}\p_{n}(\vec{r}) \),
\eeq
with $\e_F$ being the Fermi level, which in such problems with the external background like (\ref{1.0}) should be chosen at the threshold of the lower continuum ($\e_F=-1$), while $\e_{n}$ and $\p_n(\vec{r})$ are the eigenvalues and the eigenfunctions of corresponding Dirac-Coulomb (DC) spectral problem.

The essence of the  WK method is that the induced density (\ref{3.1}) is expressed via integration of the trace of the Green function for DC spectral problem along the special contours in the complex energy plane. The Green function is defined as
\beq
\label{3.2}
\(- i \,\vec{\a}\,\vec{\nabla}+V(r)+\b -\e \)G(\vec{r},\vec{r}\,' ;\e)=\d(\vec{r}-\vec{r}\,' ).
\eeq
Here it should be mentioned  that in 2+1 QED the Dirac matrices  can be chosen either in two- or four-dimensional representations. In the first case there are two inequivalent possible choices of the matrix signature ~\cite{Hosotani1993}, while in the latter  the DC spectral problem for the external source (\ref{1.0}) splits into two independent subsystems, which are related  by $m_j \to -m_j$. Therefore the degeneracy factor of the energy eigenstates with the fixed $m_j$ equals to 2 and in what follows this factor will be shown explicitly in all the expressions for $\r_{VP}(\v r)$ and $\E_{VP}$, while the DC spectral problem without any loss of generality can be considered in the two-dimensional representation with  $\a_i=\s_i$,  $\b=\s_3$.

The formal solution of (\ref{3.2}) is written as
\beq
\label{3.3}
G(\vec{r},\vec{r}\ ';\e)=\sum\limits_{n}\frac{\p_{n}(\vec{r})\p_{n}(\vec{r}\ ')^{\dagger}}{\e_{n}-\e} \ .
\eeq
\begin{figure}
	\center
	\includegraphics[scale=0.20]{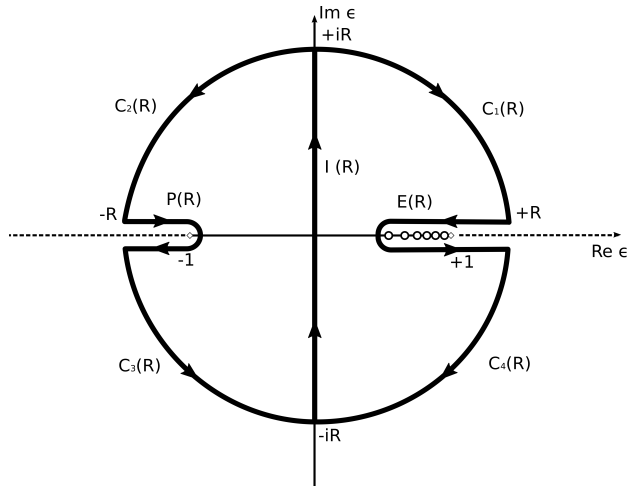}
	\caption{\small Special contours in the complex energy plane, used for representation of the vacuum charge density via contour integrals. The direction of contour integration is chosen in correspondence with (\ref{3.3}).}
	\label{pic:1}
\end{figure}
Following ~\cite{Wichmann1956}, the induced density is expressed via the integrals along the contours $P(R)$ and $E(R)$ on the first sheet of the Riemann energy surface (Fig.\ref{pic:1})

\begin{multline}\label{3.4}
\r_{VP}(\vec{r})=-\frac{|e|}{2} \lim_{R\rightarrow \infty}\( {1 \over 2\pi i}\int\limits_{P(R)}d\e\, \mathrm{Tr}G(\vec{r},\vec{r};\e) + \right. \\ \left. + {1 \over 2\pi i}\int\limits_{E(R)}d\e\, \mathrm{Tr}G(\vec{r},\vec{r};\e) \) \ .
\end{multline}
Since the DC spectral problem in the  external field (\ref{1.0}) divides into radial and angular parts via substitution
\beq
\p(\vec{r})={1 \over \sqrt{2\pi}} \begin{pmatrix}
i\p_{1}(r)\mathrm{e}^{i(m_j-1/2)\vf}\\
\p_{2}(r)\mathrm{e}^{i(m_j+1/2)\vf}
\end{pmatrix},
\eeq
for the trace of the Green function (\ref{3.3}) one obtains
\beq
\label{3.5}
\begin{aligned}
&\mathrm{Tr}G(\vec{r},\vec{r};\e) = {1 \over 2\pi}\mathrm{Tr}G(r,r;\e) = \\
&{1 \over 2\pi}\(2\, \sum\limits_{m_j= \pm 1/2, \pm 3/2,..}\mathrm{Tr}G_{m_j}(r,r;\e) \), \\
&\mathrm{Tr}G_{m_j}(r,r;\e)=\frac{1}{J_{m_j}(\e)}\p^{in}_{m_j}(r)^{\mathrm{T}} \p^{out}_{m_j}(r) \ ,
\end{aligned}
\eeq
where $\p^{in}_{m_j}(r)$ and $\p^{out}_{m_j}(r)$ are the  solutions of the radial DC  problem for the given $m_j$, which are regular at $r=0$ and $r=+\infty$ correspondingly, with $J_{m_j}(\e)$ being their Wronskian
\beq
\label{3.6}
J_{m_j}(\e)=[\p^{in}_{m_j},\p^{out}_{m_j}] \ .
\eeq
In (\ref{3.6}) and in  what follows we use the following denotation
\begin{equation*}
\[f,g\]_{a}=a\(f_{2}(a)g_{1}(a)-f_{1}(a)g_{2}(a)\) \ .
\end{equation*}
Such definition of $\tr G_{m_j}$ provides its correct normalization. It should be noted that the zeros of $J_{m_j}(\e)$, lying on the first sheet, are real-valued and correspond to the discrete spectrum, while those  on the second sheet become complex conjugate pairs and define the positions of  elastic resonances.

Proceeding further, let us construct $\tr G_{m_j}$ for the external potential (\ref{1.0}). For the given $m_j$ the radial DC  problem takes the following form
\beq
\label{3.7}
\left\lbrace\bal
&\frac{d}{d r}\p_1(r)+\frac{1/2-m_j}{r}\,\p_1(r)=(\e-V(r)+1)\p_2(r) \ ,\\
&\frac{d}{d r}\p_2(r)+\frac{1/2+m_j}{r}\,\p_2(r)=-(\e-V(r)-1)\p_1(r) \ .
\eal\right.
\eeq
For $0 \leqslant r\leqslant R_0$ the linearly independent solutions of (\ref{3.7}) are chosen in the form
\beq
\label{3.8}
\bal
\text{for $\p_1(r)$}: \quad
&\mathcal{I}_1(r)=\x I_{|m_j-1/2|}(\x r) \ , \\ &\mathcal{K}_1(r)=-\x K_{|m_j-1/2|}( \x r) \ ;\\
\text{for $\p_2(r)$}: \quad &\mathcal{I}_2(r)=\(1-\e-V_0\) I_{|m_j+1/2|}(\x r) \ ,  \\ & \mathcal{K}_2(r)=\(1-\e-V_0\) K_{|m_j+1/2|}(\x r) \ .
\eal
\eeq
In (\ref{3.8}) $I_{\nu}(z)$ and $K_{\nu}(z)$ are the modified Bessel functions of the  first and second kind, respectively,
\beq
V_0=Z \a/R_0,\quad \x=\sqrt{1-(\e+V_0)^2}, \quad \mathrm{Re} \, \x\geqslant 0 \ .
\eeq
For $r>R_0$ the fundamental pair of solutions for the system (\ref{3.7}) is taken as
\begin{multline}
\nonumber
\text{for $\p_1(r)$}: \\ \cM_{1}(r)=\frac{1+\e}{r}\[\(\vk-\n\)M_{\n-1/2,\vk}(2\g r) +  \right. \\ \left. +\(m_j+\frac{Q}{\g}\)M_{\n+1/2,\vk}(2\g r)\] \ , \\
\cW_{1}(r)=\frac{1+\e}{r}\[\(m_j-\frac{Q}{\g}\)W_{\n-1/2,\vk}(2\g r) - \right. \\ \left.  -W_{\n+1/2,\vk}(2\g r)\] \ ;
\end{multline}
\begin{multline}\label{3.9}
\text{for $\p_2(r)$}: \\ \cM_{2}(r)=\frac{\g}{r}\[\(\vk-\n\)M_{\n-1/2,\vk}(2\g r) -  \right. \\ \left. -\(m_j+\frac{Q}{\g}\)M_{\n+1/2,\vk}(2\g r)\] \ , \\
	\cW_{2}(r)=\frac{\g}{r}\[\(m_j-\frac{Q}{\g}\)W_{\n-1/2,\vk}(2\g r) +  \right. \\ \left.   + W_{\n+1/2,\vk}(2\g r)\] \ ,
\end{multline}
where $M_{b,c}(z)$ and $W_{b,c}(z)$ are the Whittaker functions ~\cite{Bateman1953},
\beq
\label{3.10}
\begin{aligned}
&Q=Z \a \ , \quad \vk=\sqrt{m_j^2-Q^2} \ , \quad  \n=\frac{\e Q}{\g} \ , \\  &\qquad \g=\sqrt{1-\e^2} \ , \qquad \mathrm{Re}\,\g\geqslant0 \ .
\end{aligned}
\eeq
Proceeding further, $\p^{in}_{m_j}(r)$ and $\p^{out}_{m_j}(r)$ are chosen as such linear combinations of the solutions (\ref{3.8}) and (\ref{3.9}), which are regular at $r=0$ and $r=+\infty$, correspondingly.

As a result, the expression for $\tr G_{m_j}$ takes the form
\begin{multline}\label{3.11}
\tr G_{m_j}(r,r;\e)=
\\
\frac{1}{\[\cI,\cK\]}\(\cI_{1}\cK_{1}+\cI_{2}\cK_{2}-\frac{\[\cK,\cW\]_{R_0}}{\[\cI,\cW\]_{R_0}}
\(\cI_{1}^{2}+\cI_{2}^{2}\)\)\tt\(R_0-r\) + \\ +
\frac{1}{\[\cM,\cW\]}\(\cM_{1}\cW_{1}+\cM_{2}\cW_{2}-\frac{\[\cI,\cM\]_{R_0}}{\[\cI,\cW\]_{R_0}}
\(\cW_{1}^{2}+\cW_{2}^{2}\)\) \times \\ \times \tt\(r-R_0\) \ ,
\end{multline}
where
\beq \label{3.13}
\[\cM,\cW\]= -4(1+\e)\g^2\frac{\G(2\vk+1)}{\G(\vk-\n)} \ , \quad [\cI,\cK]=\e+V_{0}-1 \ ,
\eeq
while the Wronskian (\ref{3.6}), which enters into the expression for $\tr G_{m_j}$ (\ref{3.5}), equals to
\beq
\label{3.14}
J_{m_j}(\e)=\[\cI,\cW\]_{R_0} \ .
\eeq

In the next step one finds the asymptotics of $\tr G_{m_j}$ on the arcs of the large circle in the upper half-plane (Fig.\ref{pic:1}) $C_1(R)$ and $C_2(R)$, where  $|\e|\to \inf$, $0<\mathrm{Arg}\, \e <\pi$:

\begin{multline}\label{3.16}
\tr G_{m_j}(r,r;\e)\to
\\
\to \frac{i}{r}+\frac{i}{2r \e^2} \(\frac{m_j^2}{r ^2}+1\)-\frac{i}{r^2 \epsilon ^3} \(\frac{m_j^2}{r} V_{0}+\frac{m_j}{2 r}+r V_{0}\)+ \\ + O\(|\e|^{-4}\) \ ,\quad r< R_0 \ ,\\
\to \frac{i}{r}+\frac{i}{2r \e^2} \(\frac{m_j^2}{r ^2}+1\)-\frac{i}{r ^2 \epsilon ^3} \(\frac{m_j^2 }{r}\frac{Q}{r}+\frac{m_j}{2r}+Q\) + \\+ O\(|\e|^{-4}\) \ ,\quad r>R_0 \ ,
\end{multline}
and on the arcs of the large circle in the lower half-plane $C_{3}(R)$ and $C_{4}(R)$, where  $|\e|\to \inf$,  $-\pi<\mathrm{Arg}\, \e <0$:
\begin{multline}
\label{3.17}
\tr G_{m_j}(r,r;\e)\to
\\
\to -\frac{i}{r}-\frac{i}{2r \e^2} \(\frac{m_j^2}{r ^2}+1\)+\frac{i}{r ^2 \e ^3} \(\frac{m_j^2}{r} V_{0}+\frac{m_j}{2 r}+r V_{0}\)+ \\ + O\(|\e|^{-4}\) \ ,\quad r< R_0, \\
\to -\frac{i}{r}-\frac{i}{2r \e^2} \(\frac{m_j^2}{r ^2}+1\)+\frac{i}{r ^2 \epsilon ^3} \(\frac{m_j^2}{r}\frac{Q}{r}+\frac{m_j}{2 r}+Q\) + \\ + O\(|\e|^{-4}\) \ ,\quad r>R_0  \ .
\end{multline}

There follows from (\ref{3.16}) and (\ref{3.17}) that the integration along the contours $P(R)$ and $E(R)$  in (\ref{3.4}) can be reduced to the imaginary axis, whence one finds the final expression  for the induced density
\beq
\label{3.18}
\r_{VP}(r)=2 \sum\limits_{m_j=1/2,\,3/2,..} \r_{VP,|m_j|}(r)\ ,
\eeq
where
\begin{multline} \label{3.18a}
\r_{VP,|m_j|}(r)=\frac{|e|}{(2 \pi)^2}\int\limits_{-\inf}^{\inf}dy\,\tr G_{|m_j|}(r,r;iy) \ , \\
\tr G_{|m_j|}(r,r;iy)=\tr G_{m_j}(r,r;iy)+\tr G_{-m_j}(r,r;iy) \ .
\end{multline}
In presence of negative discrete levels with  $-1\leqslant \e_n<0$
\begin{multline}
\label{3.19}
\r_{VP,|m_j|}(r)= {|e| \over 2\pi}\[\sum\limits_{m_j=\pm|m_j|}\sum\limits_{-1\leqslant \e_n<0}\p_{n,m_j}(r)^{\dagger}\p_{n,m_j}(r) + \right. \\ \left. + \frac{1}{2 \pi} \int\limits_{-\infty}^{+\infty}d y\, \tr G_{|m_j|}(r,r;iy)\] \ .
\end{multline}
Proceeding further, let us mention the general property of  $\tr G_{m_j}$ under the change of the sign of external field $(Q \to -Q)$ and complex conjugation
\beq
\bal\label{3.20}
&\tr G_{-m_j}(Q;r,r;\e)=-\tr G_{m_j}(-Q;r,r;-\e) \ , \\ &\quad \tr G_{m_j}(Q;r,r;\e)^{\ast}=\tr G_{m_j}(Q;r,r;\e^{\ast}) \ ,
\eal \eeq
and the direct consequence of these two properties
\beq
\label{3.21}
\tr G_{m_j} (Q;r,r; i y)^{\ast}=-\tr G_{-m_j}(-Q;r,r;i y) \ ,
\eeq
whence it follows that $\r_{VP,|m_j|}(r)$ can be expressed in terms of $ \mathrm{Re}\,\tr G_{|m_j|} (Q;r,r; i y)$ and is definitely a real function being odd in $Q$ (in the full agreement with the Furry theorem).  In the purely perturbative region the representation of $\r_{VP}(r)$ as an odd series in powers of external field (\ref{1.0}) is given by the Born series $G_{m_j}=G_{m_j}^{(0)}+G_{m_j}^{(0)} (-V) G_{m_j}^{(0)} + G_{m_j}^{(0)} (-V) G_{m_j}^{(0)} (-V) G_{m_j}^{(0)} + \dots $, whence
\beq \bal \label{f341}
& \mathrm{Re}\, \tr G_{m_j} (r,r; i y)= \\ &=\sum\limits_{k=0} \mathrm{Re}\, \tr  \[ G_{m_j}^{(0)} \(-V G_{m_j}^{(0)}\)^{2k+1}(r,r; i y) \]  \ ,
\eal \eeq
where $G_{m_j}^{(0)}$ is the free Green function of the corresponding radial Dirac equation. At the same time, in presence of negative discrete levels and especially in the overcritical region with $Z>Z_{cr,1}$ $\r_{VP}(r)$ is still an odd function in $Q$ ~\cite{Gyulassy1975}, but now the dependence  on the external field cannot be described by the series (\ref{f341}) any more,  since there appear in $\r_{VP}(r)$ essentially nonperturbative and so non-analytic in $Q$ components.

The expression for the induced  density (\ref{3.18})-(\ref{3.19}) requires renormalization, since there follows from the asymptotics of $\tr G_{m_j}$
 \begin{multline}\label{3.22}
\tr G_{m_j}(r,r;iy)\to\frac{i y}{\sqrt{1+y^2}}\frac{1}{r}+\frac{Q}{\(1+y^2\)^{3/2}}\frac{1}{r^2}+ O\({1 \over r^3}\), \\ r \to \inf \ ,
\end{multline}
that the non-renormalized $\r_{VP}(r)$ decreases for $r \to \inf$ as $1/r^2$, and so the total induced charge diverges logarithmically.

The general result, obtained in ~\cite{Gyulassy1975} via expression of $\r_{VP}(r)$ in powers of $Q$, which is valid for any number of spatial dimensions in the external fields like (\ref{1.0}), is that all the divergences of $\r_{VP}(r)$ originate from the fermionic loop with two external  lines, while the next-to-leading orders of expansion are finite. So for calculation of the renormalized induced density $\r^{ren}_{VP}$ the linear in $Q$ terms should be extracted from $\tr G_{m_j}$  (\ref{3.11}) and replaced by  $\r^{(1)}_{VP}$ (\ref{2.5c}), which is nonzero only for $|m_j|=1/2$. For these purposes one finds first the component of the induced density $\r^{(3+)}_{VP,|m_j|}(r)$, defined as
\begin{widetext} \beq \label{3.23}
\r_{VP,|m_j|}^{(3+)}(r)= {|e| \over 2\pi}\[\sum\limits_{m_j=\pm|m_j|}\sum\limits_{-1\leqslant \e_n<0}\p_{n,m_j}(r)^{\dagger}\p_{n,m_j}(r) + \frac{1}{\pi} \int\limits_{0}^{\infty}d y\,\mathrm{Re}\( \tr G_{|m_j|}(r,r;iy)-2\, \tr G^{(1)}_{m_j}(r;i y)\)\] \ ,
\eeq
where
$G^{(1)}_{m_j}=Q\left. \pd G_{m_j} /\pd Q \right|_{Q=0}$ and can be found through the first Born approximation $G_0 (-V) G_0$, which for $r\leqslant R_0$ gives
\beq \label{3.24}
\bal
\tr G^{(1)}_{m_j}(r;i y)&=\frac{Q}{(i y-1)^2}\[\(\tilde{\g}^2K^2_{|m_j-1/2|}(\tilde{\g}r)+(1-i y)^2 K^{2}_{|m_j+1/2|}(\tilde{\g}r)\)\int\limits_{0}^{r}dr'\,\frac{r'}{R_0}\(\tilde{\g}^2I^2_{|m_j-1/2|}(\tilde{\g}r')+\right.\right.\\
&\left.\left.+(1-i y)^2 I^{2}_{|m_j+1/2|}(\tilde{\g}r')\)+\(\tilde{\g}^2I^2_{|m_j-1/2|}(\tilde{\g}r)+(1-i y)^2 I^{2}_{|m_j+1/2|}(\tilde{\g}r)\)\times \right. \\
&\left.\times\left\lbrace\int\limits_{r}^{R_0}dr'\,\frac{r'}{R_0}\(\tilde{\g}^2K^2_{|m_j-1/2|}(\tilde{\g}r')+(1-i y)^2 K^{2}_{|m_j+1/2|}(\tilde{\g}r')\)+ \right. \right.\\
&\left.\left.+\int\limits_{R_0}^{\infty}dr'\,\(\tilde{\g}^2K^2_{|m_j-1/2|}(\tilde{\g}r')+(1-i y)^2 K^{2}_{|m_j+1/2|}(\tilde{\g}r')\)\right\rbrace\],
\eal
\eeq
and for $r>R_0$
\beq \label{3.24a}
\bal
\tr G^{(1)}_{m_j}(r;i y)&=\frac{Q}{(i y-1)^2}\[\(\tilde{\g}^2K^2_{|m_j-1/2|}(\tilde{\g}r)+(1-i y)^2 K^{2}_{|m_j+1/2|}(\tilde{\g}r)\)\left\lbrace\int\limits_{0}^{R_0}dr'\,\frac{r'}{R_0}\(\tilde{\g}^2I^2_{|m_j-1/2|}(\tilde{\g}r')+\right.\right.\right.\\
&\left.\left.\left.+(1-i y)^2 I^{2}_{|m_j+1/2|}(\tilde{\g}r')\)+\int\limits_{R_0}^{r}dr'\,\(\tilde{\g}^2I^2_{|m_j-1/2|}(\tilde{\g}r')+(1-i y)^2 I^{2}_{|m_j+1/2|}(\tilde{\g}r')\)\right\rbrace+ \right.\\
&\left.+\(\tilde{\g}^2I^2_{|m_j-1/2|}(\tilde{\g}r)+(1-i y)^2 I^{2}_{|m_j+1/2|}(\tilde{\g}r)\)\int\limits_{r}^{\infty}dr'\,\(\tilde{\g}^2K^2_{|m_j-1/2|}(\tilde{\g}r')+\right.\right.\\
&\left.\left.+(1-i y)^2 K^{2}_{|m_j+1/2|}(\tilde{\g}r')\)\],
\eal
\eeq\end{widetext}
where $\tilde{\g}=\sqrt{1+y^2}$. The integrals, containing in the expressions for $\tr G^{(1)}_{m_j}$ (\ref{3.24}) and (\ref{3.24a}), are not given explicitly due to their cumbersome form.

So the expression for renormalized induced  density takes the form
\beq\label{3.25}
\r^{ren}_{VP}(r)=2\[\r_{VP}^{(1)}(r)+\sum\limits_{m_j=1/2,\,3/2,..}\r_{VP,|m_j|}^{(3+)}(r)\] \ ,
\eeq
where  $\r_{VP}^{(1)}(r)$ is the perturbative induced density (\ref{2.5c}), evaluated by means of the polarization function (\ref{2.4}) in the first order of PT. Such expression for $\r^{ren}_{VP}(r)$ guarantees the vanishing  total induced charge $Q^{ren}_{VP}=\int\limits d^{2}r\,\r^{ren}_{VP}(r)$ for $Z<Z_{cr,1}$, since  $Q_{VP}^{(1)}$ is zero by construction, while the subsequent direct calculation  confirms that the contribution of $\r^{(3+)}_{VP,|m_j|}(r)$ to  $Q^{ren}_{VP}$ for $Z<Z_{cr,1}$ vanishes too. Unlike 1+1 D,  in 2+1 D such a check cannot  be performed in the purely analytical form any more  due to complexity of  expressions, containing in $\r^{(3+)}_{VP,|m_j|}(r)$. Nevertheless, it could be quite reliably performed via special combination of analytical and numerical methods (\cite{Davydov2018a}, App.B).  Moreover, it suffices to verify the disappearance of the total charge $Q_{VP}^{ren}$ not for the entire subcritical region, but only in absence of negative discrete levels. In presence of the latter,  the vanishing  total charge for $Z<Z_{cr,1}$ follows from model-independent arguments, which are based on the starting expression for the induced density (\ref{3.1}). Namely, there follows from (\ref{3.1}) that the change of the integral induced charge is possible for  $Z>Z_{cr,1}$ only, when the  discrete levels attain the lower continuum. Moreover, upon diving into the lower continuum each doubly degenerate energy level yields the change of the integral charge exactly by  $(-2|e|)$.  One of the possible correct ways to prove this statement is given in Ref.~\cite{Davydov2017}.  Let us specially mention that this effect is essentially non-perturbative and  completely included in $\r_{VP}^{(3+)}$, while  $\r_{VP}^{(1)}$ does not participate in it and still makes an exactly vanishing contribution to the total charge. Thus, the behavior of the renormalized by means of (\ref{3.25}) induced density in the non-perturbative region turns out to be indeed such that should be expected from the  general assumptions about the structure of the electron-positron (or electron-hole in our context) vacuum for $Z> Z_{cr}$ \cite{Greiner1985a, Plunien1986, Greiner2012, Rafelski2016}.

A more detailed picture of these changes in $\r^{ren}_{VP}(r)$ is quite similar to that considered in ~\cite{Greiner1985a, Plunien1986, Greiner2012, Rafelski2016} for 3+1 QED by means of U.Fano approach \cite{Fano1961}. The main result is that any discrete  level $\p_{n,m_j}(r)$, dived into the lower continuum, yields the change of the induced  density by
\beq\label{3.26}
\left. \D\r_{VP}(r)=-2|e| \p_{n,m_j}(r)^{\dagger}\p_{n,m_j}(r)\right|_{\e_n=-1} \ .
\eeq

The next point is that the renormalized induced density (\ref{3.25}) is represented by an infinite sum over $m_j$. So the convergence of this sum should be explored in detail. For these purposes let us consider the asymptotics of $\r_{VP,|m_j|}^{(3+)}(r)$ (\ref{3.23}) for large $|m_j|$.  The asymptotics of $\tr G_{m_j}(r,r;iy)$ (\ref{3.11}) for $|m_j|\to \infty$ takes the form
\begin{multline} \label{3.26a}
\tr G_{m_j}(r,r;iy)\to \\
\to \frac{iy+V_0}{|m_j|}+\frac{\mathrm{sgn}(m_j)}{2 m_j^2}+ O\(|m_j|^{-3}\), \quad r\leqslant R_0, \\
\to \frac{iy+Q/r}{|m_j|}+\frac{\mathrm{sgn}(m_j)}{2 m_j^2}+ O\(|m_j|^{-3}\), \quad r> R_0 \ ,
\end{multline}
whereas  the corresponding asymptotics of $\tr G^{(1)}_{m_j}(r;iy)$ reads
\beq \label{3.26b}
\tr G^{(1)}_{m_j}(r;iy)\to \left\lbrace \bal
&V_0\frac{1}{|m_j|}+O\(|m_j|^{-3}\), \quad &r\leqslant R_0, \\
&\frac{Q}{r}\frac{1}{|m_j|}+O\(|m_j|^{-3}\), \quad &r> R_0 \ .
\eal \right.
\eeq
Taking into account the definition of $\tr G_{|m_j|}(r,r;iy)$ (\ref{3.18a}), from (\ref{3.26a}) and (\ref{3.26b}) one obtains
\begin{multline}
\mathrm{Re}\[ \tr G_{|m_j|}(r,r;iy)-2\, \tr G^{(1)}_{m_j}(r;i y)\]=O\(|m_j|^{-3}\) \ , \\ \quad |m_j|\to \infty \ .
\end{multline}
Proceeding further, let us note that the discrete levels should rise up with increasing $|m_j|$. Therefore for any given $Q$ the additional contribution from negative discrete levels to $\r_{VP}(r)$ disappears  in the expressions (\ref{3.19}) and (\ref{3.23}) for  $|m_j|\to\infty$. Since the integral over $dy$ in (\ref{3.23}) converges uniformly (see~\cite{Davydov2018a}, App.C), there follows from  (\ref{3.26a}) that $\r_{VP,|m_j|}^{(3+)}(r)$ for $|m_j| \to \infty$ behaves as $O(|m_j|^{-3})$. So the partial series in $m_j$ converges and the renormalized induced  density $\r^{ren}_{VP}(r)$ (\ref{3.25}) is  finite everywhere up to logarithmic singularities at $r=R_0$ due to the contribution from $\r^{(1)}_{VP}(r)$.
\begin{figure*}[ht!]
\subfigure[]{\label{pic:2}
		\includegraphics[width=\columnwidth]{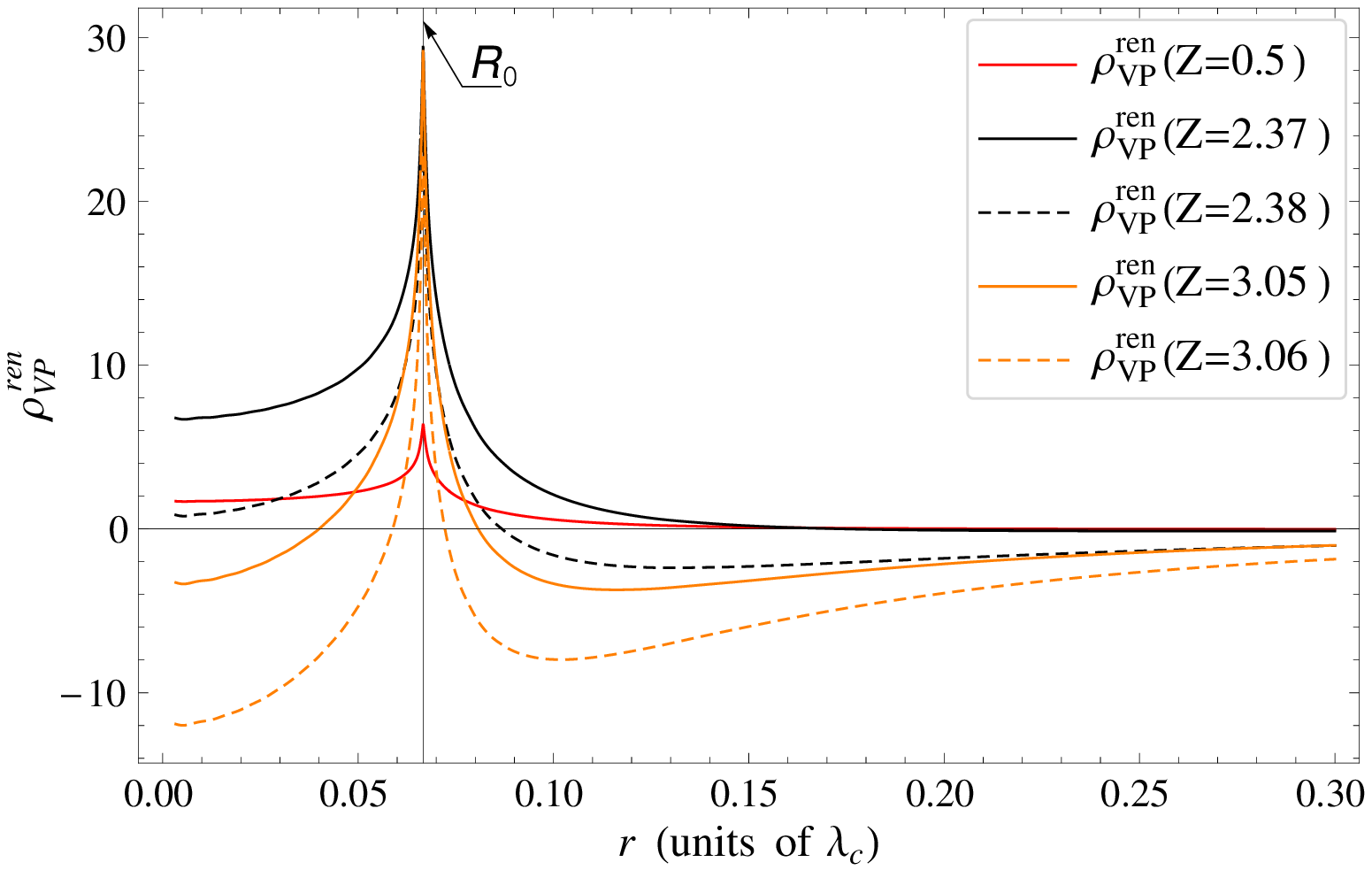}
}
\hfill
\subfigure[]{\label{pic:3}
		\includegraphics[width=\columnwidth]{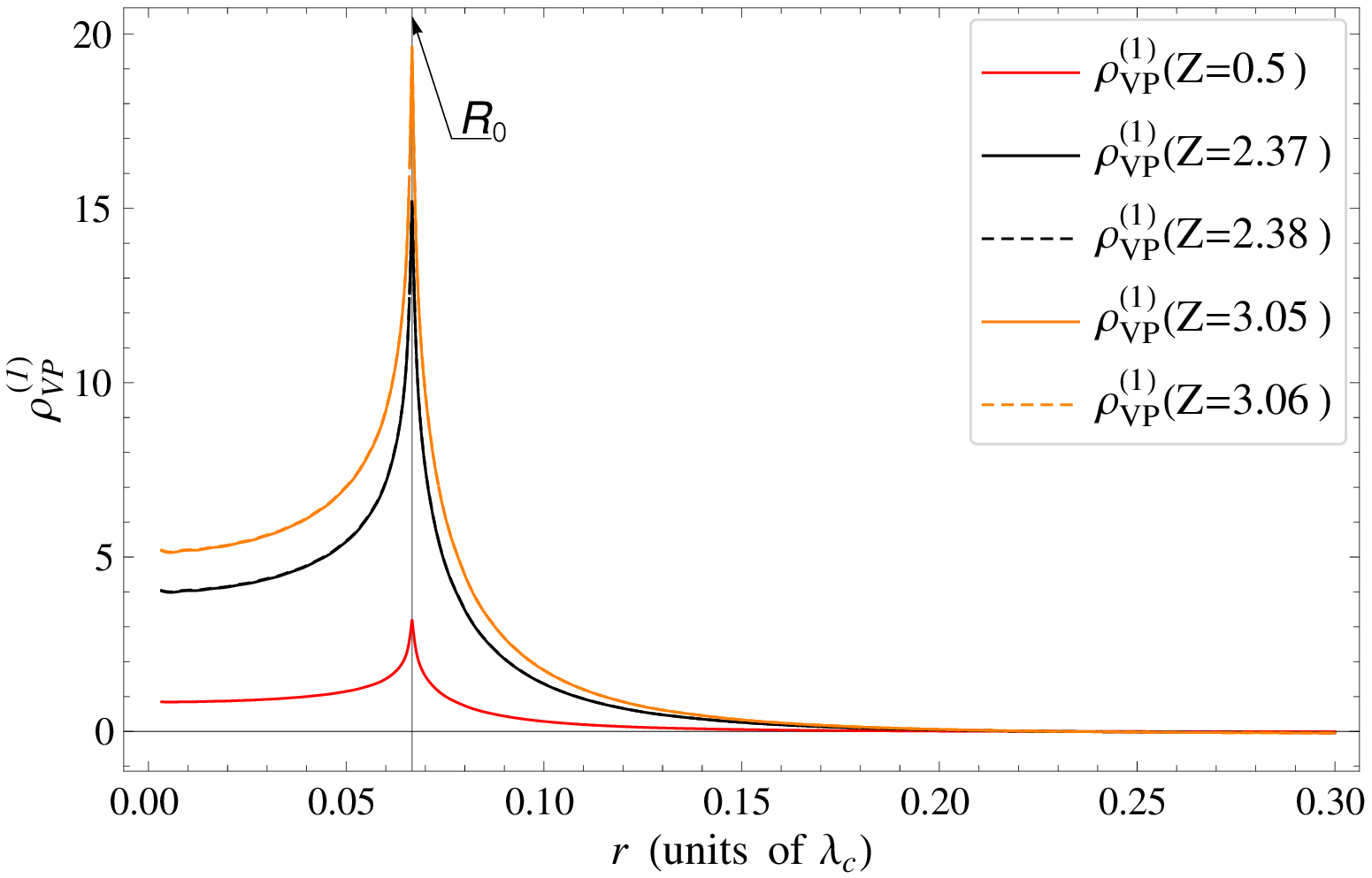}
}
\vfill
\subfigure[]{\label{pic:4}
	\includegraphics[width=\columnwidth]{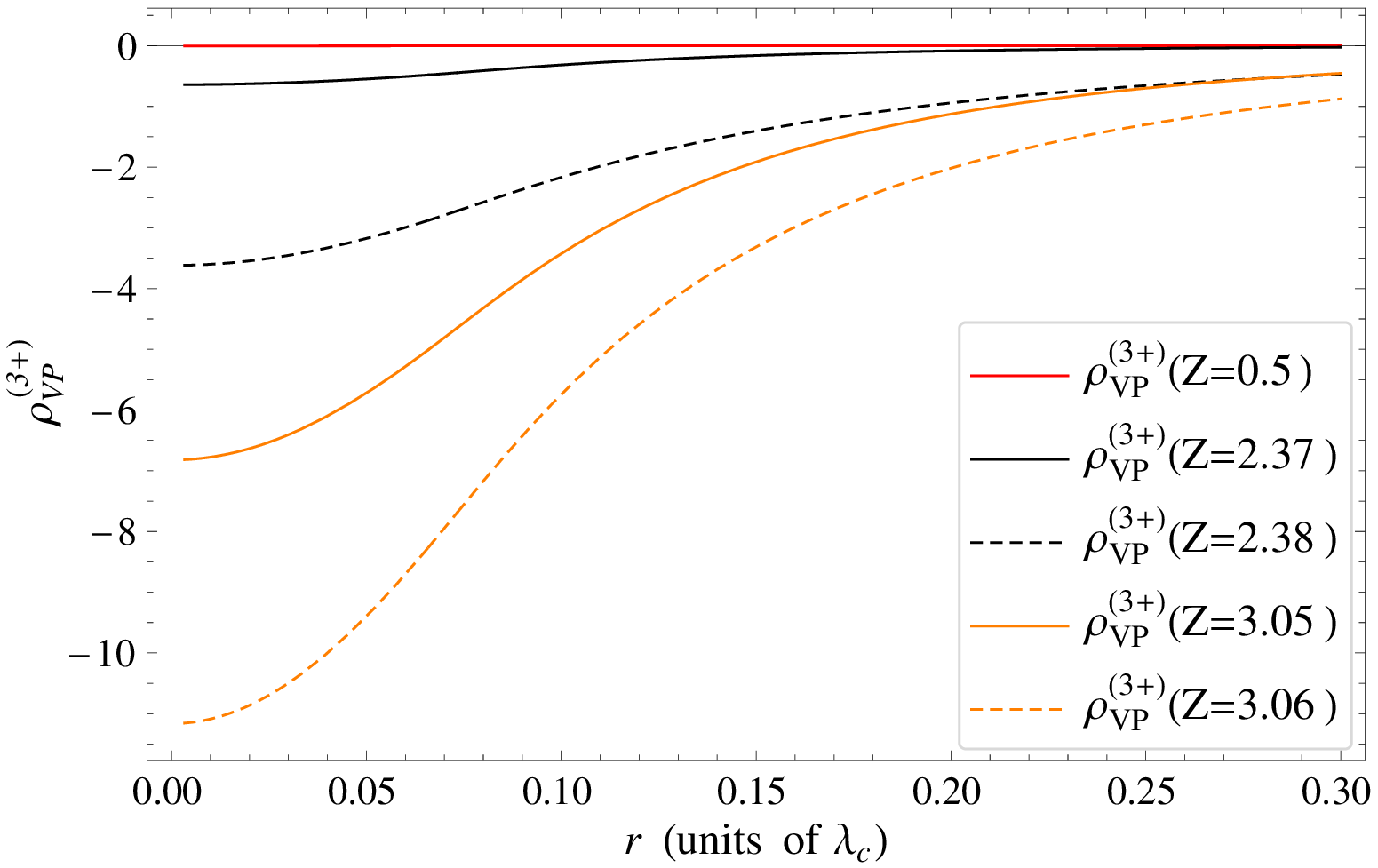}
}
\hfill
\subfigure[]{\label{pic:4a}
		\includegraphics[width=\columnwidth]{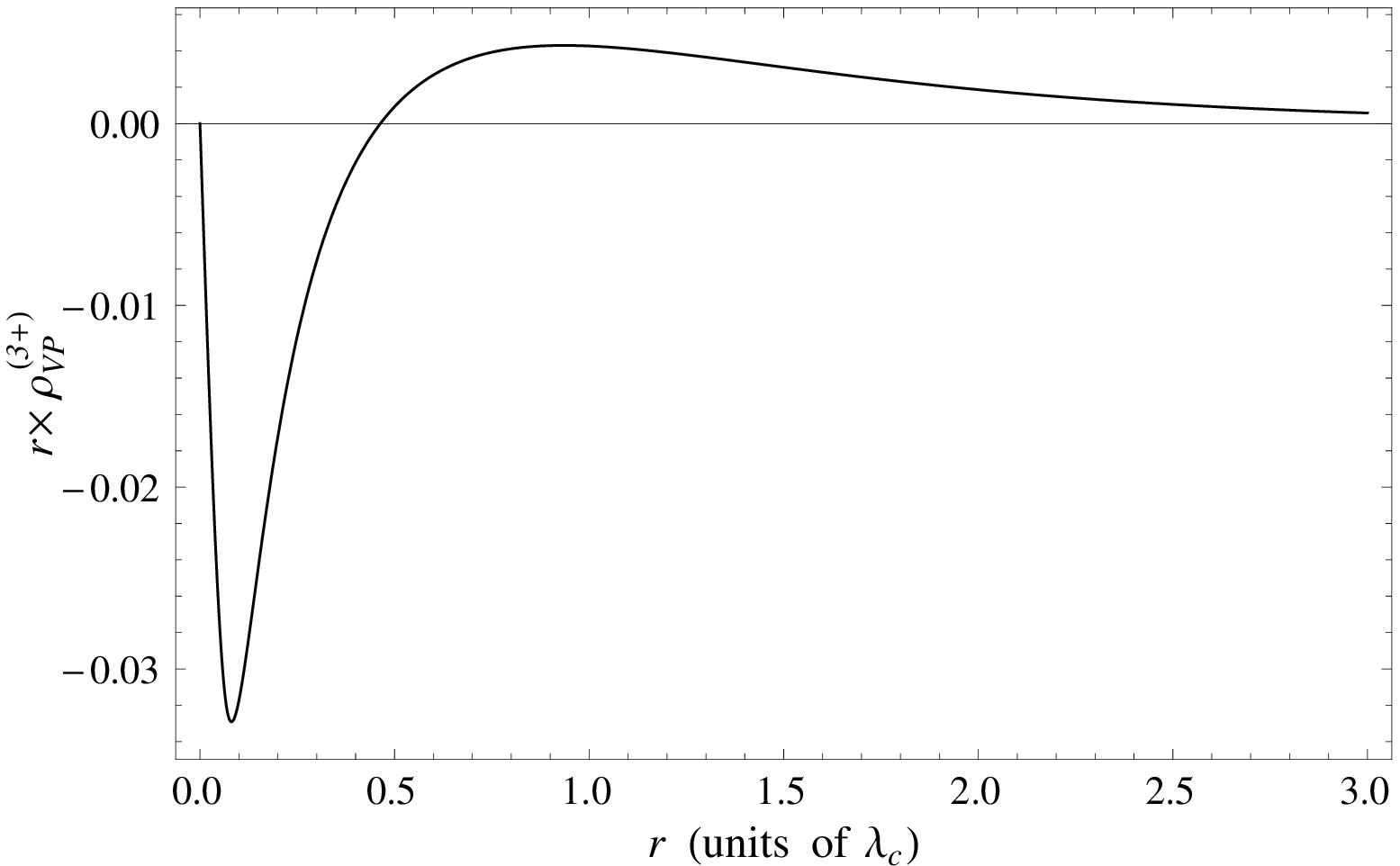}	
}
\caption{(Color online) \small ~\subref{pic:2} $\r_{VP}^{ren}(r)$, ~\subref{pic:3} $\r_{VP}^{(1)}(r)$, ~\subref{pic:4} $\r_{VP}^{(3+)}(r)$, $\a=0.4\, , R_0=1/15$ and $Z=0.5\, , 2.37\, , 2.38\, , 3.05\, , 3.06 $,  ~\subref{pic:4a} $r\times \r_{VP}^{(3+)}(r)$ on a more large  interval of variation of the radial variable for $Z=2.37$. }\label{pic:2-4}
\end{figure*}

Fig.\ref{pic:2} shows the renormalized induced  density  $\r_{VP}^{ren}(r)$ (\ref{3.25}) for the unscreened  ($R_1 \to \inf$) external potential (\ref{1.0}) in the case  $\a=0.4$, $R_0=1/15$, first in the purely perturbative regime for $Z=0.5$, thereupon for $Z=2.37$, when the first $Z_{cr,1}\simeq 2.373$ isn't reached yet, for $Z=2.38$, when the first discrete level has just dived into the lower continuum, thereon for $Z=3.05$, when the second $Z_{cr,2}\simeq 3.056$ is not reached yet, and, finally, for $Z=3.06$, i.e. just after diving of the second discrete level into the lower continuum. The critical charges are found from the transcendental equation, which is the consequence of  matching conditions for $\p^{in}_{m_j}(r)$ and $\p^{out}_{m_j}(r)$ at $r=R_0$  for $\e=-1$:
\begin{widetext}\beq\label{3.261}
\bal
 &I_{|m_j-1/2|}\(\sqrt{1-(V_0-1)^2} R_0\)\[K_{2i|\vk|-1}\(\sqrt{8 Q R_0}\)+K_{2i|\vk|+1}\(\sqrt{8 Q R_0}\)+\right.\\
&\left.+ \sqrt{\frac{2}{Q R_0}}m_j K_{2i|\vk|}\(\sqrt{8 Q R_0}\)\]+\sqrt{2(2-V_0)}\,I_{|m_j+1/2|}\(\sqrt{1-(V_0-1)^2} R_0\)K_{2i|\vk|}\(\sqrt{8 Q R_0}\)=0 \ ,
 \eal
 \eeq\end{widetext}
In (\ref{3.261}) $K_p(z)$ is the MacDonald function, which appears in the solutions of the system (\ref{3.7}) in the limit  $\e \to -1$, while $|\vk|=\sqrt{Q^2-m_j^2}$\,.

The direct numerical integration confirms that the total induced  charge for $Z=0.5, \ 2.37$ equals to zero, for $Z=2.38, 3.05$ equals to $(-2|e|)$, while for $Z=3.06$ equals to $(-4|e|)$. Fig.\ref{pic:3} displays the logarithmic singularity in $\r_{VP}^{ren}(r)$ at $r=R_0$, which originates from $ \r_{VP}^{(1)}(r)$. On the contrary, the induced density $ \r_{VP}^{(3+)}(r)=\sum_{m_j}\r_{VP,|m_j|}^{(3+)}(r) $  is a continuous function, which is shown in  Fig.\ref{pic:4}.  Fig.\ref{pic:4a} displays the weighted sign-alternating density $ r \times \r_{VP}^{(3+)}(r)$ on a sufficiently large interval of variation of the radial variable for $Z=2.37$, which confirms that the total induced charge in the subcritical region should vanish.  Note also that in the overcritical region the changes of $\r_{VP}^{ren}(r)$ with increasing $Z$ proceed not only in a step-like manner due to the formation of vacuum shells from the discrete levels diving into the lower continuum  according to (\ref{3.26}) (which is sometimes called a "real" polarization),  but also via permanent deformations in the density of states in both continua and evolution of the discrete levels with increasing $Z$ (known  as a "virtual" one). For other values of $\a$ and $R_0$ similar graphs don't change qualitatively.

Thus, the correct approach to calculation of $\r_{VP}^{ren}(r)$  for all the regions for $Z$ should be based on the expressions (\ref{3.23}) and (\ref{3.25}) with subsequent verification of the expected integer value of the total induced charge via direct  integration of $\r_{VP}^{ren}(r)$.

\section{Induced charge density for the screened Coulomb asymptotics}\label{sec:screening}

The changes start already in PT directly from  the Uehling potential. Namely, the expression (\ref{2.5}) for $A^{(1)}_{VP,0}(r)$  acquires the following additive term
\begin{multline}\label{4.1}
\D A_{VP,0}^{(1)}(r)=\frac{Z \a |e|}{4}\int\limits_{0}^{\infty}dq\,\frac{J_0(qr)}{q} \[\frac{2 }{q} + \(1-\frac{4}{q^2}\) \times \right. \\
 \left. \times \arctan \( \frac{q}{2}\)\]
\times \(2\[q R_1 J_0(q R_1)-1\] - \right. \\
 \left. - \pi q R_1\[ J_0(q R_1) \mathbf{H}_1(q R_1)- J_1(q R_1) \mathbf{H}_0(q R_1)\]\) \ ,
\end{multline}
which yields the corresponding contribution to the induced density
\begin{multline}
\label{4.2}
\D\r_{VP}^{(1)}(r)=-\frac{Z \a |e|}{16 \pi} \D_{2}\,\Bigg(\int\limits_{0}^{\infty}dq\,\frac{J_0(qr)}{q} \[\frac{2 }{q} + \(1-\frac{4}{q^2}\) \times \right. \\
 \left. \times \arctan \( \frac{q}{2}\)\]
\times \(2\[q R_1 J_0(q R_1)-1\] - \right. \\
 \left. - \pi q R_1\[ J_0(q R_1) \mathbf{H}_1(q R_1)- J_1(q R_1) \mathbf{H}_0(q R_1)\]\) \Bigg) \ .
\end{multline}
Now in (\ref{4.2}) it is not allowed to exchange the integration over $dq$ and the Laplace operator, since in contrast to the unscreened case the leading term of the integrand in (\ref{4.2}) behaves for  $q\to\inf$ as
\beq \label{4.3a}
- \frac{\cos (q (r+R_1)) + \sin(q (r-R_1))}{\sqrt{r R_1}\,q^{2}} \ ,
\eeq
which after the action of $\D_2$ gives
\beq \label{4.3b}
 \frac{\cos (q (r+ R_1)) + \sin(q (r-R_1))}{\sqrt{r R_1}} + O(1/q)  \ .
\eeq
Such behavior of the asymptotics of the integrand for $\D\r_{VP}^{(1)}(r)$ means that the induced density should reveal an even more strong singularity for $r \to  R_1$, than for $r \to  R_0$, namely
\begin{multline} \label{rho19.1}
\D \r_{VP}^{(1)}(r) \to -\frac{Z \a |e|}{8 \pi}\(
\frac{1}{2 R_1\(r-R_1\)} + \right. \\
 \left. + \frac{1}{4 R_1^2}\ln |R_1-r|  + O(1) \) \ , \quad r \to R_1 \ .
\end{multline}
Actually, the difference between the logarithmic singularity at $r=R_0$ and the power-like at $r=R_1$ in the induced density reflects the difference in behavior of the external potential  (\ref{1.0}) at these points. A more detailed analysis of the structure of singularities in $\D\r_{VP}^{(1)}(r)$ at $r=R_1$ is presented in~\cite{Sveshnikov2018a}.

It should be pointed out that for a slightly modified cutoff, which preserves the continuity of the potential at $r=R_1$, the singularity in $\r_{VP}^{(1)}(r)$ at $r=R_1$ remains a logarithmic one, as for $r=R_0$. However, for the induced charge and density the discontinuity in the external potential (\ref{1.0}) at $r=R_1$ doesn't pose any principal problems, in particular, $Q_{VP}^{(1)}$ remains zero. So we'll deal here with this type of screening, since (\ref{1.0}) looks more physical and transparent. At the same time, the perturbative one-loop vacuum polarization energy turns out to be divergent due to this discontinuity, and so calculation of the Casimir (vacuum) energy in this case requires to consider a  more soft type of screening  (see, e.g., refs.\cite{Sveshnikov2018a,Voronina2018b}).

Screening (\ref{1.0}) yields the next changes in the structure of solutions of the radial DC problem (\ref{3.7}). For $r\leqslant R_0$ the fundamental pair of solutions (\ref{3.8}) remains the same, the fundamental pair (\ref{3.9}) is now valid on the interval $R_0 < r < R_1$, while for the remaining part of the half-axis $r\geqslant R_1$ the independent solutions of (\ref{3.7}) should be chosen as
\beq\bal\label{4.5a}
\text{for $\p_1(r)$}: \quad
&\mathcal{I}_1^0(r)=\g I_{|m_j-1/2|}(\g r) \ , \\ &\mathcal{K}_1^0(r)=-\g K_{|m_j-1/2|}( \g r) \ ;
\eal\eeq \\
\beq\bal\label{4.5b}
\text{for $\p_2(r)$}: \quad &\mathcal{I}_2^0(r)=\(1-\e\) I_{|m_j+1/2|}(\g r) \ , \\  &\mathcal{K}_2^0(r)=\(1-\e\) K_{|m_j+1/2|}(\g r) \ ,
\eal\eeq
where $\g=\sqrt{1-\e^2}$.

As a result, the expression for $\tr G_{m_j}$ in the screened case takes the form
\begin{widetext}
\beq \label{4.6}
\begin{aligned}
&\tr G_{m_j}(r,r;\e)=\\
&\left\lbrace
\bal
&{1 \over \[\cI,\cK\]}\(\cI_{1}\cK_{1}+\cI_{2}\cK_{2}+{\[\cK^0,\cM\]_{R_1}\[\cW,\cK\]_{R_0}+\[\cW,\cK^0\]_{R_1}\[\cM,\cK\]_{R_0} \over \[\cK^0,\cM\]_{R_1}\[\cI,\cW\]_{R_0}+\[\cW,\cK^0\]_{R_1}\[\cI,\cM\]_{R_0}}\(\cI_{1}^{2}+\cI_{2}^{2}\)\), \quad &r\leqslant R_0 \ ,\\
&\frac{1}{\[\cM,\cW\]}\Big({\[\cK^0,\cM\]_{R_1}\[\cW,\cI\]_{R_0}+\[\cW,\cK^0\]_{R_1}\[\cI,\cM\]_{R_0} \over \[\cK^0,\cM\]_{R_1}\[\cW,\cI\]_{R_0}-\[\cW,\cK^0\]_{R_1}\[\cI,\cM\]_{R_0}}(\cM_{1}\cW_{1}+\cM_{2}\cW_{2})+\\&{\[\cK^0,\cM\]_{R_1}\[\cI,\cM\]_{R_0}\(\cW_{1}^{2}+\cW_{2}^{2}\)+\[\cW,\cK^0\]_{R_1}\[\cW,\cI\]_{R_0}\(\cM_{1}^{2}+\cM_{2}^{2}\)  \over \[\cK^0,\cM\]_{R_1}\[\cW,\cI\]_{R_0}-\[\cW,\cK^0\]_{R_1}\[\cI,\cM\]_{R_0}}\Big), \quad & R_0<r<R_1 \ ,\\
&{1 \over \[\cI^0,\cK^0\]}\(\cI_{1}^0\cK_{1}^0+\cI_{2}^0\cK_{2}^0+{\[\cI^0,\cW\]_{R_1}\[\cI,\cM\]_{R_0}+\[\cI^0,\cM\]_{R_1}\[\cW,\cI\]_{R_0} \over \[\cW,\cK^0\]_{R_1}\[\cI,\cM\]_{R_0}+\[\cM,\cK^0\]_{R_1}\[\cW,\cI\]_{R_0}}\((\cK_{1}^0)^2+(\cK_{2}^0)^2\)\), \quad &r\geqslant R_1 \ ,
\eal\right.
\end{aligned}
\eeq
\end{widetext}
where in addition to (\ref{3.13}) one has
\beq \label{4.7}
[\cI^0,\cK^0]=\e-1 \ ,
\eeq
while the Wronskian (\ref{3.6}), which enters into the expression for $\tr G_{m_j}$ (\ref{3.5}), equals now to
\beq
\label{4.8}
J_{m_j}(\e)={\[\cK^0,\cM\]_{R_1}\[\cI,\cW\]_{R_0}+\[\cW,\cK^0\]_{R_1}\[\cI,\cM\]_{R_0} \over \[\cM,\cW\]} \ .
\eeq

The asymptotics of $\tr G_{m_j}(r,r;\e)$ on the arcs of large circle (Fig.\ref{pic:1}) for $r\leqslant R_0$ and $R_0<r<R_1$  coincide with those in the problem without screening (\ref{3.16}) and (\ref{3.17}), since the terms in $\tr G_{m_j}$, depending on $R_1$, give only an exponentially decreasing contribution. For $r\geqslant R_1$ the asymptotics of $\tr G_{m_j}(r,r;\e)$ takes the form:\\
on the arcs $C_1(R)$ and $C_2(R)$ in the upper half-plane, where  $|\e|\to \inf$ \ , $0<\mathrm{Arg}\, \e <\pi \ ,$
\begin{multline}\label{4.9}
\tr G_{m_j}(r,r;\e)\to
\\
\frac{i}{r}+\frac{i}{2r \e^2} \(\frac{m_j^2}{r ^2}+1\)-\frac{im_j}{2r^3 \e^3} +O\(|\e|^{-4}\),\quad r\geqslant R_1 \ ;
\end{multline}
on the arcs of large circle in the lower half-plane $C_{3}(R)$ and $C_{4}(R)$, where  $|\e|\to \inf$,  $-\pi<\mathrm{Arg}\, \e <0 \ ,$
\begin{multline}\label{4.10}
\tr G_{m_j}(r,r;\e)\to
\\
-\frac{i}{r}-\frac{i}{2r \e^2} \(\frac{m_j^2}{r ^2}+1\)+\frac{im_j}{2r^3 \e^3}+O\(|\e|^{-4}\),\quad r\geqslant R_1 \ .
\end{multline}

Again, there follows from the asymptotics of $\tr G_{m_j}(r,r;\e)$ on the arcs of the large circle that the integration along the contours $P(R)$ and $E(R)$  in (\ref{3.4}) can be reduced to the imaginary axis, whence one finds the same final expression  for the vacuum  density as in the unscreened case (\ref{3.18})-(\ref{3.19}) with the same properties (\ref{3.20})-(\ref{f341}).

However, the asymptotics of $\tr G_{m_j}(r,r;\e)$ for $r \to \inf$ undergoes significant changes, caused by the different structure (\ref{4.5a}),(\ref{4.5b}) of solutions of the system (\ref{3.7}) for $r\geqslant R_1$, namely
\begin{multline}
\label{4.11}
\tr G_{m_j}(r,r;iy) \to  \\ \frac{i y}{\sqrt{1+y^2}}\frac{1}{r}+{m_j\(1-im_jy\) \over 2\(1+y^2\)^{3/2}} {1\over r^3}+O\({1\over r^5}\) \ , \quad r\to \inf \ .
\end{multline}
There follows from (\ref{4.11}) that in the case of finite $R_1$ for any $m_j$ the total induced charge  is finite from the outset. Nevertheless,  the induced density  requires  renormalization, since the non-renormalized total induced charge doesn't vanish in the subcritical region. For these purposes we define once more the component  $\r_{VP,|m_j|}^{(3+)}(r)$ by the same expression (\ref{3.23}), where $\tr G^{(1)}_{m_j}(r;i y)$ should now be replaced by:
\\ \begin{widetext}
for $r\leqslant R_0$
\beq \label{4.12}
\bal
\tr G^{(1)}_{m_j}(r;i y)&=\frac{Q}{(i y-1)^2}\[\(\tilde{\g}^2K^2_{|m_j-1/2|}(\tilde{\g}r)+(1-i y)^2 K^{2}_{|m_j+1/2|}(\tilde{\g}r)\)\int\limits_{0}^{r}dr'\,\frac{r'}{R_0}\(\tilde{\g}^2I^2_{|m_j-1/2|}(\tilde{\g}r')+\right.\right.\\
&\left.\left.+(1-i y)^2 I^{2}_{|m_j+1/2|}(\tilde{\g}r')\)+\(\tilde{\g}^2I^2_{|m_j-1/2|}(\tilde{\g}r)+(1-i y)^2 I^{2}_{|m_j+1/2|}(\tilde{\g}r)\)\times \right. \\
&\left.\times\left\lbrace\int\limits_{r}^{R_0}dr'\,\frac{r'}{R_0}\(\tilde{\g}^2K^2_{|m_j-1/2|}(\tilde{\g}r')+(1-i y)^2 K^{2}_{|m_j+1/2|}(\tilde{\g}r')\)+ \right. \right.\\
&\left.\left.+\int\limits_{R_0}^{R_1}dr'\,\(\tilde{\g}^2K^2_{|m_j-1/2|}(\tilde{\g}r')+(1-i y)^2 K^{2}_{|m_j+1/2|}(\tilde{\g}r')\)\right\rbrace\] \ ,
\eal
\eeq
for $R_0<r<R_1$
\beq \label{4.12a}
\bal
\tr G^{(1)}_{m_j}(r;i y)&=\frac{Q}{(i y-1)^2}\[\(\tilde{\g}^2K^2_{|m_j-1/2|}(\tilde{\g}r)+(1-i y)^2 K^{2}_{|m_j+1/2|}(\tilde{\g}r)\)\left\lbrace\int\limits_{0}^{R_0}dr'\,\frac{r'}{R_0}\(\tilde{\g}^2I^2_{|m_j-1/2|}(\tilde{\g}r')+\right.\right.\right.\\
&\left.\left.\left.+(1-i y)^2 I^{2}_{|m_j+1/2|}(\tilde{\g}r')\)+\int\limits_{R_0}^{r}dr'\,\(\tilde{\g}^2I^2_{|m_j-1/2|}(\tilde{\g}r')+(1-i y)^2 I^{2}_{|m_j+1/2|}(\tilde{\g}r')\)\right\rbrace+ \right.\\
&\left.+\(\tilde{\g}^2I^2_{|m_j-1/2|}(\tilde{\g}r)+(1-i y)^2 I^{2}_{|m_j+1/2|}(\tilde{\g}r)\)\int\limits_{r}^{R_1}dr'\,\(\tilde{\g}^2K^2_{|m_j-1/2|}(\tilde{\g}r')+\right.\right.\\
&\left.\left.+(1-i y)^2 K^{2}_{|m_j+1/2|}(\tilde{\g}r')\)\] \ ,
\eal
\eeq
for $r\geqslant R_1$
\beq \label{4.12c}
\bal
\tr G^{(1)}_{m_j}(r;i y)&=\frac{Q}{(i y-1)^2}\[\(\tilde{\g}^2K^2_{|m_j-1/2|}(\tilde{\g}r)+(1-i y)^2 K^{2}_{|m_j+1/2|}(\tilde{\g}r)\)\left\lbrace\int\limits_{0}^{R_0}dr'\,\frac{r'}{R_0}\(\tilde{\g}^2I^2_{|m_j-1/2|}(\tilde{\g}r')+\right.\right.\right.\\
&\left.\left.\left.+(1-i y)^2 I^{2}_{|m_j+1/2|}(\tilde{\g}r')\)+\int\limits_{R_0}^{R_1}dr'\,\(\tilde{\g}^2I^2_{|m_j-1/2|}(\tilde{\g}r')+(1-i y)^2 I^{2}_{|m_j+1/2|}(\tilde{\g}r')\)\right\rbrace\]  \ ,
\eal
\eeq
where $\tilde{\g}=\sqrt{1+y^2}$, and introduce the renormalized $\r^{ren}_{VP}(r)$ via expression
\beq\label{4.13}
\r^{ren}_{VP}(r)= 2\[\r_{VP}^{(1)}(r)+\D\r_{VP}^{(1)}(r)+\sum\limits_{m_j=1/2,\,3/2,..}\r_{VP,|m_j|}^{(3+)}(r)\] \ ,
\eeq\end{widetext}
with  $\r_{VP}^{(1)}(r)$ being   the  perturbative induced density (\ref{2.5c}) for the unscreened case, while $\D \r_{VP}^{(1)}(r)$ is the additional contribution (\ref{4.2}), caused by finite $R_1$.

Qualitatively, the behavior of the renormalized induced density (\ref{4.13})  looks like in the unscreened case ($R_1\to\infty$): for $Z<Z_{cr,1}$ the total induced charge vanishes exactly, each discrete level $\p_{n,m_j}(r)$ by diving into the lower continuum yields the change in the induced density described by (\ref{3.26}), while the total induced charge loses an amount equal to $(-2|e|)$.
To demonstrate the correspondence between the screened  and unscreened cases, in Figs.3a-d there are shown the components of the induced density $\r_{VP,|m_j|}^{(3+)}(r)$ for the case $\a=0.4$, $R_0=1/15$, $|m_j|=3/2$ and  $R_1 = \inf\, , 10\, R_0\, , 5\, R_0\, , 2\, R_0$, correspondingly. Each plot contains the induced density before and just after diving of the first discrete level into the lower continuum. Moreover, from Figs.3a-d it is clearly seen that for decreasing $R_1$ the components of the induced density $\r_{VP,|m_j|}^{(3+)}(r)$ localize in the region $r\sim R_0$ due to  contraction of the Coulomb well.
\begin{figure*}[ht!!]
\subfigure[]{\label{pic:5}
		\includegraphics[width=\columnwidth]{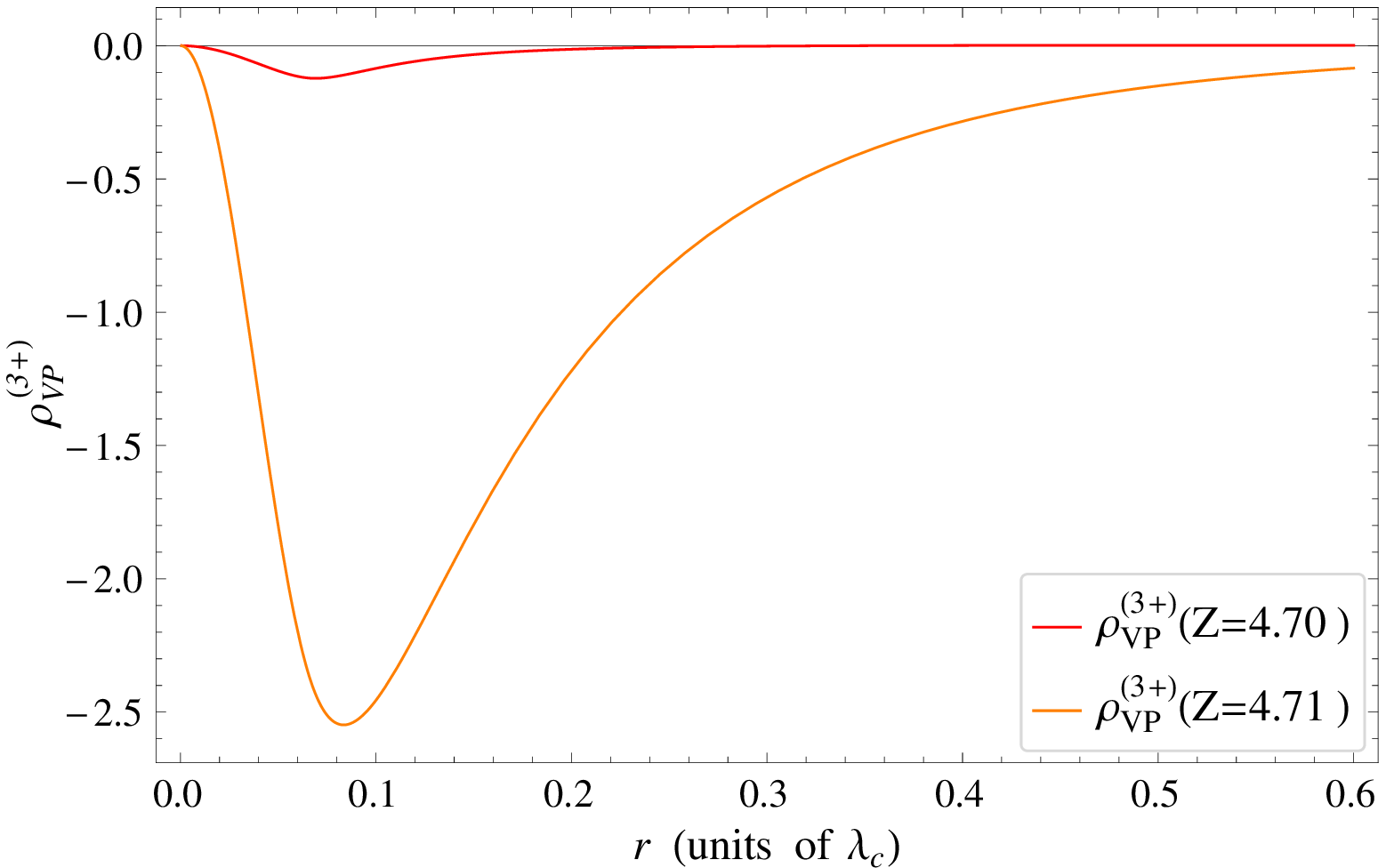}
}
\hfill
\subfigure[]{\label{pic:6}
		\includegraphics[width=\columnwidth]{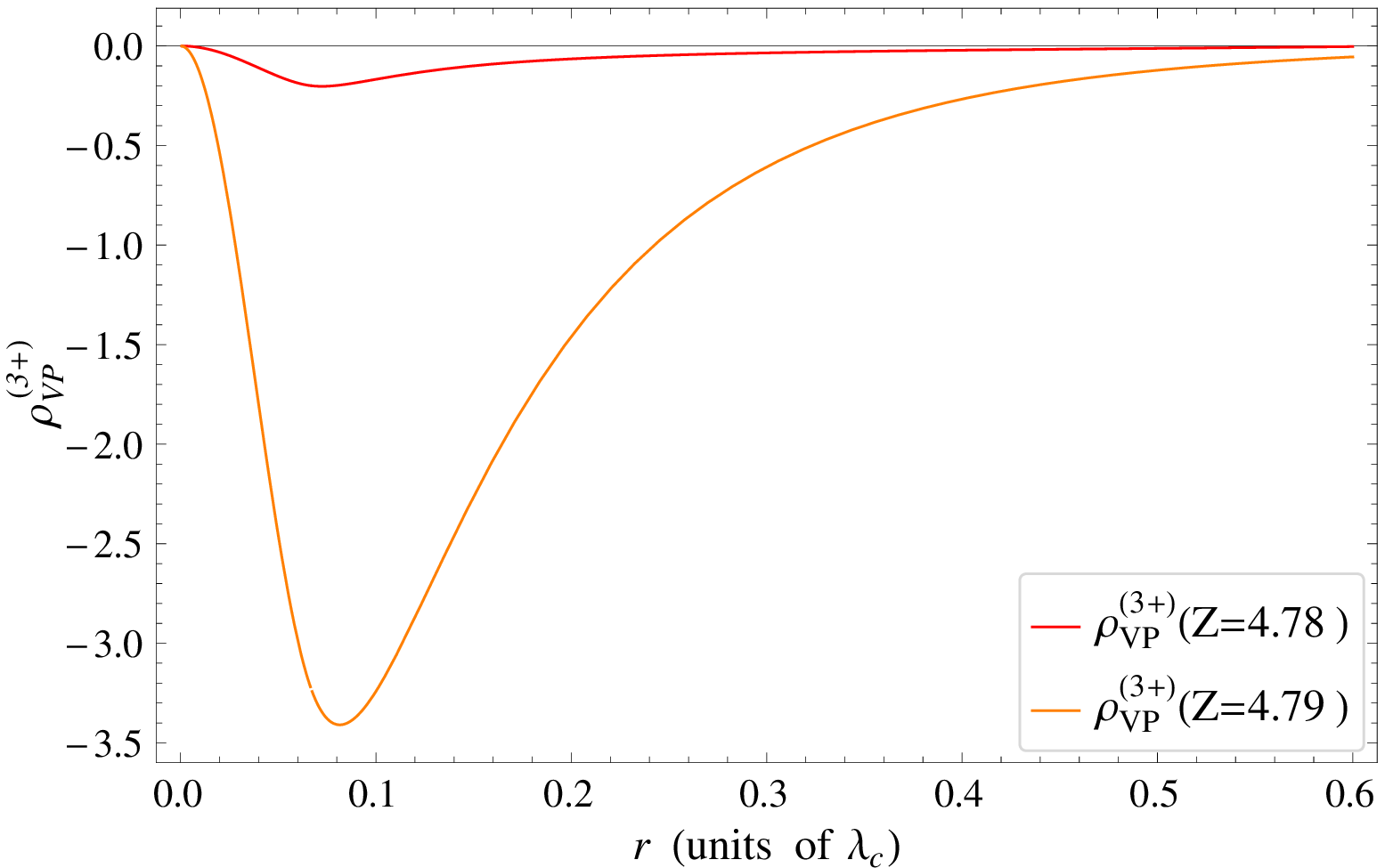}
}
\vfill
\subfigure[]{\label{pic:7}
	\includegraphics[width=\columnwidth]{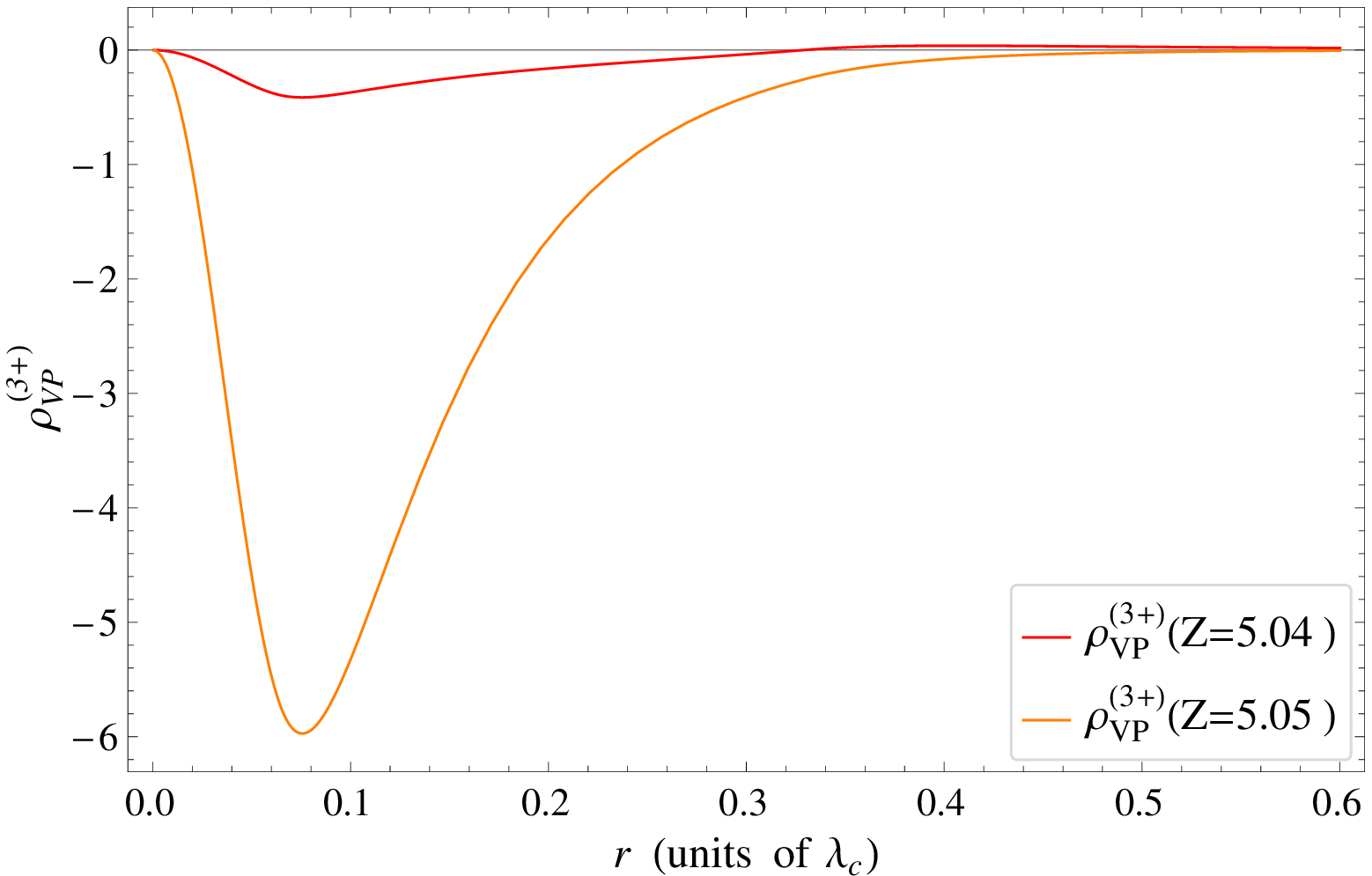}
}
\hfill	
\subfigure[]{\label{pic:8}
	\includegraphics[width=\columnwidth]{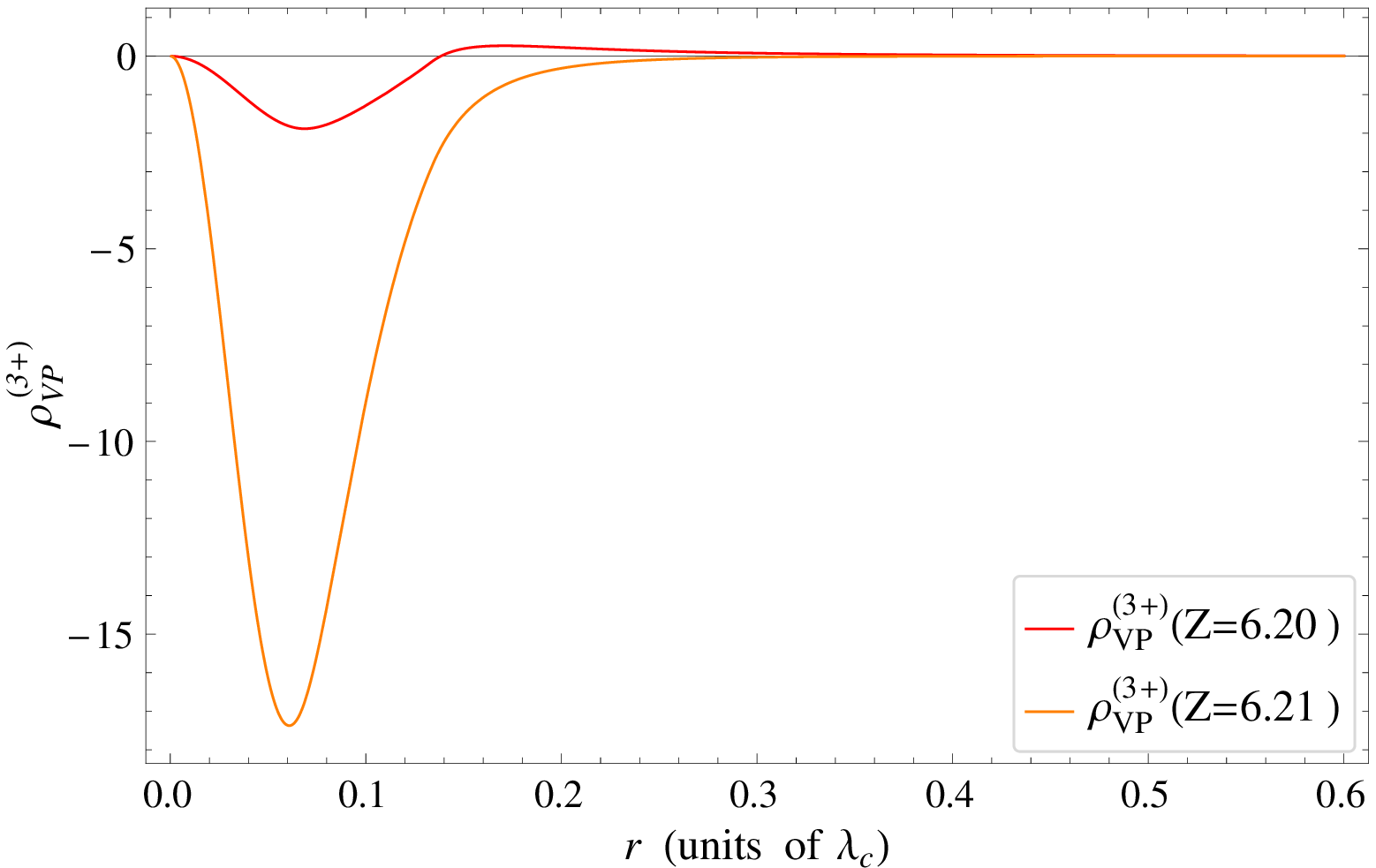}
}
\caption{(Color online) \small $ \r_{VP,3/2}^{(3+)}(r) $ for $\a=0.4$ , $R_0=1/15$ and   ~\subref{pic:5} $R_1\to\infty$, ~\subref{pic:6} $R_1=10R_0$, ~\subref{pic:7} $R_1=5R_0$, ~\subref{pic:8} $R_1=2R_0$. }
\end{figure*}

\section{Peculiar effects in  the screened two-dimensional case}\label{sec:peculiareffects}

In the screened two-dimensional case  the behavior of levels at thresholds of both continua reveals some peculiar features, which are quite different from the unscreened one and absent in one- or three-dimensional DC systems. As in the previous Section, we'll consider them here for the screened potential (\ref{1.0}).

It would be instructive to start directly with the  DC spectral problem (\ref{3.7}) at the lower threshold. For $\e=-1$ the system (\ref{3.7}) takes the from
\beq
\label{1cr}
\left\lbrace\bal
&\frac{d}{d r}\p_1(r)+\frac{1/2-m_j}{r}\,\p_1(r)=-V(r)\p_2(r) \ ,\\
&\frac{d}{d r}\p_2(r)+\frac{1/2+m_j}{r}\,\p_2(r)=(2+V(r))\p_1(r) \ .
\eal\right.
\eeq
For $r\leq R_0$ the solutions of (\ref{1cr}) can be represented as
\beq
\label{sol1in}
\begin{aligned}
	&\p_1(r)=\sqrt{V_0}\, I_{|m_j-1/2|}\(\sqrt{V_0(2-V_0)}\, r\),
	\\&\p_2(r)=\sqrt{2-V_0}\, I_{|m_j+1/2|}\(\sqrt{V_0(2-V_0)}\, r\) \ ,
\end{aligned}
\eeq
while for $R_0 <r< R_1$  they should be written as follows
\beq
\label{sol1mid}
\begin{aligned}
	&\p_1(r)=\sqrt{2Q \over r} \(\l K_{2i|\vk|}\(\sqrt{8Q r}\)+  I_{2i|\vk|}\(\sqrt{8Q r}\)\) \ ,
	\\&\p_2(r)=2I_{1+2i|\vk|}\(\sqrt{8Q r}\)+ \\ & + \(i|\vk|-m_j\)\sqrt{ 2 \over Qr }I_{2i|\vk|}\(\sqrt{8Q r}\) - \\& -\l \(2K_{1+2i|\vk|}\(\sqrt{8Q r}\)-\right. \\ & \left. - \(i|\vk|-m_j\)\sqrt{ 2 \over Qr }K_{2i|\vk|}\(\sqrt{8Q r}\)\)\ .
\end{aligned}
\eeq
The coefficient $\l$ is determined by matching the solutions (\ref{sol1in}) and (\ref{sol1mid}) at $r=R_0$.

The new circumstance, which is specific to the screened two-dimensional case, is that for $r\geq R_1$ the solutions of  (\ref{1cr}) turn out to be the power-like ones, which degree separates the channels with $|m_j|=1/2$, $m_j=-3/2$ and the others into two essentially  different groups. Namely, for $m_j \geq 3/2$ and $m_j \leq -5/2$ the solutions of (\ref{1cr}) in the region $r\geq R_1$ up to a common normalization factor take the form
\beq\label{sol1out}
\begin{aligned}
	m_j \geq 3/2 \ : \ &\p_1(r)=0, \,  \p_2(r)=r^{-(m_j+1/2)},
	\\ m_j \leq -5/2 \ : \ &\p_1(r)=r^{m_j-1/2}, \, \p_2(r)={r^{m_j+1/2} \over m_j+1/2} \ ,
\end{aligned}
\eeq
which leads to normalizable discrete levels at $\e=-1$. For these channels the corresponding critical charges are found via matching  at  $r=R_1$, what gives
\beq
\label{critZadsol}
W_{1,m_j}^-=0
\eeq
for $m_j \geq 3/2$ and
\beq
\label{critZbcsol}
W_{2,|m_j|}^-=0 \ ,
\eeq
for $m_j \leq -5/2$, where $(m_j>0)$
\begin{widetext}
\beq
\begin{aligned}
W_{1,m_j}^\mp  = & K_{2i|\vk|}\(\sqrt{\pm 8QR_1}\) \times \\ & \times
\Bigg(I_{m_j-1/2}(\sqrt{\pm V_0(2\mp V_0)}\,R_0)\(2I_{1+2i|\vk|}(\sqrt{\pm 8QR_0}) \pm \(i|\vk|-m_j\)\,\sqrt{ 2 \over \pm QR_0 }\,I_{2i|\vk|}\(\sqrt{\pm 8QR_0}\)\)-\\& - \sqrt{2\(2\mp V_0\)}\,I_{m_j+1/2}\(\sqrt{\pm V_0(2\mp V_0)}\,R_0\)I_{2i|\vk|}\(\sqrt{\pm 8QR_0}\)\Bigg) + \\ & + I_{2i|\vk|}\(\sqrt{\pm 8QR_1}\) \times \\ & \times \Bigg(I_{m_j-1/2}(\sqrt{\pm V_0(2\mp V_0)}\,R_0) \Big(2K_{1+2i|\vk|}(\sqrt{\pm 8QR_0}) \mp \(i|\vk|-m_j\)\,\sqrt{ 2 \over \pm QR_0 }\,K_{2i|\vk|}\(\sqrt{\pm 8QR_0}\)\Big)+ \\& +\sqrt{2\(2\mp V_0\)}\,I_{m_j+1/2}\(\sqrt{\pm V_0(2\mp V_0)}\,R_0\)  K_{2i|\vk|}\(\sqrt{\pm 8QR_0}\)\Bigg) \ ,
\end{aligned}
\eeq
\beq
\begin{aligned}
W_{2,m_j}^\mp = &\(\sqrt{\pm2Q}K_{2i|\vk|}\(\sqrt{\pm8QR_1}\)-{m_j-1/2 \over \sqrt{R_1}}\(2K_{1+2i|\vk|}(\sqrt{\pm8QR_1}) \mp \(i|\vk|+m_j\)\sqrt{ 2 \over \pm QR_1 }K_{2i|\vk|}\(\sqrt{\pm 8QR_1}\)\)\)\times \\ &\times \Bigg(I_{m_j+1/2}(\sqrt{\pm V_0(2\mp V_0)}\,R_0)\(2I_{1+2i|\vk|}(\sqrt{\pm 8QR_0}) \pm \(i|\vk|+m_j\)\sqrt{ 2 \over \pm QR_0 }I_{2i|\vk|}\(\sqrt{\pm 8QR_0}\)\)-\\& - \sqrt{2\(2\mp V_0\)}I_{m_j-1/2}\(\sqrt{\pm V_0(2\mp V_0)}\,R_0\) I_{2i|\vk|}\(\sqrt{\pm 8QR_0}\)\Bigg) + \\ & + \Bigg(\sqrt{\pm 2Q}I_{2i|\vk|}\(\sqrt{\pm 8QR_1}\)+{m_j-1/2 \over \sqrt{R_1}}\,\Big(2I_{1+2i|\vk|}(\sqrt{\pm 8QR_1})\pm\(i|\vk|+m_j\)\sqrt{ 2 \over \pm QR_1 }I_{2i|\vk|}\(\sqrt{\pm 8QR_0}\)\Big)\Bigg) \times \\ & \times \Bigg(I_{m_j+1/2}(\sqrt{\pm V_0(2\mp V_0)}\,R_0) \Big(2K_{1+2i|\vk|}(\sqrt{\pm 8QR_0}) \mp \(i|\vk|+m_j\)\sqrt{ 2 \over \pm QR_0 }K_{2i|\vk|}\(\sqrt{\pm 8QR_0}\)\Big)+ \\& +\sqrt{2\(2\mp V_0\)}\,I_{m_j-1/2}\(\sqrt{\pm V_0(2\mp V_0)}\,R_0\)\, K_{2i|\vk|}\(\sqrt{\pm 8QR_0}\) \Bigg) \ .
\end{aligned}
\eeq
\end{widetext}

At  the same time, for $|m_j|=1/2$ and $m_j=-3/2$ the system (\ref{1cr}) doesn't possess normalizable solutions at the lower threshold, since for $r\geq R_1$ one finds
\beq\label{sol1out1}
\begin{aligned}
	m_j = 1/2 \ : \ &\p_1(r)=0 \ ,  \quad \p_2(r)=A/r \ ,
	\\ m_j = -1/2 \ : \ &\p_1(r)=0 \ , \quad \p_2(r)=B \ ,
	\\ m_j = -3/2 \ : \ &\p_1(r)=C/r^2 \ , \quad \p_2(r)=-C/r \ .
\end{aligned}
\eeq

Moreover, the solutions (\ref{sol1out1}) cannot be interpreted as the scattering states at the lower threshold too, since in the latter case both components of the WF should  demonstrate the behavior typical for the cylindrical waves, namely $\sim 1/\sqrt{r}$. Remarkably enough, such form of solutions (\ref{sol1out1}) for $r>R_1$ is a specific feature of two spatial dimensions. In the one-dimensional case the corresponding solutions at both thresholds are the scattering states with vanishing wavenumber. In 3 spatial dimensions, on the contrary, at the lower threshold the electronic WF for any orbital number $l$ contains only one non-vanishing component, which behaves like $O(r^{-(l+2)})$ and so all the states  belong to the discrete spectrum, whereas at the upper threshold the $s$-wave is the scattering state, while the others again belong to the discrete spectrum.

In Figs.4a,b there are shown the components $\p_1(r)$ and $\p_2(r)$ of the electronic WF, corresponding to levels with $m_j=\pm 1/2$, lying very close to the threshold, namely $\e=-0.99999999999999999999762$ ($Z=3.808194785685109813175$) for $m_j=1/2$ and $\e=-0.99999999999999999999773$ ($Z=5.57$)) for $m_j=-1/2$. Figs.4c,d represent  $\p_1(r)$ and $\p_2(r)$  of levels with $m_j=\pm 3/2$ for $\e=-0.999999113$ ($Z=6.2059331$) and $\e=-0.999999118$ ($Z=6.38159669$)), respectively.

\begin{figure*}[ht!]
\subfigure[]{\label{pic:9}
		\includegraphics[width=\columnwidth]{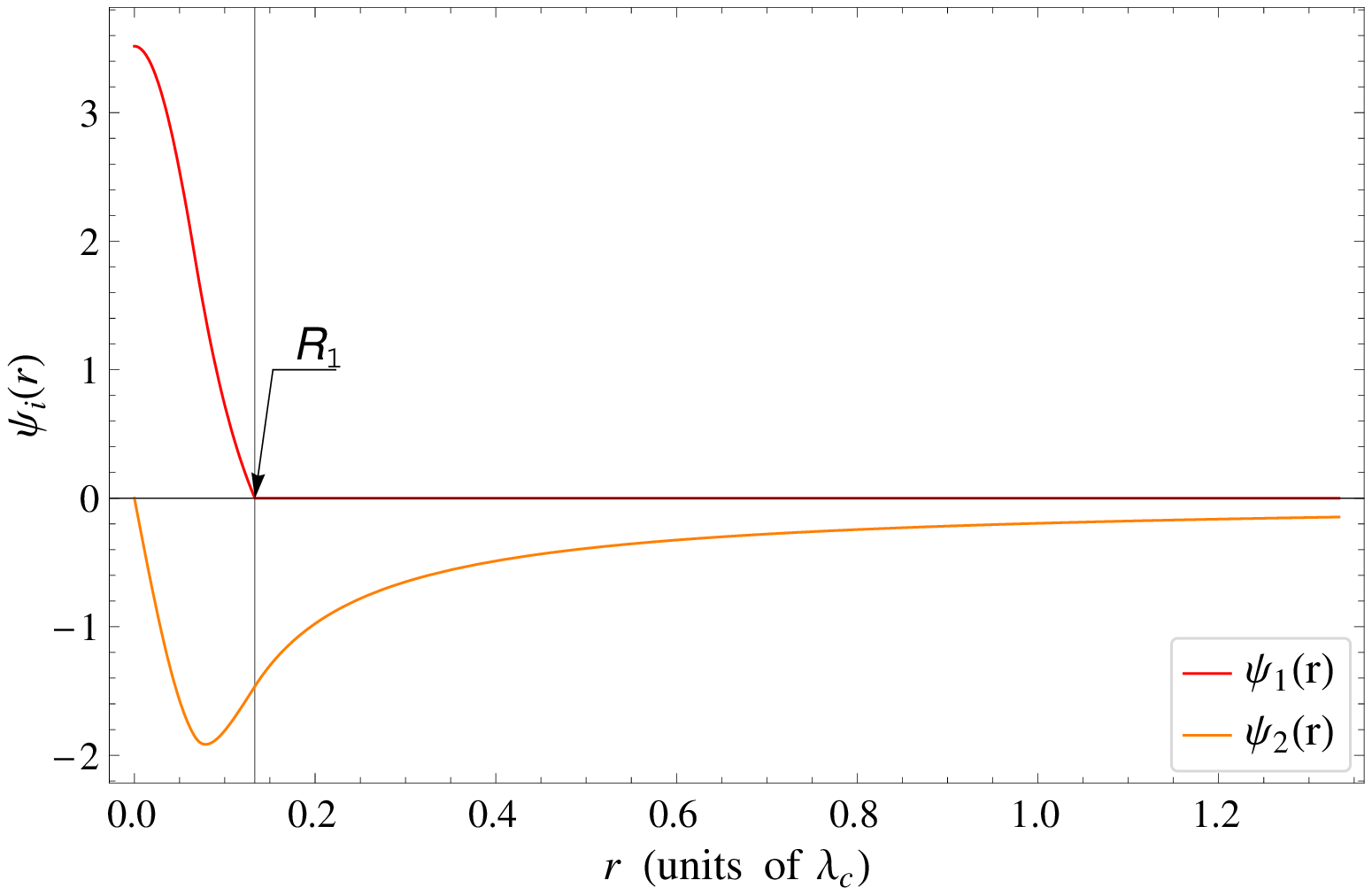}
}
\hfill
\subfigure[]{\label{pic:10}
		\includegraphics[width=\columnwidth]{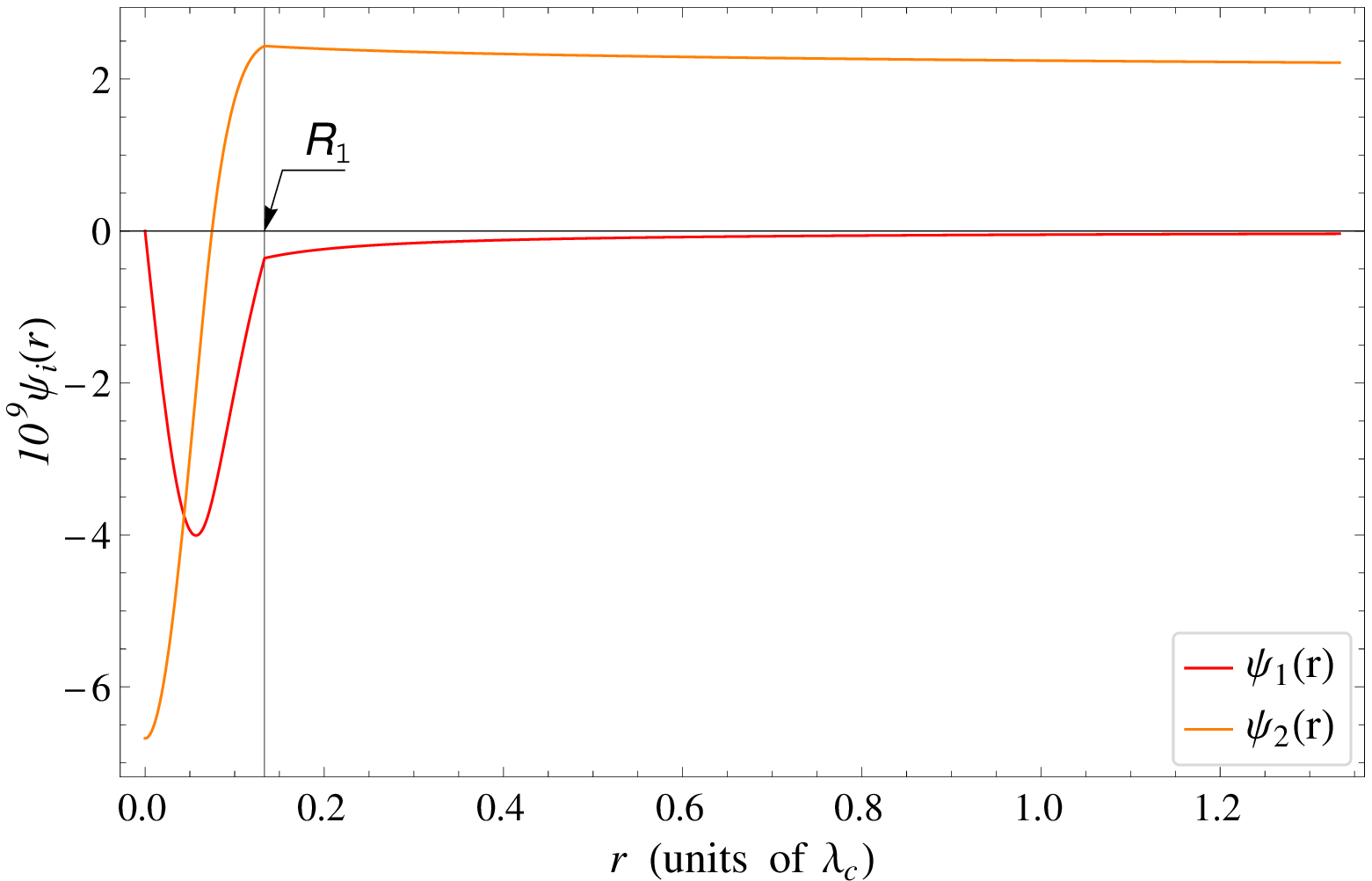}
}
\vfill
\subfigure[]{\label{pic:11}
	\includegraphics[width=\columnwidth]{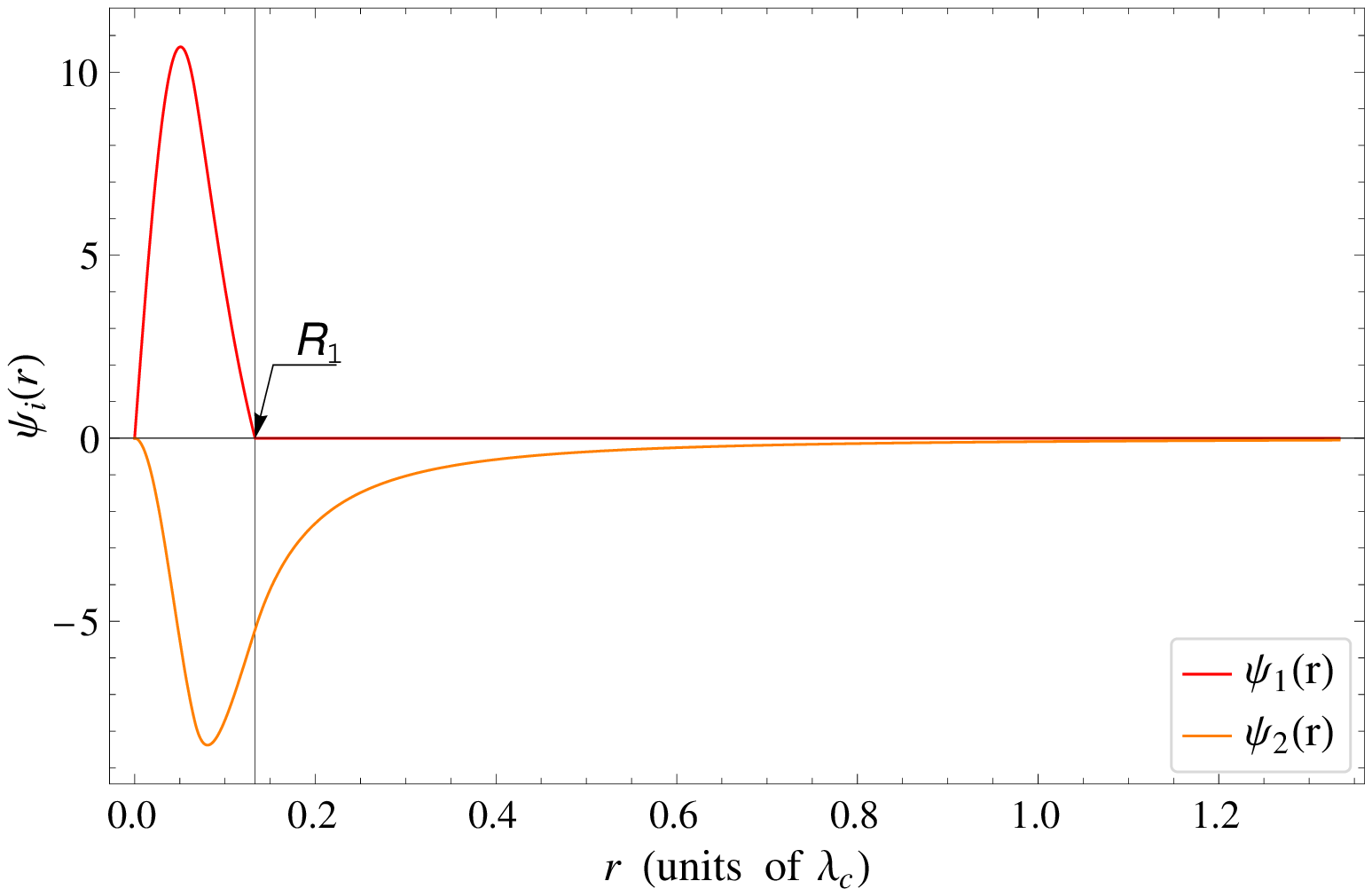}
}
\hfill
\subfigure[]{\label{pic:12}
		\includegraphics[width=\columnwidth]{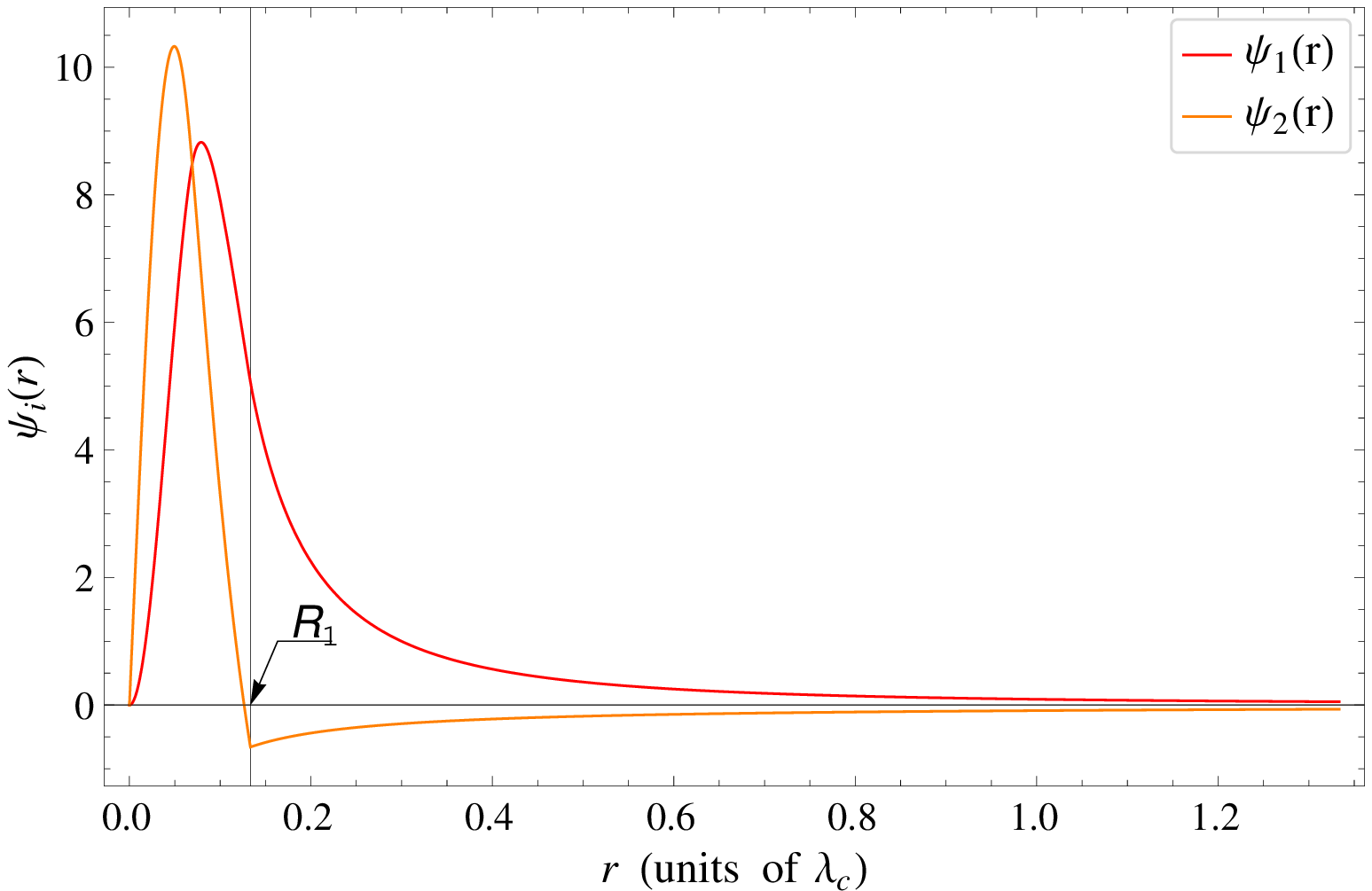}
}
\caption{(Color online) \small   $\p_1(r)$ and  $\p_2(r)$ for $\a=0.4$, $R_0=1/15$, $R_1=2R_0$ and ~\subref{pic:9} $m_j=1/2$, $Z=3.808194785685109813175$, ~\subref{pic:10} $m_j=-1/2$, $Z=5.57$, ~\subref{pic:11} $m_j=3/2$, $Z=6.2059331$, ~\subref{pic:12} $m_j=-3/2$, $Z=6.38159669$. }\label{pic:9-12}
\end{figure*}

Although in this case we deal with non-normalizable solutions, the equations for the corresponding critical charges can be also found via matching  the solutions  (\ref{sol1mid}) and (\ref{sol1out1}) at  $r=R_1$, what gives:\\
\begin{widetext}
for $m_j=1/2$
\begin{multline}\label{critad}
\hbox{Im}\Bigg[ \Bigg(Q\sqrt{1-{2\over V_0}}\,J_{-2 i |\vk|}(-2 \sqrt{-2 Q R_0})\, J_1(\sqrt{V_0(V_0-2)}\,R_0) + \\
 + \left( \sqrt{-2 Q R_0}\,J_{1-2 i |\vk|}(-2 \sqrt{-2 Q R_0}) - (1/2 + i|\vk|) J_{-2 i |\vk|}(-2 \sqrt{-2 Q R_0})\right)\,J_0 (\sqrt{V_0(V_0-2)}\,R_0)\Bigg)\times \\
 \times J_{2 i |\vk|}(2 \sqrt{-2 Q R_1})\Bigg] = 0 \ ,
\end{multline}
for $m_j=-1/2$
\begin{multline}\label{critbc}
\hbox{Im}\Bigg[\Bigg( Q\sqrt{1-{2\over V_0}}\,J_{-2 i |\vk|}(-2 \sqrt{-2 Q R_0})\, J_1 (\sqrt{V_0(V_0-2)}\,R_0) - \\
 -\left( \sqrt{-2 Q R_0}\,J_{1-2 i |\vk|}(-2 \sqrt{-2 Q R_0}) + (1/2 - i|\vk|)\, J_{-2 i |\vk|}(-2 \sqrt{-2 Q R_0})\right)J_1 (\sqrt{V_0(V_0-2)}\,R_0)\Bigg) \times \\
 \times J_{2 i |\vk|}(2 \sqrt{-2 Q R_1})\Bigg]  = 0 \ ,
\end{multline}
and for $m_j=-3/2$
\begin{multline}
\hbox{Im}\Bigg[ \Bigg(Q\sqrt{1-{2\over V_0}}\,J_{-2 i |\vk|}(-2 \sqrt{-2 Q R_0}) J_1 (\sqrt{V_0(V_0-2)}\,R_0) -\\
 -\left( \sqrt{-2 Q R_0}\,J_{1-2 i |\vk|}(-2 \sqrt{-2 Q R_0}) + (3/2 - i|\vk|) J_{-2 i |\vk|}(-2 \sqrt{-2 Q R_0})\right)\,J_2 (\sqrt{V_0(V_0-2)}\,R_0)\Bigg)\times \\
 \times \left( (Q R_1+3/2+i|\vk|)\,J_{2 i |\vk|}(2 \sqrt{-2 Q R_1}) - \sqrt{-2 Q R_1}\,J_{1+2 i |\vk|}(2 \sqrt{-2 Q R_1})\right) \Bigg]=0 \ .
\end{multline}
The validity of these equations can be easily verified by taking  the limit $\e \to -1$ in the equations for discrete levels with $-1< \e <1$ and $|m_j|<Q$ for the screened potential (\ref{1.0})
\begin{multline}
\label{4.23}
\mathrm{Im}\Big[\Big(K_{m_j+1/2}(\g R_1)\((m_j+ Q/\g)\F(b,c,2\g R_1)+b \F(b+1,c,2 \g R_1)\)+\\K_{m_j-1/2}(\g R_1)\(-(m_j+ Q/\g)\F(b,c,2\g R_1)+b \F(b+1,c,2 \g R_1)\)\Big)\\
\times\Big(\sqrt{(\e+V_0+1)(1-\e)}\,J_{m_j-1/2}(\z R_0)\,\(-(Q/\g+m_j)\F^*(b,c,2\g R_0)+b^*\, \F^*(b+1,c,2\g R_0)\)\\
+\sqrt{(\e+V_0-1)(1+\e)}\,J_{m_j+1/2}(\z R_0)\,\((Q/\g+m_j)\F^*(b,c,2\g R_0)+b^*\, \F^*(b+1,c,2\g R_0)\)\Big)\Big]=0 \ .
\end{multline}
\end{widetext}

The mirror-symmetrical situation takes place now at the threshold of the upper continuum, since in the screened case the condensation of levels at $\e \to 1$ disappears and the total number of discrete levels becomes finite. The whole difference is the change in signs of $m_j$.
Namely, for  $\e=1$ the system (\ref{3.7}) takes the form
\beq
\label{1ncr}
\left\lbrace\bal
&\frac{d}{d r}\p_1(r)+\frac{1/2-m_j}{r}\,\p_1(r)=(2-V(r))\p_2(r) \ ,\\
&\frac{d}{d r}\p_2(r)+\frac{1/2+m_j}{r}\,\p_2(r)=V(r)\p_1(r) \ .
\eal\right.
\eeq
For $r\leq R_0$ the solutions of (\ref{1ncr}) are chosen as
\beq
\label{sol1nin}
\begin{aligned}
	&\p_1(r)=\sqrt{V_0+2}\,I_{|m_j-1/2|}\(\sqrt{-V_0(2+V_0)}\,r\) \ ,
	\\&\p_2(r)=\sqrt{V_0}\,I_{|m_j+1/2|}\(\sqrt{-V_0(2+V_0)}\,r\) \ ,
\end{aligned}
\eeq
while for $R_0 <r< R_1$  they should be written as follows
\beq
\label{sol1nmid}
\begin{aligned}
	&\p_1(r)=2I_{1+2i|\vk|}\(\sqrt{-8Q r}\)+ \\ & + \(i|\vk|-m_j\)\,\sqrt{ 2 \over -Qr }\,I_{2i|\vk|}\(\sqrt{-8Q r}\) - \\ & - \l \(2K_{1+2i|\vk|}\(\sqrt{-8Q r}\)- \right. \\ & \left.  - \(i|\vk|-m_j\)\sqrt{ 2 \over -Qr }\,K_{2i|\vk|}\(\sqrt{-8Q r}\)\) \ ,
\\&
	\p_2(r)=\sqrt{-2Q \over r} \(\l K_{2i|\vk|}\(\sqrt{-8Q r}\)+I_{2i|\vk|}\(\sqrt{-8Q r}\)\) \ ,
\end{aligned}
\eeq
where the coefficient  $\l$ quite similar to the lower threshold is found via matching of solutions (\ref{sol1nin}) and (\ref{sol1nmid}) at the point  $r=R_0$.

Again, for $r\geq R_1$ the solutions of  (\ref{1ncr}) turn out to be the power-like ones, whose degree separates now the channels with $|m_j|=1/2$, $m_j=3/2$ and the others into two essentially  different groups. Namely, for $m_j \geq 5/2$ and $m_j \leq -3/2$ for the solutions of (\ref{1ncr}) in the region $r\geq R_1$ up to a common normalization factor one finds
\beq\label{sol1nout}
\begin{aligned}
	m_j \geq 5/2 \ : \ &\p_1(r)={r^{1/2-m_j} \over 1/2 - m_j}, \quad \p_2(r)=r^{-m_j-1/2} \ ,
	\\ m_j \leq -3/2 \ : \ &\p_1(r)=r^{m_j-1/2}, \,  \p_2(r)=0 \ .
\end{aligned}
\eeq
These solutions give rise to normalizable discrete levels at $\e=1$. The corresponding ``upper critical'' charges, when the virtual levels, descending to the threshold of the upper continuum from above, transform into the real ones, can be found from equations, which are derived by matching the solutions of (\ref{1ncr}) at the point $r=R_1$, namely
\beq
\label{critZadnsol}
W_{2,m_j}^+=0
\eeq
for $m_j \geq 5/2$ and
\beq
\label{critZbcnsol}
W_{1,|m_j|}^+=0
\eeq
for $m_j \leq -3/2$.

For $|m_j|=1/2$ and $m_j=3/2$ the system (\ref{1ncr}) doesn't possess normalizable solutions at the upper threshold, which could be classified either as discrete levels or as the scattering states, since for $r\geq R_1$ the latter take the form
\beq\label{sol1nout1}
\begin{aligned}
	m_j = 1/2 \ : \ &\p_1(r)=B \ , \quad  \p_2(r)=0 \ ,
	\\ m_j = -1/2 \ : \ &\p_1(r)=A/r \ , \quad \p_2(r)=0 \ ,
	\\ m_j = 3/2 \ : \  &\p_1(r)=-C/r \ , \quad \p_2(r)=C/r^2 \ .
\end{aligned}
\eeq
The corresponding equations, defining the upper critical charges, are found now by the same procedure as for the lower ones and read:\\
\begin{widetext}
for $m_j=1/2$
\begin{multline}\label{critnad}
\hbox{Im}\Bigg[\Bigg( Q\sqrt{1+{2\over V_0}}\,J_{-2 i |\vk|}(2 \sqrt{2 Q R_0})\, J_0(\sqrt{V_0(V_0+2)}\,R_0)\, + \\
 +\left( \sqrt{2 Q R_0}\,J_{1-2 i |\vk|}(2 \sqrt{2 Q R_0}) - (1/2 - i|\vk|) J_{-2 i |\vk|}(2 \sqrt{2 Q R_0})\right)  J_1(\sqrt{V_0(V_0+2)}\,R_0)\Bigg)\, J_{2 i |\vk|}(2 \sqrt{2 Q R_1})\Bigg] =0 \ ,
\end{multline}
for $m_j=-1/2$
\begin{multline}\label{critnbc}
\hbox{Im}\Bigg[ \Bigg(-Q\sqrt{1+{2\over V_0}}\,J_{-2 i |\vk|}(2 \sqrt{2 Q R_0})\, J_1(\sqrt{V_0(V_0+2)}\,R_0)\, + \\
 + \left( \sqrt{2 Q R_0}\,J_{1-2 i |\vk|}(2 \sqrt{2 Q R_0}) + (1/2 + i|\vk|)\, J_{-2 i |\vk|}(2 \sqrt{2 Q R_0})\right) J_0(\sqrt{V_0(V_0+2)}R_0)\Bigg)\, J_{2 i |\vk|}(2 \sqrt{2 Q R_1})\Bigg]=0 \ ,
\end{multline}
while for $m_j=3/2$
\begin{multline}
\hbox{Im}\Bigg[ \Bigg(Q\sqrt{1+{2\over V_0}}\,J_{-2 i |\vk|}(2 \sqrt{2 Q R_0})\, J_1(\sqrt{V_0(V_0+2)}\,R_0)\, + \\
 + \left( \sqrt{2 Q R_0}\,J_{1-2 i |\vk|}(2 \sqrt{2 Q R_0}) - (3/2 - i|\vk|) J_{-2 i |\vk|}(2 \sqrt{2 Q R_0})\right)\,J_2 (\sqrt{V_0(V_0+2)}\,R_0)\Bigg) \times \\
 \times \left( (Q R_1-3/2-i|\vk|)\,J_{2 i |\vk|}(2 \sqrt{2 Q R_1}) + \sqrt{2 Q R_1}\,J_{1+2 i |\vk|}(2 \sqrt{2 Q R_1})\right)\Bigg]=0 \ .
\end{multline}
\end{widetext}

The peculiar feature of the channel $|m_j|=1/2$ is a substantial difference in behavior of discrete levels in this channel by approaching both thresholds, not only between this channel and the others, but also between $m_j=1/2$ and $m_j=-1/2$. Moreover, the last difference turns out to be the most impressive. Figs.5a,b represent the evolution of existing discrete levels on the interval $0< Z <10$ by growing $Z$ for $\a=0.4$, $R_0=1/15$, $R_1=2R_0$ and $|m_j|=1/2\, , 3/2$. The vertical dashed lines denote the positions of the lower critical charges, when the levels approach the lower threshold, while the dotted ones --- the upper ones, i.e. the moments of transformation of virtual levels into the real ones at the upper threshold. Fig.5a corresponds to $|m_j|=1/2$, while Fig.5b to $|m_j|=3/2$.
\begin{figure*}[ht!]
\subfigure[]{\label{pic:13}
		\includegraphics[width=\columnwidth]{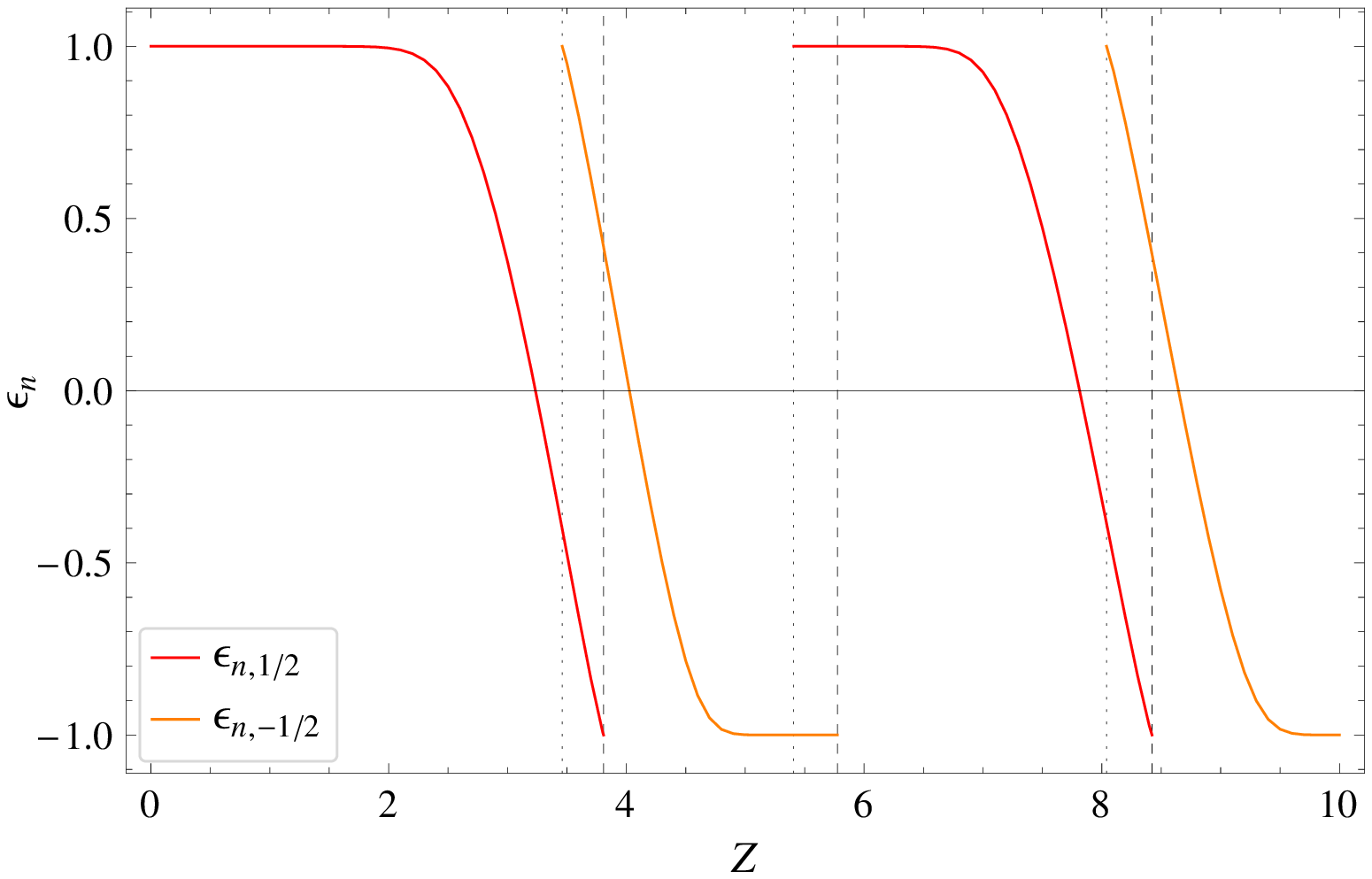}
}
\hfill
\subfigure[]{\label{pic:14}
		\includegraphics[width=\columnwidth]{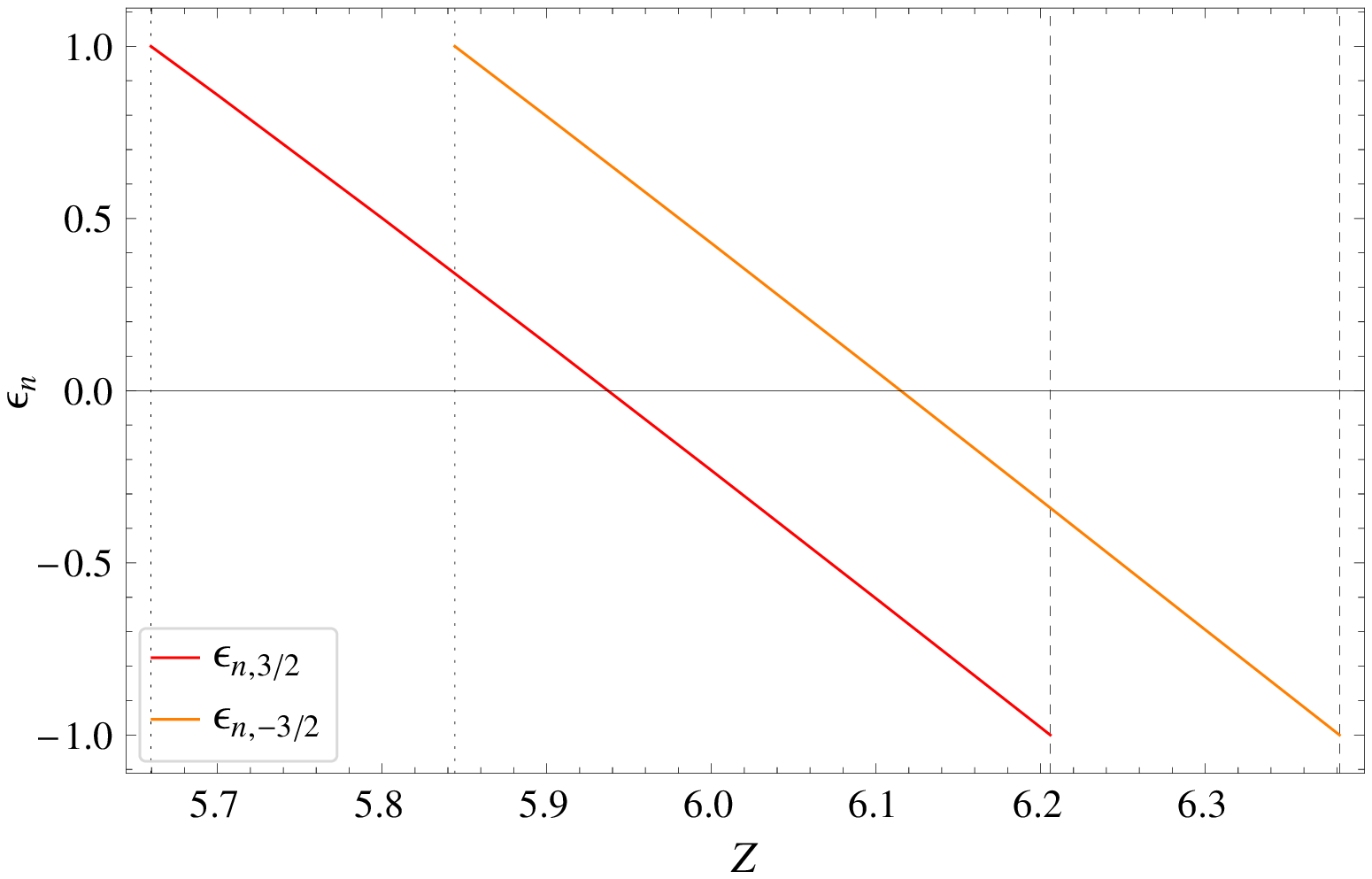}
}
\caption{(Color online) \small  The evolution of discrete levels by growing $Z$  for $\a=0.4$, $R_0=1/15$, $R_1=2R_0$ and ~\subref{pic:13} $|m_j|=1/2$ \ ,  ~\subref{pic:14} $|m_j|=3/2$. The vertical dashed lines denote the positions of the lower critical charges, when the levels approach the lower threshold, while the dotted ones --- the moments of transformation of virtual levels into the real ones at the upper threshold.}\label{pic:13-14}
\end{figure*}

Such impressive difference in behavior of levels with $m_j=\pm 1/2$ near the thresholds can be well understood with account of  expansion of corresponding equations for discrete levels with $-1< \e <1$ and $|m_j|<Q$ (\ref{4.23}) for $\e \to \pm 1$. Namely, for $\e \to -1$ the expansion of the corresponding equation for the levels with $m_j=1/2$ gives
\beq
\label{eqAD}
A_0 \(Q/\g^2 + m_j/\g\)  + C_0 \ln\g + D_0 =0,
\eeq
with $\g=\sqrt{1-\e^2}$, while the coefficients $A_0 \ , B_0 \ , C_0 \ , D_0$ are determined by the following expressions
\beq
A_0(Q,R_1)={\sqrt{8 V_0}\over R_1}\, \hbox{Im}\left[B_0(Q) J_{-2i|\vk|}(-2\sqrt{-2 Q R_1}) \right],
\eeq
\begin{widetext}
\begin{multline}
B_0(Q)=Q\sqrt{1-{2\over V_0}}\,J_{2i|\vk|}(2\sqrt{-2 Q R_0})\,J_1(R_0\sqrt{(V_0-2)V_0})-\\
-(\sqrt{-2 Q R_0}\,J_{1+2i|\vk|}(2\sqrt{-2 Q R_0})+(1/2-i|\vk|)\,J_{2i|\vk|}(2\sqrt{-2 Q R_0}))\,J_0(R_0\sqrt{(V_0-2)V_0}) \ ,
\end{multline}
\begin{multline}
C_0(Q,R_1)=\hbox{Im}\Big[ B_0(Q)\big(  Q R_1 J_{-2i|\vk|}(-2\sqrt{-2 Q R_1})-\\
-(\sqrt{-2 Q R_1}\,J_{1-2i|\vk|}(-2\sqrt{-2 Q R_1})-(1/2+i|\vk|)\, J_{-2i|\vk|}(-2\sqrt{-2 Q R_1})\big)\Big] \ ,
\end{multline}
\begin{multline}
D_0(Q,R_1)=-{1\over \sqrt{2 Q R_1}}{1\over 6 R_1}\hbox{Im}\Bigg[
{1\over 2}\sqrt{-2 Q R_1}J_{-2 i \vk}(-2\sqrt{-2 Q R_1})\times\\
\times\Big\{\Big[2\Big(3+7 Q^2+2 i \vk(3+4 Q^2)\Big)J_{1+2i\vk}(2\sqrt{-2 Q R_0})-\\
-\Big(6-4 Q^2+3(1/2-i\vk)V_0+4 Q R_0(1+3 V_0)\Big)\sqrt{-2 Q R_0}J_{2i\vk}(2\sqrt{-2 Q R_0})
\Big]J_0(R_0\sqrt{V_0(V_0-2)})+\\
+\Big[
4\Big(6+2 Q R_0-4 Q^2(V_0-1)-3 V_0\Big)J_{1+2i\vk}(2\sqrt{-2 Q R_0})-\\
-\Big(2+2 V_0-3V_0^2+4i\vk(2V_0-1)\Big)\sqrt{-2 Q R_0}J_{2i\vk}(2\sqrt{-2 Q R_0})
\Big]{Q J_1(R_0\sqrt{V_0(V_0-2)})\over \sqrt{V_0(V_0-2)}}
\Big\}+\\
+i\sqrt{2 V_0}B_0(Q)\Big(2 R_1(3+4 Q^2-2 Q R_1)J_{1-2i\vk}(-2\sqrt{-2 Q R_1})-\\
-(3 Q(1+R_1^2)+2 R_1(i\vk-1))\sqrt{-2 Q R_1}J_{-2i\vk}(-2\sqrt{-2 Q R_1})
\Big)
\Bigg]+C_0(Q,R_1)\left(\g_E+\ln(R_1/2)\right) \ ,
\end{multline}
with $\g_E$ being the EulerGamma.
\end{widetext}
The critical charges in this case coincide with zeros of the function $A_0(Q,R_1)$, since at the lower threshold  $\gamma\to 0$ and so the equation (\ref{eqAD}), multiplied by $\gamma^2$, takes the form $Q A_0 + m_j A_0\gamma + C_0 \gamma^2 \ln\gamma + D_0 \gamma^2 =0$, whence it follows that  $A_0(Q,R_1)$ should vanish. In  its turn, the equation  $A_0(Q,R_1)=0$ is completely equivalent to the equation (\ref{critad}) for the corresponding lower critical charges (up to the complex conjugation).

To the contrary, for levels with $m_j=-1/2$  the expansion of the corresponding equation (\ref{4.23})  for $\e \to -1$ by keeping the leading terms containing $\gamma^{-1}\ln\gamma$ and $\gamma^{-1}$ and omitting the next-to-leading ones, starting from $\ln\gamma$, yields an equation, which allows for a simple analytic solution of the form
\beq
\label{eqBCfull}
\e=-\sqrt{1-C^2(Q,R_1)} \ ,
\eeq
with the function $C(Q,R_1)$ being determined by the following expression
\begin{widetext}
\begin{multline}
\label{eqBC}
C(Q,R_1)={2\over R_1}\exp\left({\hbox{Im}\left[B(Q)(\sqrt{-2 Q R_1}\,J_{1-2i|\vk|}(-2\sqrt{-2 Q R_1})+(m_j-i|\vk|)J_{-2i|\vk|}(-2\sqrt{-2 Q R_1}))\right]   \over 2 Q R_1 \hbox{Im}\left[B(Q)\, J_{-2i|\vk|}\,(-2\sqrt{-2 Q R_1})\right]}-\g_E \right) \ ,
\end{multline}
\begin{multline}
B(Q)=Q\sqrt{1-{2\over V_0}}\,J_{2i|\vk|}(2\sqrt{-2 Q R_0})\,J_0(R_0\sqrt{(V_0-2)V_0})+\\
+(\sqrt{-2 Q R_0}\,J_{1+2i|\vk|}(2\sqrt{-2 Q R_0})-(1/2+i|\vk|)\,J_{2i|\vk|}(2\sqrt{-2 Q R_0}))J_1(R_0\sqrt{(V_0-2)V_0}) \ .
\end{multline}
\end{widetext}
In this case the lower critical charges correspond to the zeros of the denominator in the exponent in the expression for $C(Q,R_1)$ (\ref{eqBC}). Again, the latter precisely coincides with the equation for critical charges in this channel (\ref{critbc}) (up to the complex conjugation).

The origin of such apparent difference in behavior of levels with $m_j=\pm 1/2$ by approaching the  thresholds of continua, which is clearly seen in Fig.5a, lyes in the different structure of approximate equations (\ref{eqAD}) and (\ref{eqBCfull}) and hence, of their solutions. Namely, the r.h.s. of (\ref{eqBCfull}) contains an exponent in the function $C(Q,R_1)$, which decreases very rapidly, when $Z$ approaches the corresponding $Z_{cr}$. As a consequence, for $R_1=2 R_0$ in the vicinity of the first $Z_{cr}=5.7757028739...$ in the channel with $m_j=-1/2$ the level  very quickly takes on values close to  $\e = -1$. In particular, for $Z=Z_{cr}-1/10$ one obtains  $1+\e \simeq 10^{-45}$, for $Z=Z_{cr}-1/100$ the level lyes  much closer   $1+\e \simeq 10^{-474}$, while for $Z=Z_{cr}-1/1000$ the difference should be already estimated as $1+\e \simeq 10^{-4757}$. To the contrary, for $m_j=1/2$ with the same screening in the vicinity of the first $Z_{cr}=3.8081947856...$ in the channel for $Z=Z_{cr}-1/100$ the position of the level is estimated as $1+\e \simeq 10^{-2}$, while for $Z=Z_{cr}-1/1000$ one finds only $1+\e \simeq 10^{-3}$. This is the reason, why the slopes of curves for levels with $m_j=\pm 1/2$ by approaching the thresholds turn out to be substantially different. Indeed here, in these estimates, the benefit of approximate equations (\ref{eqAD}) and (\ref{eqBCfull}) is manifested most clearly, since they allow to monitor the position of the levels even in the case when the latter are located extremely close, e.g., $\simeq -1+10^{-4757}$, to the threshold.

It is worth to note that such exponentially slow approach of levels in the channel with $m_j=-1/2$ to the lower threshold  and with $m_j=1/2$ to the upper one reflects in fact the well-known feature of the two-dimensional non-relativistic  quantum-mechanical well, in which at least one exponentially  shallow discrete level $\e_{0,1/2}$ exists for arbitrary  small well parameters  in the partial channel with $m_j=1/2$  (the only condition is the convergence of the integral $\int \! dr \ r U(r)$) \cite{LL}.  In our DC problem the   position of such a level  for small $Z$ near the upper threshold is defined by the relation, quite similar to (\ref{eqBCfull})-(\ref{eqBC})
\beq
\label{ADnear1}
\e_{0,1/2}=\sqrt{1-\wtC^2(Q,R_1)},
\eeq
where
\begin{widetext}
\beq\bal
&\wtC(Q,R_1)=  {2\over R_1}  \exp\left[{ F(-\vk, R_1)\wtB(-\vk)-F(\vk, R_1)\wtB(\vk)\over 2 Q R_1 (J_{2\vk}(2\sqrt{2 Q R_1}) \wtB(-\vk) - J_{-2\vk}(2\sqrt{2 Q R_1}) \wtB(\vk))}-\g_E \right] \ ,
\eal\eeq
\beq\bal
&F(\vk, R_1)=  \sqrt{2 Q R_1} J_{1-2\vk}(2\sqrt{2 Q R_1})-(1/2-\vk)J_{-2\vk}(2\sqrt{2 Q R_1}) \ ,
\eal \eeq
\begin{multline}
	\wtB(\vk)=Q\sqrt{1+{2\over V_0}}J_{2\vk}(2\sqrt{2 Q R_0})J_0 (R_0\sqrt{(V_0+2)V_0})+
\\
	+(\sqrt{2 Q R_0}J_{1+2\vk}(2\sqrt{2 Q R_0})-(1/2+\vk)J_{2\vk}(2\sqrt{2 Q R_0}))J_1 (R_0\sqrt{(V_0+2)V_0}) \ .
\end{multline}
\end{widetext}
For $R_0=1/15$, $R_1=2 R_0$, $\alpha=0.4$ the behavior of this level on the interval $0 < Z \alpha < m_j$ is shown in Fig.6a, while in Fig.6b its dependence on the cutoff $R_1$ on the interval $R_0 < R_1 < 500 R_0$ is given for $Z=1/10$. For $R_1 \to \inf$  it transforms into the lowest  discrete level in this partial channel for the unscreened case with the limiting value $\e_{0,1/2} \simeq 0.996828726314219 \dots$.
\begin{figure*}[ht!]
\subfigure[]{\label{pic:21}
		\includegraphics[width=\columnwidth]{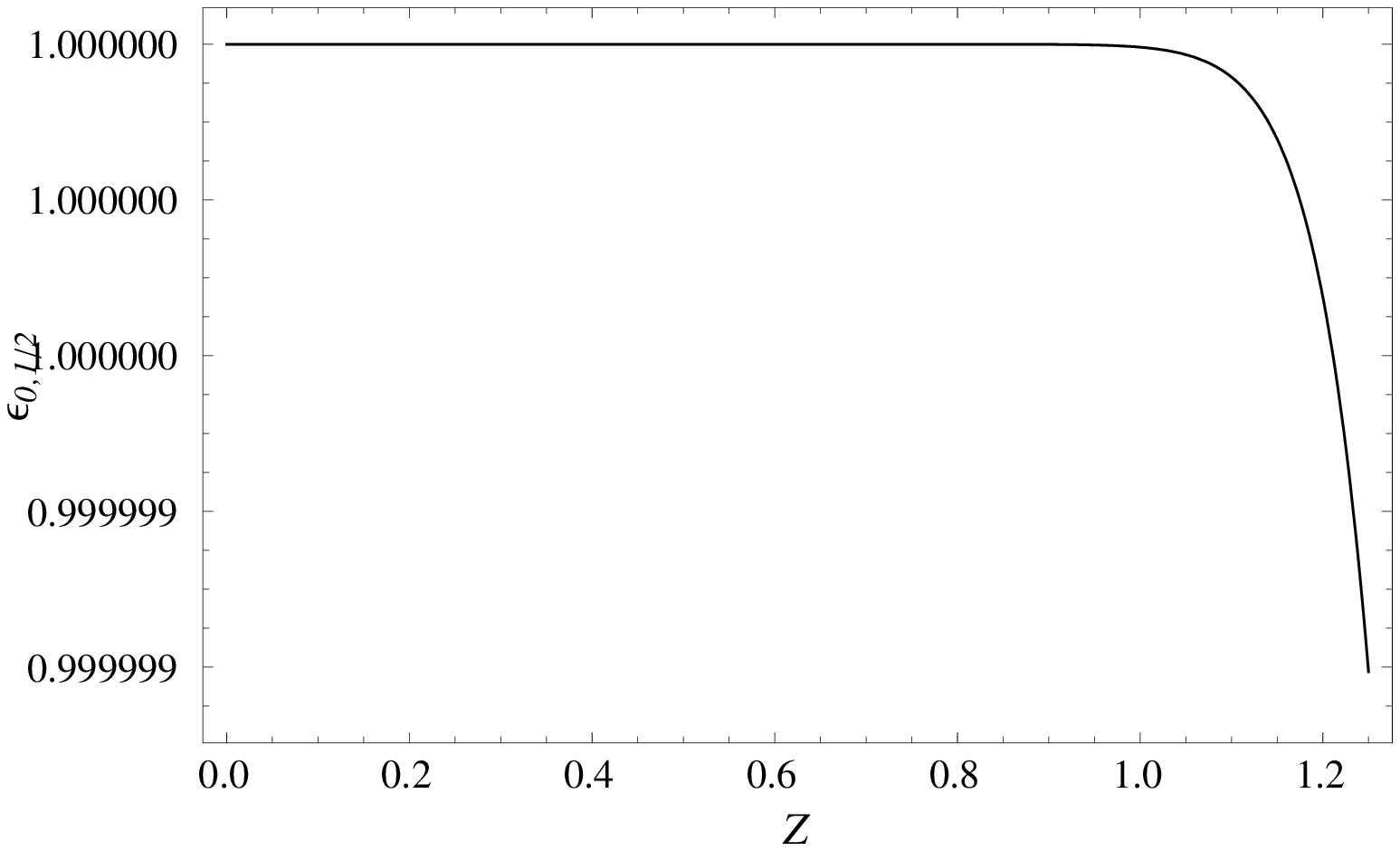}
}
\hfill
\subfigure[]{\label{pic:22}
		\includegraphics[width=\columnwidth]{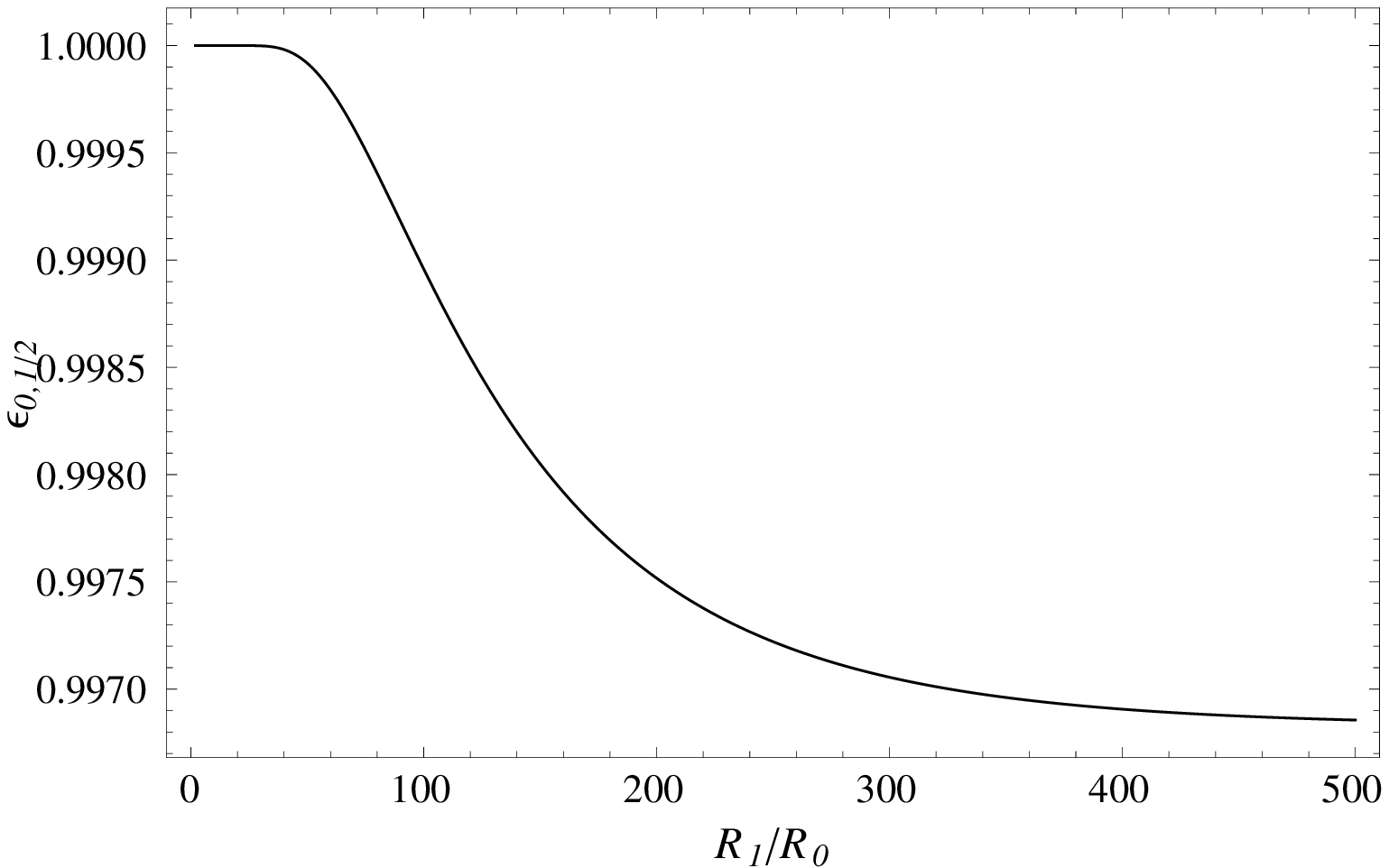}
}
	\caption{ \small ~\subref{pic:21}: The behavior of the first discrete level $\e_{0,1/2}$ as a function of the source charge $Z$ for $\a=0.4$, $R_0=1/15$, $R_1=2R_0$;~\subref{pic:22}: the dependence of the level $\e_{0,1/2}$ on the cutoff $R_1$ for $\a=0.4$, $R_0=1/15$ and $Z=1/10$.} \label{pic:21-22}
\end{figure*}
From Figs.6a,b there follows that for $Z$ very close to zero there exists always one discrete level $\e_{0,1/2}$ with $m_j=1/2$, which with decreasing $Z$ tends very rapidly to 1. In particular, for our standard set of parameters  $R_0=1/15$, $R_1=2 R_0$, $\alpha=0.4$ and $Z=1/10$ one has $\e_{0,1/2} \simeq 1-10^{-107}$, whereas for  $Z=1/100$ it lyes already at $ \simeq 1- 10^{-1084}$. Indeed such  behavior is demonstrated by other levels with $m_j=1/2$ just after creation at the upper threshold and with  $m_j=-1/2$ by approaching the lower one for $Q>|m_j|$.

Another peculiar feature of the channels with  $|m_j|=1/2$ and $|m_j|=3/2$ is the behavior of the induced density by crossing the lower $Z_{cr}$, since in this case the Fano rule in the form (\ref{3.26}) for $\D\r_{VP}(r)$ doesn't work. Nevertheless, the jump in the density by crossing the lower threshold can be well  understood by taking into account that the non-normalizable solutions (\ref{sol1out1}) and (\ref{sol1nout1}) appear as  the limiting behavior of corresponding discrete levels for $\e \to -1$, and so instead of (\ref{3.26}) we get
\beq\label{3.26a}
\D\r_{VP}(r)=-2|e|\lim_{\e_n \to -1} \p_{n,m_j}(r)^{\dagger}\p_{n,m_j}(r) \ .
\eeq
In particular, the jump in the induced density distribution by diving of the corresponding level with  $|m_j|=1/2$ and $m_j=-3/2$ into the lower continuum turns out to be the (improper) limit of normalizable distributions spread over the whole half-axis $0 \leqslant r \leqslant \inf $, which carries with an amount $(-2 |e|)$ of the total induced charge. The latter result is well verified via direct numerical calculation. The behavior of the jumps in the induced density depending on $\D Z$ by crossing the lower threshold are shown in Figs.7a,c,e for $m_j=1/2$ and in Fig.7b,d,f for $m_j=-3/2$. In these figures the difference between induced densities taken at $Z_{cr}-\D Z/2$ and $Z_{cr}+\D Z/2$ for decreasing $\D Z$ is presented. As it was already stated in Section 3, the jumps in the induced density by levels diving into the lower continuum represent themselves the essentially non-perturbative effect,   completely included in $\r_{VP}^{(3+)}$, while  $\r_{VP}^{(1)}$ does not participate in it and still makes an exactly vanishing contribution to the total induced charge. To demonstrate the effect of spreading of jumps in the induced density at the threshold more clearly, in Figs.7c,d and in Figs.7e,f the weighted densities $r \times \r_{VP,m_j}^{(3+)}(r)$ and $r^2 \times \r_{VP,m_j}^{(3+)}(r)$ are used. From Figs.7 it should be clear that the less is $\D Z$ by crossing the threshold, the   more spread is the jump in the induced density distribution, but the loss of  amount $(-2 |e|)$ of the total induced charge remains unchanged.
\begin{figure*}[ht!]
\subfigure[]{\label{pic:15}
		\includegraphics[width=\columnwidth]{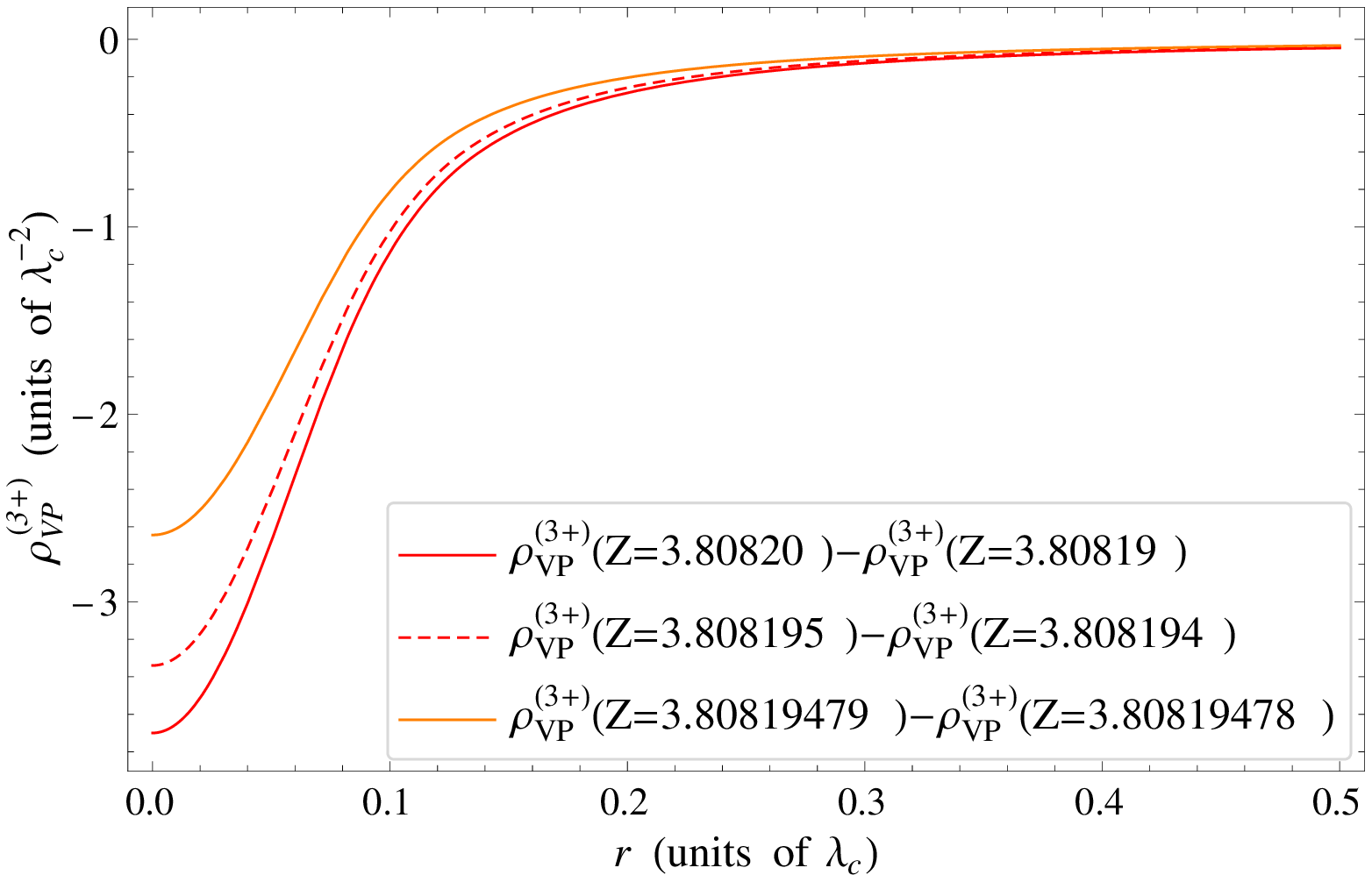}
}
\hfill
\subfigure[]{\label{pic:16}
		\includegraphics[width=\columnwidth]{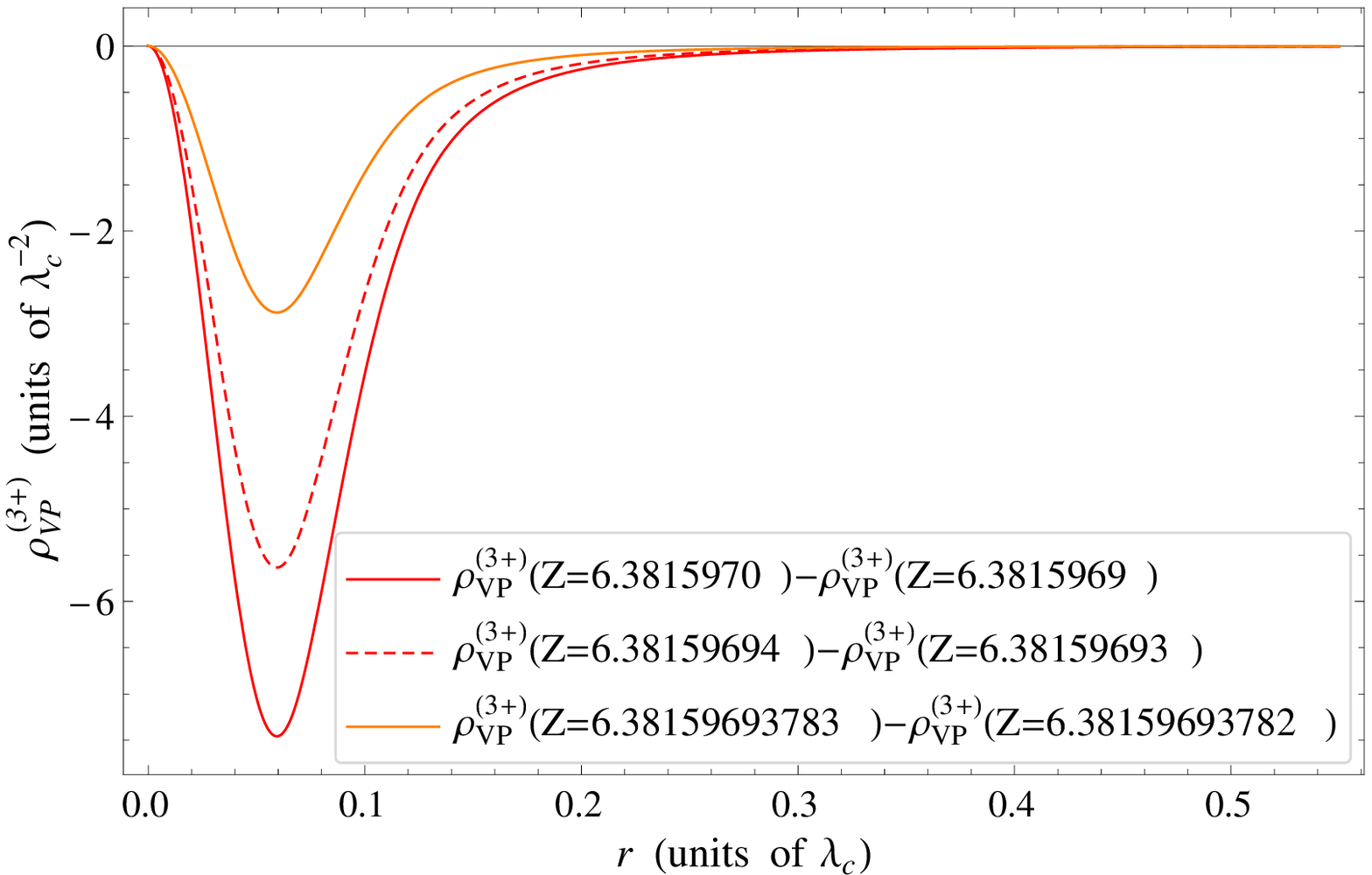}
}
\vfill
\subfigure[]{\label{pic:17}
		\includegraphics[width=\columnwidth]{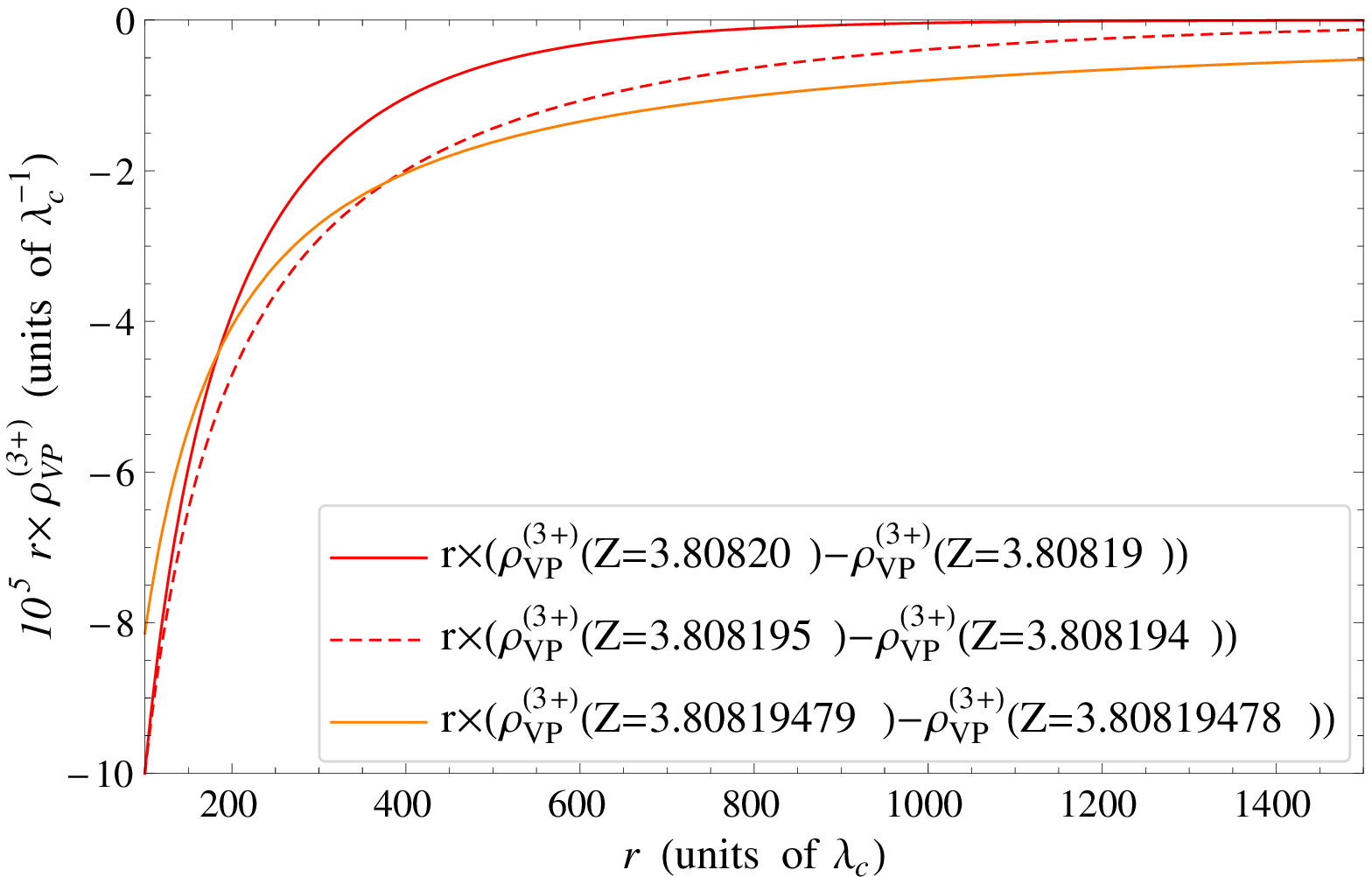}
}
\hfill
\subfigure[]{\label{pic:18}
		\includegraphics[width=\columnwidth]{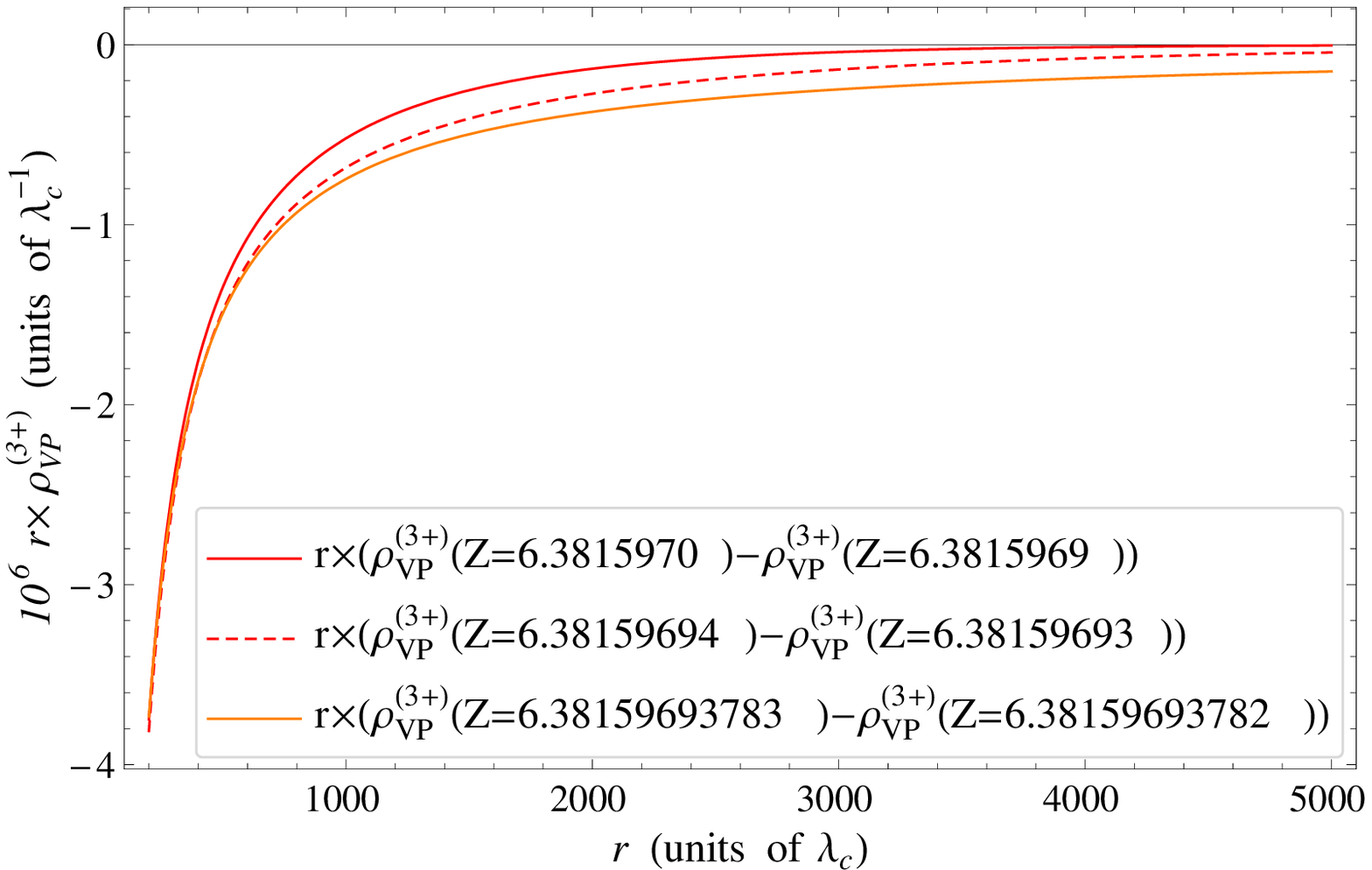}
}
\vfill
\subfigure[]{\label{pic:19}
		\includegraphics[width=\columnwidth]{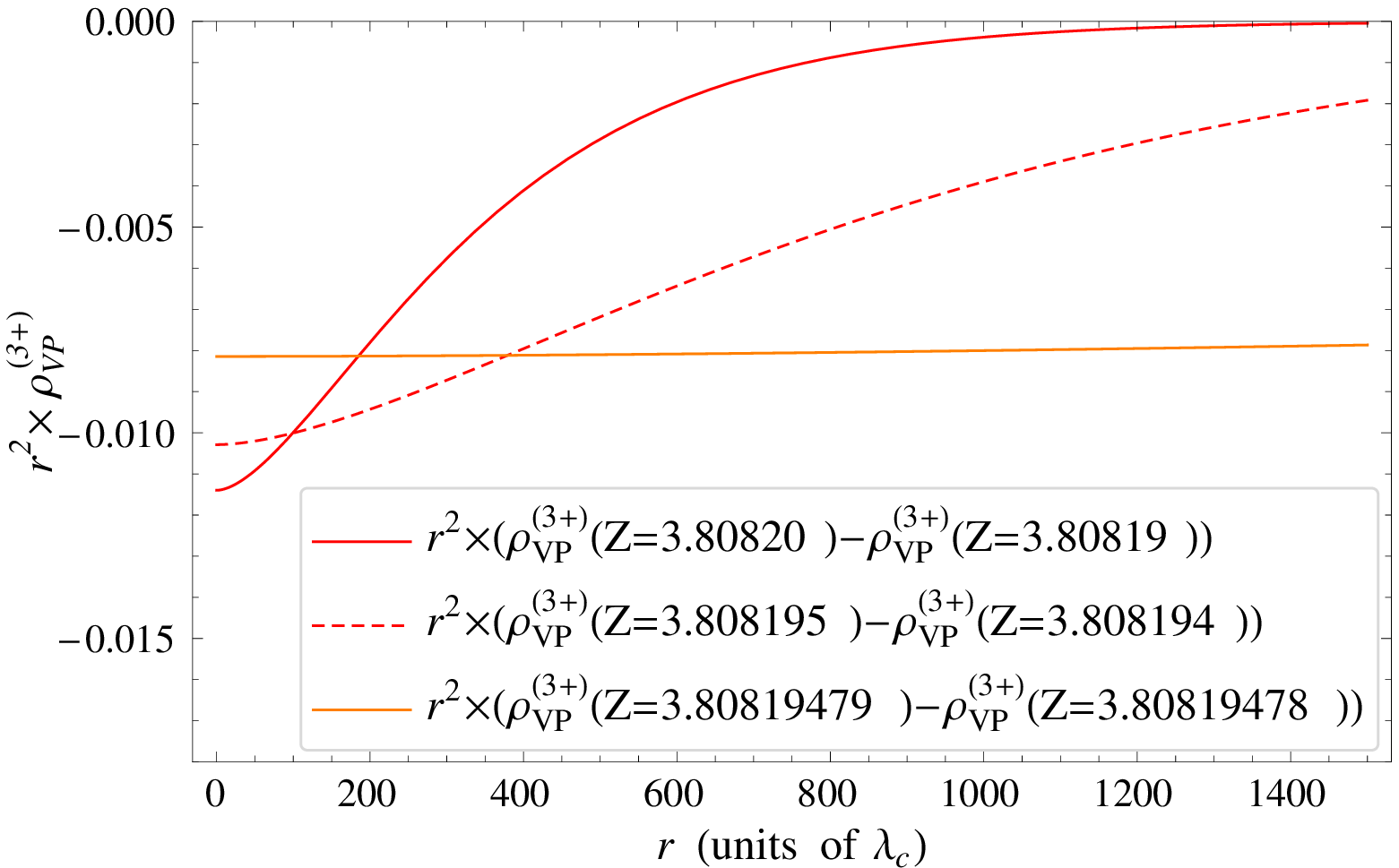}
}
\hfill
\subfigure[]{\label{pic:20}
		\includegraphics[width=\columnwidth]{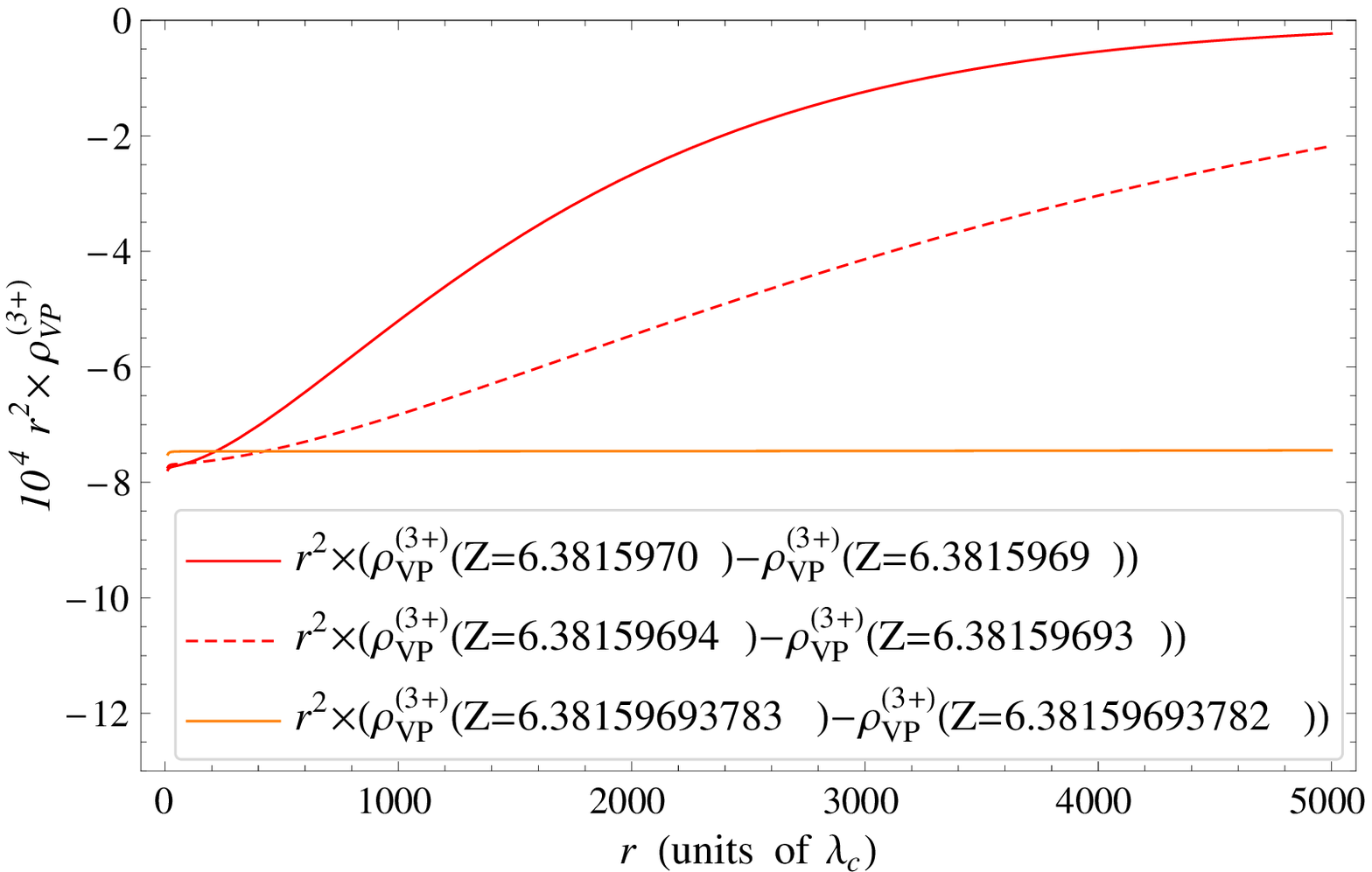}
}
\caption{(Color online) \small  The jumps in the induced density depending on $\D Z$ by crossing the lower threshold for $\a=0.4$, $R_0=1/15$, $R_1=2R_0$ ~\subref{pic:15},~\subref{pic:17},~\subref{pic:19}: for $m_j=1/2$ and ~\subref{pic:16},~\subref{pic:18},~\subref{pic:20}: for $m_j=-3/2$.}\label{pic:15-20}
\end{figure*}

\section{Conclusion}

Thus, in this work the  vacuum polarization  effects in the  2+1-dimensional strongly-coupled QED, caused by diving of levels  into the lower continuum, have been considered in terms of the renormalized vacuum density $\r_{VP}^{ren}(\v r)$. The 2+1-dimensional case differs significantly from the one-dimensional one first of all in that  $\r_{VP}^{ren}(\v r)$ is represented now by an infinite  series in the rotational quantum number $m_j$, and so there appears an additional problem of its convergence.  As it was shown in \cite{Davydov2018a}, this problem can be successfully solved by renormalization of the vacuum density within PT, i.e. via regularization of the solely divergent  fermionic loop with two external lines. Simultaneously, the integral vacuum charge vanishes automatically in the subcritical region, and is changed  by $(-2|e|)$  upon diving of each subsequent doubly degenerate  discrete level into the lower continuum. Such behavior of  $\r_{VP}^{ren}(\v r)$ in the overcritical region confirms once more the assumption of the neutral vacuum transmutation  into the charged one under such conditions (\cite{Greiner1985a, Plunien1986, Ruffini2010, Greiner2012, Rafelski2016} and refs. therein).

It should be noted also that in the most of works cited above ~\cite{Katsnelson2006b, Biswas2007, Pereira2007, Shytov2007, Fogler2007,   Kotov2008, Terekhov2008, Milstein2010,  Nishida2014, Khalilov2017} the impurity potentials are considered without any kind  of screening at large distances from the Coulomb source. However, in fact in such systems there should definitely exist a finite $R_1>R_0$, beyond which the influence of the impurity charge  will be negligibly small.
And although the concrete form of the screened potential could be more smooth (e.g., an exponential one), already the considered peculiar effects by screening  in the form of the simplest shielding via vertical wall for the lowest partial channels $|m_j|=1/2\, , 3/2$ deserve special interest, since indeed the levels with such rotational numbers should  first dive into the lower continuum. The most intriguing circumstance here is that the jump in the induced density by crossing the corresponding $Z_{cr}$ can be very weakly expressed, since the change in the induced density distribution should  be evenly spread over the entire surface of the sample. This effect should be especially remarkable for the levels with $m_j=-1/2$, since in this case such spreading of density will take place for a whole interval $\D Z \sim 1$ before reaching the corresponding $Z_{cr}$ (see Fig.5a). Therefore, recording such a change in the induced density  can be significantly hampered, and the only reliable way of confirming the effect is to measure directly the corresponding change in the total induced charge. Such measurement, however, poses additional problems. Moreover,  even for the first normalizable discrete levels at the lower threshold with $m_j=3/2\, , \pm 5/2\, , \dots $ that provide just the required degree of decrease of the electronic WF at the radial infinity, the jump in the induced density in the screened case will also be significantly smeared over the surface of the sample. So in the screened case the study of such critical effects in terms of the induced density could run into serious difficulties.

This circumstance, however, doesn't mean that the critical effects cannot be observed in the screened case at all, rather it indicates the need to study the Casimir effects. Although the most of works cited above treats the induced charge density $\r_{VP}(\v r)$ as the main polarization observable, the Casimir (vacuum) energy $\E_{VP}$  turns out to be not less informative and in many respects complementary to $\r_{VP}(\v r)$. Moreover, compared to $\r_{VP}(\v r)$, the main non-perturbative effects, which appear in the vacuum polarization for $Z > Z_{cr,1}$ due to levels diving into the lower continuum, show up in the behavior of $\E_{VP}$  even more clear, demonstrating explicitly their possible role in the  overcritical region~\cite{Davydov2017, Sveshnikov2017, Voronina2017,Davydov2018b}. Namely, with growing $Z$ in the overcritical region $\E_{VP}^{ren}(Z)$ falls into the region of large negative values with a rate that depends on the total number  of discrete levels, dived into the lower continuum. In 2+1 D   the growth rate of this number   turns out to be sufficiently higher than in 1+1 D due to the contributions from  different partial channels.  The estimate, obtained in \cite{Davydov2018a}, shows that this growth rate should be not less than $\sim Z^{2.17}$.  As a result, in the 2+1 D toy models considered in \cite{Davydov2018b, Sveshnikov2018a},    $\E_{VP}^{ren}(Z)$ turns out to be negative estimated as $\sim -  Z^3/R_0\, $.  Such decrease of $\E_{VP}$ could significantly affect the basic properties of the considered graphene-like DC system upon doping by charged impurities with $Z >Z^{\ast} \sim O(10)$, leading to  a special type of affinity between the impurities and the graphene plane. However, a rigorous evaluation of this effect  requires for a substantial amount of additional calculations and numerical tools and will be reported in a separate paper.

\twocolumngrid

\bibliography{biblio/VP2DG}

\begin{thebibliography}{42}%
\makeatletter
\providecommand \@ifxundefined [1]{%
 \@ifx{#1\undefined}
}%
\providecommand \@ifnum [1]{%
 \ifnum #1\expandafter \@firstoftwo
 \else \expandafter \@secondoftwo
 \fi
}%
\providecommand \@ifx [1]{%
 \ifx #1\expandafter \@firstoftwo
 \else \expandafter \@secondoftwo
 \fi
}%
\providecommand \natexlab [1]{#1}%
\providecommand \enquote  [1]{``#1''}%
\providecommand \bibnamefont  [1]{#1}%
\providecommand \bibfnamefont [1]{#1}%
\providecommand \citenamefont [1]{#1}%
\providecommand \href@noop [0]{\@secondoftwo}%
\providecommand \href [0]{\begingroup \@sanitize@url \@href}%
\providecommand \@href[1]{\@@startlink{#1}\@@href}%
\providecommand \@@href[1]{\endgroup#1\@@endlink}%
\providecommand \@sanitize@url [0]{\catcode `\\12\catcode `\$12\catcode
  `\&12\catcode `\#12\catcode `\^12\catcode `\_12\catcode `\%12\relax}%
\providecommand \@@startlink[1]{}%
\providecommand \@@endlink[0]{}%
\providecommand \url  [0]{\begingroup\@sanitize@url \@url }%
\providecommand \@url [1]{\endgroup\@href {#1}{\urlprefix }}%
\providecommand \urlprefix  [0]{URL }%
\providecommand \Eprint [0]{\href }%
\providecommand \doibase [0]{http://dx.doi.org/}%
\providecommand \selectlanguage [0]{\@gobble}%
\providecommand \bibinfo  [0]{\@secondoftwo}%
\providecommand \bibfield  [0]{\@secondoftwo}%
\providecommand \translation [1]{[#1]}%
\providecommand \BibitemOpen [0]{}%
\providecommand \bibitemStop [0]{}%
\providecommand \bibitemNoStop [0]{.\EOS\space}%
\providecommand \EOS [0]{\spacefactor3000\relax}%
\providecommand \BibitemShut  [1]{\csname bibitem#1\endcsname}%
\let\auto@bib@innerbib\@empty
%</preamble>
\bibitem [{\citenamefont {Castro~Neto}\ \emph {et~al.}(2009)\citenamefont
  {Castro~Neto}, \citenamefont {Guinea}, \citenamefont {Peres}, \citenamefont
  {Novoselov},\ and\ \citenamefont {Geim}}]{CastroNeto2009}%
  \BibitemOpen
  \bibfield  {author} {\bibinfo {author} {\bibfnamefont {A.~H.}\ \bibnamefont
  {Castro~Neto}}, \bibinfo {author} {\bibfnamefont {F.}~\bibnamefont {Guinea}},
  \bibinfo {author} {\bibfnamefont {N.~M.~R.}\ \bibnamefont {Peres}}, \bibinfo
  {author} {\bibfnamefont {K.~S.}\ \bibnamefont {Novoselov}}, \ and\ \bibinfo
  {author} {\bibfnamefont {A.~K.}\ \bibnamefont {Geim}},\ }\href {\doibase
  10.1103/RevModPhys.81.109} {\bibfield  {journal} {\bibinfo  {journal} {Rev.
  Mod. Phys.}\ }\textbf {\bibinfo {volume} {81}},\ \bibinfo {pages} {109}
  (\bibinfo {year} {2009})}\BibitemShut {NoStop}%
\bibitem [{\citenamefont {Reed}\ \emph {et~al.}(2010)\citenamefont {Reed},
  \citenamefont {Uchoa}, \citenamefont {Joe}, \citenamefont {Gan},
  \citenamefont {Casa}, \citenamefont {Fradkin},\ and\ \citenamefont
  {Abbamonte}}]{Reed2010}%
  \BibitemOpen
  \bibfield  {author} {\bibinfo {author} {\bibfnamefont {J.~P.}\ \bibnamefont
  {Reed}}, \bibinfo {author} {\bibfnamefont {B.}~\bibnamefont {Uchoa}},
  \bibinfo {author} {\bibfnamefont {Y.~I.}\ \bibnamefont {Joe}}, \bibinfo
  {author} {\bibfnamefont {Y.}~\bibnamefont {Gan}}, \bibinfo {author}
  {\bibfnamefont {D.}~\bibnamefont {Casa}}, \bibinfo {author} {\bibfnamefont
  {E.}~\bibnamefont {Fradkin}}, \ and\ \bibinfo {author} {\bibfnamefont
  {P.}~\bibnamefont {Abbamonte}},\ }\href {\doibase 10.1126/science.1190920}
  {\bibfield  {journal} {\bibinfo  {journal} {Science}\ }\textbf {\bibinfo
  {volume} {330}},\ \bibinfo {pages} {805} (\bibinfo {year}
  {2010})}\BibitemShut {NoStop}%
\bibitem [{\citenamefont {Shytov}\ \emph {et~al.}(2009)\citenamefont {Shytov},
  \citenamefont {Rudner}, \citenamefont {Gu}, \citenamefont {Katsnelson},\ and\
  \citenamefont {Levitov}}]{Shytov2009}%
  \BibitemOpen
  \bibfield  {author} {\bibinfo {author} {\bibfnamefont {A.}~\bibnamefont
  {Shytov}}, \bibinfo {author} {\bibfnamefont {M.}~\bibnamefont {Rudner}},
  \bibinfo {author} {\bibfnamefont {N.}~\bibnamefont {Gu}}, \bibinfo {author}
  {\bibfnamefont {M.}~\bibnamefont {Katsnelson}}, \ and\ \bibinfo {author}
  {\bibfnamefont {L.}~\bibnamefont {Levitov}},\ }\href {\doibase
  https://doi.org/10.1016/j.ssc.2009.02.043} {\bibfield  {journal} {\bibinfo
  {journal} {Solid State Communications}\ }\textbf {\bibinfo {volume} {149}},\
  \bibinfo {pages} {1087 } (\bibinfo {year} {2009})}\BibitemShut {NoStop}%
\bibitem [{\citenamefont {Shytov}\ \emph {et~al.}(2007)\citenamefont {Shytov},
  \citenamefont {Katsnelson},\ and\ \citenamefont {Levitov}}]{Shytov2007}%
  \BibitemOpen
  \bibfield  {author} {\bibinfo {author} {\bibfnamefont {A.~V.}\ \bibnamefont
  {Shytov}}, \bibinfo {author} {\bibfnamefont {M.~I.}\ \bibnamefont
  {Katsnelson}}, \ and\ \bibinfo {author} {\bibfnamefont {L.~S.}\ \bibnamefont
  {Levitov}},\ }\href {\doibase 10.1103/PhysRevLett.99.236801} {\bibfield
  {journal} {\bibinfo  {journal} {Phys. Rev. Lett.}\ }\textbf {\bibinfo
  {volume} {99}},\ \bibinfo {pages} {236801} (\bibinfo {year}
  {2007})}\BibitemShut {NoStop}%
\bibitem [{\citenamefont {{Katsnelson}}\ \emph {et~al.}(2006)\citenamefont
  {{Katsnelson}}, \citenamefont {{Novoselov}},\ and\ \citenamefont
  {{Geim}}}]{Katsnelson2006a}%
  \BibitemOpen
  \bibfield  {author} {\bibinfo {author} {\bibfnamefont {M.~I.}\ \bibnamefont
  {{Katsnelson}}}, \bibinfo {author} {\bibfnamefont {K.~S.}\ \bibnamefont
  {{Novoselov}}}, \ and\ \bibinfo {author} {\bibfnamefont {A.~K.}\ \bibnamefont
  {{Geim}}},\ }\href {\doibase 10.1038/nphys384} {\bibfield  {journal}
  {\bibinfo  {journal} {Nature Physics}\ }\textbf {\bibinfo {volume} {2}},\
  \bibinfo {pages} {620} (\bibinfo {year} {2006})}\BibitemShut {NoStop}%
\bibitem [{\citenamefont {Gusynin}\ and\ \citenamefont
  {Sharapov}(2005)}]{Gusynin2005}%
  \BibitemOpen
  \bibfield  {author} {\bibinfo {author} {\bibfnamefont {V.~P.}\ \bibnamefont
  {Gusynin}}\ and\ \bibinfo {author} {\bibfnamefont {S.~G.}\ \bibnamefont
  {Sharapov}},\ }\href {\doibase 10.1103/PhysRevLett.95.146801} {\bibfield
  {journal} {\bibinfo  {journal} {Phys. Rev. Lett.}\ }\textbf {\bibinfo
  {volume} {95}},\ \bibinfo {pages} {146801} (\bibinfo {year}
  {2005})}\BibitemShut {NoStop}%
\bibitem [{\citenamefont {Giesbers}\ \emph {et~al.}(2008)\citenamefont
  {Giesbers}, \citenamefont {Zeitler}, \citenamefont {Katsnelson},
  \citenamefont {Ponomarenko}, \citenamefont {Mohiuddin},\ and\ \citenamefont
  {Maan}}]{Giesbers2008}%
  \BibitemOpen
  \bibfield  {author} {\bibinfo {author} {\bibfnamefont {A.}~\bibnamefont
  {Giesbers}}, \bibinfo {author} {\bibfnamefont {U.}~\bibnamefont {Zeitler}},
  \bibinfo {author} {\bibfnamefont {M.}~\bibnamefont {Katsnelson}}, \bibinfo
  {author} {\bibfnamefont {M.}~\bibnamefont {Ponomarenko}}, \bibinfo {author}
  {\bibfnamefont {T.}~\bibnamefont {Mohiuddin}}, \ and\ \bibinfo {author}
  {\bibfnamefont {J.}~\bibnamefont {Maan}},\ }\href {\doibase
  10.1016/j.physe.2007.08.111} {\bibfield  {journal} {\bibinfo  {journal}
  {Physica E}\ }\textbf {\bibinfo {volume} {40}},\ \bibinfo {pages} {1089}
  (\bibinfo {year} {2008})}\BibitemShut {NoStop}%
\bibitem [{\citenamefont {Cobaleda}\ \emph {et~al.}(2011)\citenamefont
  {Cobaleda}, \citenamefont {Rossella}, \citenamefont {Pezzini}, \citenamefont
  {Diez}, \citenamefont {Bellani}, \citenamefont {Maude},\ and\ \citenamefont
  {Blake}}]{Cobaleda2011}%
  \BibitemOpen
  \bibfield  {author} {\bibinfo {author} {\bibfnamefont {C.}~\bibnamefont
  {Cobaleda}}, \bibinfo {author} {\bibfnamefont {F.}~\bibnamefont {Rossella}},
  \bibinfo {author} {\bibnamefont {Pezzini}}, \bibinfo {author} {\bibfnamefont
  {E.}~\bibnamefont {Diez}}, \bibinfo {author} {\bibfnamefont {V.}~\bibnamefont
  {Bellani}}, \bibinfo {author} {\bibfnamefont {D.}~\bibnamefont {Maude}}, \
  and\ \bibinfo {author} {\bibfnamefont {P.}~\bibnamefont {Blake}},\ }\href
  {\doibase 10.1016/j.physe.2011.10.007} {\bibfield  {journal} {\bibinfo
  {journal} {Physica E}\ }\textbf {\bibinfo {volume} {44}},\ \bibinfo {pages}
  {530} (\bibinfo {year} {2011})}\BibitemShut {NoStop}%
\bibitem [{\citenamefont {Shytov}\ \emph {et~al.}(2008)\citenamefont {Shytov},
  \citenamefont {Rudner},\ and\ \citenamefont {Levitov}}]{Shytov2008}%
  \BibitemOpen
  \bibfield  {author} {\bibinfo {author} {\bibfnamefont {A.~V.}\ \bibnamefont
  {Shytov}}, \bibinfo {author} {\bibfnamefont {M.~S.}\ \bibnamefont {Rudner}},
  \ and\ \bibinfo {author} {\bibfnamefont {L.~S.}\ \bibnamefont {Levitov}},\
  }\href {\doibase 10.1103/PhysRevLett.101.156804} {\bibfield  {journal}
  {\bibinfo  {journal} {Phys. Rev. Lett.}\ }\textbf {\bibinfo {volume} {101}},\
  \bibinfo {pages} {156804} (\bibinfo {year} {2008})}\BibitemShut {NoStop}%
\bibitem [{\citenamefont {He}\ \emph {et~al.}(2013)\citenamefont {He},
  \citenamefont {Chu},\ and\ \citenamefont {He}}]{He2013}%
  \BibitemOpen
  \bibfield  {author} {\bibinfo {author} {\bibfnamefont {W.-Y.}\ \bibnamefont
  {He}}, \bibinfo {author} {\bibfnamefont {Z.-D.}\ \bibnamefont {Chu}}, \ and\
  \bibinfo {author} {\bibfnamefont {L.}~\bibnamefont {He}},\ }\href {\doibase
  10.1103/PhysRevLett.111.066803} {\bibfield  {journal} {\bibinfo  {journal}
  {Phys. Rev. Lett.}\ }\textbf {\bibinfo {volume} {111}},\ \bibinfo {pages}
  {066803} (\bibinfo {year} {2013})}\BibitemShut {NoStop}%
\bibitem [{\citenamefont {Greiner}\ \emph {et~al.}(1985)\citenamefont
  {Greiner}, \citenamefont {M\"uller},\ and\ \citenamefont
  {Rafelski}}]{Greiner1985a}%
  \BibitemOpen
  \bibfield  {author} {\bibinfo {author} {\bibfnamefont {W.}~\bibnamefont
  {Greiner}}, \bibinfo {author} {\bibfnamefont {B.}~\bibnamefont {M\"uller}}, \
  and\ \bibinfo {author} {\bibfnamefont {J.}~\bibnamefont {Rafelski}},\ }\href
  {http://link.springer.com/book/10.1007/978-3-642-82272-8} {\emph {\bibinfo
  {title} {Quantum Electrodynamics of Strong Fields}}},\ \bibinfo {edition}
  {2nd}\ ed.\ (\bibinfo  {publisher} {Springer},\ \bibinfo {address} {Berlin},\
  \bibinfo {year} {1985})\BibitemShut {NoStop}%
\bibitem [{\citenamefont {Plunien}\ \emph {et~al.}(1986)\citenamefont
  {Plunien}, \citenamefont {M\"uller},\ and\ \citenamefont
  {Greiner}}]{Plunien1986}%
  \BibitemOpen
  \bibfield  {author} {\bibinfo {author} {\bibfnamefont {G.}~\bibnamefont
  {Plunien}}, \bibinfo {author} {\bibfnamefont {B.}~\bibnamefont {M\"uller}}, \
  and\ \bibinfo {author} {\bibfnamefont {W.}~\bibnamefont {Greiner}},\ }\href
  {\doibase 10.1016/0370-1573(86)90020-7} {\bibfield  {journal} {\bibinfo
  {journal} {Phys. Rep.}\ }\textbf {\bibinfo {volume} {134}},\ \bibinfo {pages}
  {87 } (\bibinfo {year} {1986})}\BibitemShut {NoStop}%
\bibitem [{\citenamefont {Ruffini}\ \emph {et~al.}(2010)\citenamefont
  {Ruffini}, \citenamefont {Vereshchagin},\ and\ \citenamefont
  {Xue}}]{Ruffini2010}%
  \BibitemOpen
  \bibfield  {author} {\bibinfo {author} {\bibfnamefont {R.}~\bibnamefont
  {Ruffini}}, \bibinfo {author} {\bibfnamefont {G.}~\bibnamefont
  {Vereshchagin}}, \ and\ \bibinfo {author} {\bibfnamefont {S.-S.}\
  \bibnamefont {Xue}},\ }\href {\doibase 10.1016/j.physrep.2009.10.004}
  {\bibfield  {journal} {\bibinfo  {journal} {Phys. Rep.}\ }\textbf {\bibinfo
  {volume} {487}},\ \bibinfo {pages} {1 } (\bibinfo {year} {2010})}\BibitemShut
  {NoStop}%
\bibitem [{\citenamefont {Greiner}\ and\ \citenamefont
  {Reinhardt}(2009)}]{Greiner2012}%
  \BibitemOpen
  \bibfield  {author} {\bibinfo {author} {\bibfnamefont {W.}~\bibnamefont
  {Greiner}}\ and\ \bibinfo {author} {\bibfnamefont {J.}~\bibnamefont
  {Reinhardt}},\ }\href {\doibase /10.1007/978-3-540-87561-1} {\emph {\bibinfo
  {title} {Quantum Electrodynamics}}},\ \bibinfo {edition} {4th}\ ed.\
  (\bibinfo  {publisher} {Springer-Verlag Berlin Heidelberg},\ \bibinfo {year}
  {2009})\BibitemShut {NoStop}%
\bibitem [{\citenamefont {Rafelski}\ \emph {et~al.}(2017)\citenamefont
  {Rafelski}, \citenamefont {Kirsch}, \citenamefont {M\"uller}, \citenamefont
  {Reinhardt},\ and\ \citenamefont {Greiner}}]{Rafelski2016}%
  \BibitemOpen
  \bibfield  {author} {\bibinfo {author} {\bibfnamefont {J.}~\bibnamefont
  {Rafelski}}, \bibinfo {author} {\bibfnamefont {J.}~\bibnamefont {Kirsch}},
  \bibinfo {author} {\bibfnamefont {B.}~\bibnamefont {M\"uller}}, \bibinfo
  {author} {\bibfnamefont {J.}~\bibnamefont {Reinhardt}}, \ and\ \bibinfo
  {author} {\bibfnamefont {W.}~\bibnamefont {Greiner}},\ }\enquote {\bibinfo
  {title} {Probing {QED} {Vacuum} with {Heavy} {Ions}},}\ in\ \href {\doibase
  10.1007/978-3-319-44165-8_17} {\emph {\bibinfo {booktitle} {New {Horizons} in
  {Fundamental} {Physics}}}},\ \bibinfo {series and number} {{FIAS}
  {Interdisciplinary} {Science} {Series}}\ (\bibinfo  {publisher} {Springer},\
  \bibinfo {year} {2017})\ pp.\ \bibinfo {pages} {211--251}\BibitemShut
  {NoStop}%
\bibitem [{\citenamefont {Wang}\ \emph {et~al.}(2013)\citenamefont {Wang},
  \citenamefont {Wong}, \citenamefont {Shytov}, \citenamefont {Brar},
  \citenamefont {Choi}, \citenamefont {Wu}, \citenamefont {Tsai}, \citenamefont
  {Regan}, \citenamefont {Zettl}, \citenamefont {Kawakami}, \citenamefont
  {Louie}, \citenamefont {Levitov},\ and\ \citenamefont {Crommie}}]{Wang2013}%
  \BibitemOpen
  \bibfield  {author} {\bibinfo {author} {\bibfnamefont {Y.}~\bibnamefont
  {Wang}}, \bibinfo {author} {\bibfnamefont {D.}~\bibnamefont {Wong}}, \bibinfo
  {author} {\bibfnamefont {A.~V.}\ \bibnamefont {Shytov}}, \bibinfo {author}
  {\bibfnamefont {V.~W.}\ \bibnamefont {Brar}}, \bibinfo {author}
  {\bibfnamefont {S.}~\bibnamefont {Choi}}, \bibinfo {author} {\bibfnamefont
  {Q.}~\bibnamefont {Wu}}, \bibinfo {author} {\bibfnamefont {H.-Z.}\
  \bibnamefont {Tsai}}, \bibinfo {author} {\bibfnamefont {W.}~\bibnamefont
  {Regan}}, \bibinfo {author} {\bibfnamefont {A.}~\bibnamefont {Zettl}},
  \bibinfo {author} {\bibfnamefont {R.~K.}\ \bibnamefont {Kawakami}}, \bibinfo
  {author} {\bibfnamefont {S.~G.}\ \bibnamefont {Louie}}, \bibinfo {author}
  {\bibfnamefont {L.~S.}\ \bibnamefont {Levitov}}, \ and\ \bibinfo {author}
  {\bibfnamefont {M.~F.}\ \bibnamefont {Crommie}},\ }\href {\doibase
  10.1126/science.1234320} {\bibfield  {journal} {\bibinfo  {journal}
  {Science}\ }\textbf {\bibinfo {volume} {340}},\ \bibinfo {pages} {734}
  (\bibinfo {year} {2013})}\BibitemShut {NoStop}%
\bibitem [{\citenamefont {Katsnelson}(2006)}]{Katsnelson2006b}%
  \BibitemOpen
  \bibfield  {author} {\bibinfo {author} {\bibfnamefont {M.~I.}\ \bibnamefont
  {Katsnelson}},\ }\href {\doibase 10.1103/PhysRevB.74.201401} {\bibfield
  {journal} {\bibinfo  {journal} {Phys. Rev. B}\ }\textbf {\bibinfo {volume}
  {74}},\ \bibinfo {pages} {201401} (\bibinfo {year} {2006})}\BibitemShut
  {NoStop}%
\bibitem [{\citenamefont {Biswas}\ \emph {et~al.}(2007)\citenamefont {Biswas},
  \citenamefont {Sachdev},\ and\ \citenamefont {Son}}]{Biswas2007}%
  \BibitemOpen
  \bibfield  {author} {\bibinfo {author} {\bibfnamefont {R.~R.}\ \bibnamefont
  {Biswas}}, \bibinfo {author} {\bibfnamefont {S.}~\bibnamefont {Sachdev}}, \
  and\ \bibinfo {author} {\bibfnamefont {D.~T.}\ \bibnamefont {Son}},\ }\href
  {\doibase 10.1103/PhysRevB.76.205122} {\bibfield  {journal} {\bibinfo
  {journal} {Phys. Rev. B}\ }\textbf {\bibinfo {volume} {76}},\ \bibinfo
  {pages} {205122} (\bibinfo {year} {2007})}\BibitemShut {NoStop}%
\bibitem [{\citenamefont {Pereira}\ \emph {et~al.}(2007)\citenamefont
  {Pereira}, \citenamefont {Nilsson},\ and\ \citenamefont
  {Castro~Neto}}]{Pereira2007}%
  \BibitemOpen
  \bibfield  {author} {\bibinfo {author} {\bibfnamefont {V.~M.}\ \bibnamefont
  {Pereira}}, \bibinfo {author} {\bibfnamefont {J.}~\bibnamefont {Nilsson}}, \
  and\ \bibinfo {author} {\bibfnamefont {A.~H.}\ \bibnamefont {Castro~Neto}},\
  }\href {\doibase 10.1103/PhysRevLett.99.166802} {\bibfield  {journal}
  {\bibinfo  {journal} {Phys. Rev. Lett.}\ }\textbf {\bibinfo {volume} {99}},\
  \bibinfo {pages} {166802} (\bibinfo {year} {2007})}\BibitemShut {NoStop}%
\bibitem [{\citenamefont {Kotov}\ \emph {et~al.}(2008)\citenamefont {Kotov},
  \citenamefont {Pereira},\ and\ \citenamefont {Uchoa}}]{Kotov2008}%
  \BibitemOpen
  \bibfield  {author} {\bibinfo {author} {\bibfnamefont {V.~N.}\ \bibnamefont
  {Kotov}}, \bibinfo {author} {\bibfnamefont {V.~M.}\ \bibnamefont {Pereira}},
  \ and\ \bibinfo {author} {\bibfnamefont {B.}~\bibnamefont {Uchoa}},\ }\href
  {\doibase 10.1103/PhysRevB.78.075433} {\bibfield  {journal} {\bibinfo
  {journal} {Phys. Rev. B}\ }\textbf {\bibinfo {volume} {78}},\ \bibinfo
  {pages} {075433} (\bibinfo {year} {2008})}\BibitemShut {NoStop}%
\bibitem [{\citenamefont {Terekhov}\ \emph {et~al.}(2008)\citenamefont
  {Terekhov}, \citenamefont {Milstein}, \citenamefont {Kotov},\ and\
  \citenamefont {Sushkov}}]{Terekhov2008}%
  \BibitemOpen
  \bibfield  {author} {\bibinfo {author} {\bibfnamefont {I.~S.}\ \bibnamefont
  {Terekhov}}, \bibinfo {author} {\bibfnamefont {A.~I.}\ \bibnamefont
  {Milstein}}, \bibinfo {author} {\bibfnamefont {V.~N.}\ \bibnamefont {Kotov}},
  \ and\ \bibinfo {author} {\bibfnamefont {O.~P.}\ \bibnamefont {Sushkov}},\
  }\href {\doibase 10.1103/PhysRevLett.100.076803} {\bibfield  {journal}
  {\bibinfo  {journal} {Phys. Rev. Lett.}\ }\textbf {\bibinfo {volume} {100}},\
  \bibinfo {pages} {076803} (\bibinfo {year} {2008})}\BibitemShut {NoStop}%
\bibitem [{\citenamefont {Nishida}(2014)}]{Nishida2014}%
  \BibitemOpen
  \bibfield  {author} {\bibinfo {author} {\bibfnamefont {Y.}~\bibnamefont
  {Nishida}},\ }\href {\doibase 10.1103/PhysRevB.90.165414} {\bibfield
  {journal} {\bibinfo  {journal} {Phys. Rev. B}\ }\textbf {\bibinfo {volume}
  {90}},\ \bibinfo {pages} {165414} (\bibinfo {year} {2014})}\BibitemShut
  {NoStop}%
\bibitem [{\citenamefont {Khalilov}\ and\ \citenamefont
  {Mamsurov}(2017)}]{Khalilov2017}%
  \BibitemOpen
  \bibfield  {author} {\bibinfo {author} {\bibfnamefont {V.~R.}\ \bibnamefont
  {Khalilov}}\ and\ \bibinfo {author} {\bibfnamefont {I.~V.}\ \bibnamefont
  {Mamsurov}},\ }\href {\doibase 10.1016/j.physletb.2017.03.052} {\bibfield
  {journal} {\bibinfo  {journal} {Phys. Lett. B}\ }\textbf {\bibinfo {volume}
  {769}},\ \bibinfo {pages} {152} (\bibinfo {year} {2017})}\BibitemShut
  {NoStop}%
\bibitem [{\citenamefont {Fogler}\ \emph {et~al.}(2007)\citenamefont {Fogler},
  \citenamefont {Novikov},\ and\ \citenamefont {Shklovskii}}]{Fogler2007}%
  \BibitemOpen
  \bibfield  {author} {\bibinfo {author} {\bibfnamefont {M.~M.}\ \bibnamefont
  {Fogler}}, \bibinfo {author} {\bibfnamefont {D.~S.}\ \bibnamefont {Novikov}},
  \ and\ \bibinfo {author} {\bibfnamefont {B.~I.}\ \bibnamefont {Shklovskii}},\
  }\href {\doibase 10.1103/PhysRevB.76.233402} {\bibfield  {journal} {\bibinfo
  {journal} {Phys. Rev. B}\ }\textbf {\bibinfo {volume} {76}},\ \bibinfo
  {pages} {233402} (\bibinfo {year} {2007})}\BibitemShut {NoStop}%
\bibitem [{\citenamefont {Milstein}\ and\ \citenamefont
  {Terekhov}(2010)}]{Milstein2010}%
  \BibitemOpen
  \bibfield  {author} {\bibinfo {author} {\bibfnamefont {A.~I.}\ \bibnamefont
  {Milstein}}\ and\ \bibinfo {author} {\bibfnamefont {I.~S.}\ \bibnamefont
  {Terekhov}},\ }\href {\doibase 10.1103/PhysRevB.81.125419} {\bibfield
  {journal} {\bibinfo  {journal} {Phys. Rev. B}\ }\textbf {\bibinfo {volume}
  {81}},\ \bibinfo {pages} {125419} (\bibinfo {year} {2010})}\BibitemShut
  {NoStop}%
\bibitem [{\citenamefont {Pereira}\ \emph {et~al.}(2008)\citenamefont
  {Pereira}, \citenamefont {Kotov},\ and\ \citenamefont
  {Castro~Neto}}]{Pereira2008}%
  \BibitemOpen
  \bibfield  {author} {\bibinfo {author} {\bibfnamefont {V.~M.}\ \bibnamefont
  {Pereira}}, \bibinfo {author} {\bibfnamefont {V.~N.}\ \bibnamefont {Kotov}},
  \ and\ \bibinfo {author} {\bibfnamefont {A.~H.}\ \bibnamefont
  {Castro~Neto}},\ }\href {\doibase 10.1103/PhysRevB.78.085101} {\bibfield
  {journal} {\bibinfo  {journal} {Phys. Rev. B}\ }\textbf {\bibinfo {volume}
  {78}},\ \bibinfo {pages} {085101} (\bibinfo {year} {2008})}\BibitemShut
  {NoStop}%
\bibitem [{\citenamefont {Goerbig}(2011)}]{Goerbig2011}%
  \BibitemOpen
  \bibfield  {author} {\bibinfo {author} {\bibfnamefont {M.~O.}\ \bibnamefont
  {Goerbig}},\ }\href {\doibase 10.1103/RevModPhys.83.1193} {\bibfield
  {journal} {\bibinfo  {journal} {Rev. Mod. Phys.}\ }\textbf {\bibinfo {volume}
  {83}},\ \bibinfo {pages} {1193} (\bibinfo {year} {2011})}\BibitemShut
  {NoStop}%
\bibitem [{\citenamefont {Sadeghi}\ \emph {et~al.}(2016)\citenamefont
  {Sadeghi}, \citenamefont {Sangtarash},\ and\ \citenamefont
  {Lambert}}]{Sadeghi2015}%
  \BibitemOpen
  \bibfield  {author} {\bibinfo {author} {\bibfnamefont {H.}~\bibnamefont
  {Sadeghi}}, \bibinfo {author} {\bibfnamefont {S.}~\bibnamefont {Sangtarash}},
  \ and\ \bibinfo {author} {\bibfnamefont {C.}~\bibnamefont {Lambert}},\ }\href
  {\doibase 10.1016/j.physe.2015.09.005} {\bibfield  {journal} {\bibinfo
  {journal} {Physica E}\ }\textbf {\bibinfo {volume} {82}},\ \bibinfo {pages}
  {12 } (\bibinfo {year} {2016})}\BibitemShut {NoStop}%
\bibitem [{\citenamefont {Davydov}\ \emph
  {et~al.}(2018{\natexlab{a}})\citenamefont {Davydov}, \citenamefont
  {Sveshnikov},\ and\ \citenamefont {Voronina}}]{Davydov2018a}%
  \BibitemOpen
  \bibfield  {author} {\bibinfo {author} {\bibfnamefont {A.}~\bibnamefont
  {Davydov}}, \bibinfo {author} {\bibfnamefont {K.}~\bibnamefont {Sveshnikov}},
  \ and\ \bibinfo {author} {\bibfnamefont {Y.}~\bibnamefont {Voronina}},\
  }\href {\doibase 10.1142/S0217751X18500045} {\bibfield  {journal} {\bibinfo
  {journal} {Int. J. Mod. Phys. A}\ }\textbf {\bibinfo {volume} {33}},\
  \bibinfo {pages} {1850004} (\bibinfo {year}
  {2018}{\natexlab{a}})}\BibitemShut {NoStop}%
\bibitem [{\citenamefont {Mohr}\ \emph {et~al.}(1998)\citenamefont {Mohr},
  \citenamefont {Plunien},\ and\ \citenamefont {Soff}}]{Mohr1998}%
  \BibitemOpen
  \bibfield  {author} {\bibinfo {author} {\bibfnamefont {P.~J.}\ \bibnamefont
  {Mohr}}, \bibinfo {author} {\bibfnamefont {G.}~\bibnamefont {Plunien}}, \
  and\ \bibinfo {author} {\bibfnamefont {G.}~\bibnamefont {Soff}},\ }\href
  {\doibase 10.1016/S0370-1573(97)00046-X} {\bibfield  {journal} {\bibinfo
  {journal} {Phys. Rep.}\ }\textbf {\bibinfo {volume} {293}},\ \bibinfo {pages}
  {227 } (\bibinfo {year} {1998})}\BibitemShut {NoStop}%
\bibitem [{\citenamefont {Davydov}\ \emph
  {et~al.}(2018{\natexlab{b}})\citenamefont {Davydov}, \citenamefont
  {Sveshnikov},\ and\ \citenamefont {Voronina}}]{Davydov2018b}%
  \BibitemOpen
  \bibfield  {author} {\bibinfo {author} {\bibfnamefont {A.}~\bibnamefont
  {Davydov}}, \bibinfo {author} {\bibfnamefont {K.}~\bibnamefont {Sveshnikov}},
  \ and\ \bibinfo {author} {\bibfnamefont {Y.}~\bibnamefont {Voronina}},\
  }\href {\doibase 10.1142/S0217751X18500057} {\bibfield  {journal} {\bibinfo
  {journal} {Int. J. Mod. Phys. A}\ }\textbf {\bibinfo {volume} {33}},\
  \bibinfo {pages} {1850005} (\bibinfo {year}
  {2018}{\natexlab{b}})}\BibitemShut {NoStop}%
\bibitem [{\citenamefont {Wichmann}\ and\ \citenamefont
  {Kroll}(1956)}]{Wichmann1956}%
  \BibitemOpen
  \bibfield  {author} {\bibinfo {author} {\bibfnamefont {E.~H.}\ \bibnamefont
  {Wichmann}}\ and\ \bibinfo {author} {\bibfnamefont {N.~M.}\ \bibnamefont
  {Kroll}},\ }\href {\doibase 10.1103/PhysRev.101.843} {\bibfield  {journal}
  {\bibinfo  {journal} {Phys. Rev.}\ }\textbf {\bibinfo {volume} {101}},\
  \bibinfo {pages} {843} (\bibinfo {year} {1956})}\BibitemShut {NoStop}%
\bibitem [{\citenamefont {Hosotani}(1993)}]{Hosotani1993}%
  \BibitemOpen
  \bibfield  {author} {\bibinfo {author} {\bibfnamefont {Y.}~\bibnamefont
  {Hosotani}},\ }\href {\doibase https://doi.org/10.1016/0370-2693(93)90822-Y}
  {\bibfield  {journal} {\bibinfo  {journal} {Phys. Lett. B}\ }\textbf
  {\bibinfo {volume} {319}},\ \bibinfo {pages} {332 } (\bibinfo {year}
  {1993})}\BibitemShut {NoStop}%
\bibitem [{\citenamefont {Bateman}\ and\ \citenamefont
  {Erdelyi}(1953)}]{Bateman1953}%
  \BibitemOpen
  \bibfield  {author} {\bibinfo {author} {\bibfnamefont {H.}~\bibnamefont
  {Bateman}}\ and\ \bibinfo {author} {\bibfnamefont {A.}~\bibnamefont
  {Erdelyi}},\ }\href@noop {} {\emph {\bibinfo {title} {Higher Transcendental
  Functions}}},\ Vol.\ \bibinfo {volume} {1-2}\ (\bibinfo  {publisher} {Mc
  Graw-Hill, New York},\ \bibinfo {year} {1953})\BibitemShut {NoStop}%
\bibitem [{\citenamefont {Gyulassy}(1975)}]{Gyulassy1975}%
  \BibitemOpen
  \bibfield  {author} {\bibinfo {author} {\bibfnamefont {M.}~\bibnamefont
  {Gyulassy}},\ }\href {\doibase 10.1016/0375-9474(75)90554-0} {\bibfield
  {journal} {\bibinfo  {journal} {Nucl. Phys. A}\ }\textbf {\bibinfo {volume}
  {244}},\ \bibinfo {pages} {497 } (\bibinfo {year} {1975})}\BibitemShut
  {NoStop}%
\bibitem [{\citenamefont {Davydov}\ \emph {et~al.}(2017)\citenamefont
  {Davydov}, \citenamefont {Sveshnikov},\ and\ \citenamefont
  {Voronina}}]{Davydov2017}%
  \BibitemOpen
  \bibfield  {author} {\bibinfo {author} {\bibfnamefont {A.}~\bibnamefont
  {Davydov}}, \bibinfo {author} {\bibfnamefont {K.}~\bibnamefont {Sveshnikov}},
  \ and\ \bibinfo {author} {\bibfnamefont {Y.}~\bibnamefont {Voronina}},\
  }\href {\doibase 10.1142/S0217751X17500543} {\bibfield  {journal} {\bibinfo
  {journal} {Int. J. Mod. Phys. A}\ }\textbf {\bibinfo {volume} {32}},\
  \bibinfo {pages} {1750054} (\bibinfo {year} {2017})}\BibitemShut {NoStop}%
\bibitem [{\citenamefont {Fano}(1961)}]{Fano1961}%
  \BibitemOpen
  \bibfield  {author} {\bibinfo {author} {\bibfnamefont {U.}~\bibnamefont
  {Fano}},\ }\href {\doibase 10.1103/PhysRev.124.1866} {\bibfield  {journal}
  {\bibinfo  {journal} {Phys. Rev.}\ }\textbf {\bibinfo {volume} {124}},\
  \bibinfo {pages} {1866} (\bibinfo {year} {1961})}\BibitemShut {NoStop}%
\bibitem [{\citenamefont {Sveshnikov}\ \emph {et~al.}(2018)\citenamefont
  {Sveshnikov}, \citenamefont {Voronina}, \citenamefont {Davydov},\ and\
  \citenamefont {Grashin}}]{Sveshnikov2018a}%
  \BibitemOpen
  \bibfield  {author} {\bibinfo {author} {\bibfnamefont {K.}~\bibnamefont
  {Sveshnikov}}, \bibinfo {author} {\bibfnamefont {Y.}~\bibnamefont
  {Voronina}}, \bibinfo {author} {\bibfnamefont {A.}~\bibnamefont {Davydov}}, \
  and\ \bibinfo {author} {\bibfnamefont {P.}~\bibnamefont {Grashin}},\
  }\href@noop {} {\bibfield  {journal} {\bibinfo  {journal} {submitted to
  Theor. Math. Phys.}\ } (\bibinfo {year} {2018})}\BibitemShut {NoStop}%
\bibitem [{\citenamefont {Voronina}\ \emph {et~al.}(2018)\citenamefont
  {Voronina}, \citenamefont {Sveshnikov}, \citenamefont {Davydov},\ and\
  \citenamefont {Grashin}}]{Voronina2018b}%
  \BibitemOpen
  \bibfield  {author} {\bibinfo {author} {\bibfnamefont {Y.}~\bibnamefont
  {Voronina}}, \bibinfo {author} {\bibfnamefont {K.}~\bibnamefont
  {Sveshnikov}}, \bibinfo {author} {\bibfnamefont {A.}~\bibnamefont {Davydov}},
  \ and\ \bibinfo {author} {\bibfnamefont {P.}~\bibnamefont {Grashin}},\
  }\href@noop {} {\bibfield  {journal} {\bibinfo  {journal} {arXiv:1802.05336
  [cond-mat.mes-hall]}\ } (\bibinfo {year} {2018})}\BibitemShut {NoStop}%
\bibitem [{\citenamefont {Landau}\ and\ \citenamefont {Lifshitz}(1981)}]{LL}%
  \BibitemOpen
  \bibfield  {author} {\bibinfo {author} {\bibfnamefont {L.}~\bibnamefont
  {Landau}}\ and\ \bibinfo {author} {\bibfnamefont {E.}~\bibnamefont
  {Lifshitz}},\ }\enquote {\bibinfo {title} {Quantum mechanics non-relativistic
  theory, third edition: Volume 3 (course of theoretical physics) (vol. 3) 3rd
  edition},}\ \ (\bibinfo  {publisher} {Pergamon Press},\ \bibinfo {address}
  {NY},\ \bibinfo {year} {1981})\ pp.\ \bibinfo {pages} {1--673}\BibitemShut
  {NoStop}%
\bibitem [{\citenamefont {Voronina}\ \emph
  {et~al.}(2017{\natexlab{a}})\citenamefont {Voronina}, \citenamefont
  {Davydov},\ and\ \citenamefont {Sveshnikov}}]{Sveshnikov2017}%
  \BibitemOpen
  \bibfield  {author} {\bibinfo {author} {\bibfnamefont {Y.}~\bibnamefont
  {Voronina}}, \bibinfo {author} {\bibfnamefont {A.}~\bibnamefont {Davydov}}, \
  and\ \bibinfo {author} {\bibfnamefont {K.}~\bibnamefont {Sveshnikov}},\
  }\href {\doibase 10.1134/S004057791711006X} {\bibfield  {journal} {\bibinfo
  {journal} {Theor. Math. Phys.}\ }\textbf {\bibinfo {volume} {193}},\ \bibinfo
  {pages} {1647} (\bibinfo {year} {2017}{\natexlab{a}})}\BibitemShut {NoStop}%
\bibitem [{\citenamefont {Voronina}\ \emph
  {et~al.}(2017{\natexlab{b}})\citenamefont {Voronina}, \citenamefont
  {Davydov},\ and\ \citenamefont {Sveshnikov}}]{Voronina2017}%
  \BibitemOpen
  \bibfield  {author} {\bibinfo {author} {\bibfnamefont {Y.}~\bibnamefont
  {Voronina}}, \bibinfo {author} {\bibfnamefont {A.}~\bibnamefont {Davydov}}, \
  and\ \bibinfo {author} {\bibfnamefont {K.}~\bibnamefont {Sveshnikov}},\
  }\href {\doibase 10.1134/S1547477117050144} {\bibfield  {journal} {\bibinfo
  {journal} {Phys. Part. Nucl. Lett.}\ }\textbf {\bibinfo {volume} {14}},\
  \bibinfo {pages} {698 } (\bibinfo {year} {2017}{\natexlab{b}})}\BibitemShut
  {NoStop}%
\end{thebibliography}%
\end{document}